\documentstyle[12pt,epsf,styles]{article}
\def\theequation{\thesection.\arabic{equation}}

\makeglossary
\makeindex
\begin{document}
\title{ Quantum Chrono-Topology of Nuclear and Sub-Nuclear Reactions}
\vspace{4cm}
\author{\\
\\
C. Syros and  C. Schulz-Mirbach$ ^\ast$\\
University of Patras \\
Laboratory of Nuclear Technology\\
P.O. Box 1418\\
261 10 Patras, Greece \\
email: C.Syros@upatras.gr \\
 \\
$ ^\ast$ 
Arbeitsbereich Hochfrequenztechnik \\
Technische Universit\"at Hamburg-Harburg \\
D-21071 Hamburg-Harburg, Germany\\
email: c.schulz-mirbach@tu-harburg.d400.de 
\\
\\
\normalsize Presented at the \\
7th Symposium of the Hellenic Nuclear Physics Society\\
Athens  24 - 25 May 1996.
}
\date{ }
\maketitle
\thispagestyle{empty}

\begin{abstract}
A quantum time topological space is developed and applied to solve some problems
about quantum theory.
It is disconnected and satifies the separation axioms of ${\cal T}_4$. The
degree of disconnectedness of the time-space is a decreasing function of the
number of simultaneous or almost simultaneous  fundamental interactions.
The disconnectedness of the $\kappa\times  \lambda_{\kappa}$-fold time-space,
${\cal T}_4^{\kappa\lambda_{\kappa}}$, imparts a quantization to
$\kappa \times  \lambda_{\kappa}$-fold space-time,
$\bar{M}^4_{\kappa\lambda_{\kappa}}$,  and induces it's topology.
In  this topology the ${\bf U}+ {\bf R}$ Penrose dynamics is implemented
by means of a time evolution operator, ${\cal U}$, in QFT. $\cal U$ is
unitary or non-unitary,
depending on the type of quantization of the field action-integral.
${\cal U}$ allows to find the Boltzmann factor in QFT in
the above space-time.
From an elementary solution of the Liouville equation  the quantization of
the time follows and the  Planck constant  has been calculated.
Compatibility between time-reversal and irreversibility is spontaneously 
obtained.
The renormalization of the field action-integral follows from quantization.
The solution of the  measurement problem and the wave function reduction 
have been deduced in the framework of the Schroedinger theory.
The Schroedinger cat's paradoxon and the paradoxon of the
wave paket decay have been resolved.
\end{abstract}
\newpage
 
\pagestyle{plain}
\abovedisplayskip16pt
\belowdisplayskip16pt
\abovedisplayshortskip0pt
\belowdisplayshortskip16pt
\pagenumbering{arabic}
\tableofcontents
\setcounter{equation}{0}
\setcounter{table}{0}
\setcounter{figure}{0}
\newpage

\section{Introduction}

   Indications that systems of atomic, nuclear  and sub-nuclear  particles  
cannot have the topology of the Newtonian time ($\Rset^1$) were available as early 
as in 1974 \cite{1}.   
It became clear that quanta  need not  take  notice of  any observers 
except  interactions with other quanta or particles. 
What is time for the quanta?   What is its nature?  
Why should quanta  take notice of  the time defined  by observers
of their behavior?   
If  at  all, should not  particles  have their  own  times?   
The question about the time is a very old one. 
Nevertheless, a definitive answer has not been obtained  sofar.

The increasing number of paradoxes in theoretical physics  generally 
and in nuclear and sub-nuclear theory in particular, while the 
experimental quantum techniques become more and more sophisticated and 
of higher accuracy, impo\-ses increasingly the view that something  
about  the fundamental physical con\-cepts should  be revised. 

   A systematic examination showed that most uncertainty in physics 
is associated with  the  time concept.   
This variable, being  interwoven with  the space through relativity,
imposes  its topology to the space-time, and  it determines, 
in this way, the evolution  in the nuclear and sub-nuclear 
interactions among others.

   We propose a  new space-time topology, the Chrono-topology.  
It is based on  the concept of the  Interaction Proper-time Neighbourhood,
$\tau$, (IPN). The space-time topology on the quantum level is 
determined by the number of the inter\-acting  particles  in every 
particular  system. For small numbers of  interacting particles the 
new time-space is a ${\cal T}_4$ topological space.   

The new space-times,  $\bar{M^4_{\kappa\lambda}}$ being in general 
$\kappa \times \lambda$-fold   in time, are defined 
as the Cartesian products, $\Rset^3\times{\cal T}^{\kappa\lambda}_4$, of 
${\cal T}^{\kappa\lambda}_4$. 
The latter is a $\kappa\times\lambda$-fold, 
discon\-nected topological space satisfying the separation axioms of a 
${\cal T}_4$  space whose elements are the interaction 
proper-time neighbourhoods, $\{\tau_\lambda |\forall \lambda \in \Zset^+\}$.

Our  $\bar{M^4}$ is not  related to Hawking's space-time foam \cite{23} neither 
as to the cell magnitude nor as to its creation process. 
While Hawking explains, as does Wheeler \cite{wheeler}
the space-time foam creation  by means of 
the field fluctuations in Planck  time scale, our $\bar{M^4}$  is due to 
changes of physical observables caused by fundamental interactions and 
mapped into a definite IPN  in each one case.

Although we do not discuss General Relativity in this work, we feel, 
knowing the results of our chrono-topology, that General Relativity, 
being based on the
Newtonian time topology, is {\it per construction} a non-quantizable theory, 
and that  a reformulation of the field equations in the framework of 
chrono-topology may lead to a quantum theory of space-time whose average will 
give for macroscopic space-time neighorhoods the Einstein field equations of 
gravity.

\subsection{Hunting the trace of the time}

Many famous authors have been occupied with the answering the question 
about the nature of  time as:  Aristotle \cite{2}, Newton \cite{3}, 
Kant \cite{4}, Bergson \cite{5}, and many others.  
The  searching  for  the  meaning of the time by the above 
Researchers and Philosophers was rather of  a  knowledge  
theoretical character, such that no direct physical judgement - except 
a logical one - of the practical applicability to modern problems in physics 
was possible. Also, Eddington \cite{6}, Whitehead \cite{7}, Einstein
\cite{8}, Dirac \cite{9}, 
and particularly Prigogine \cite{10}, Wheeler   \cite{wheeler} and others  
have  shown a deep concern  in the elucidation of the nature and properties of time.
Their search for the meaning of time  was  of  such  a  character  
that  the results obtained on the basis of its structure were accessible 
to a certain extent to a kind of physical verification.  

   It is extremely interesting to verify, after a debate of long decades,
Einstein's terrifically strong insight: 
Now we know  that, in fact, ''Gott wuerfelt nicht'' - God does not play 
dice in matters of quantum theory. 

   It will be shown that quantum mechanics is, in fact, {\it per se } not a 
statistical theo\-ry inside an IPN. 
The statistical character is imposed on the wave function by the topology of 
the space-time $\bar{M^4_{\kappa \lambda_{\kappa}}}$, 
not by  the Minkowski space-time,
$M^4$.  

Meanwhile, new problems appeared mainly in theoretical physics which are 
not solvable in the frame of the current understanding of time's nature. 
Bell \cite{12}, Hawking \cite{13}, Penrose \cite{14}, Unruh \cite{15},
Stamp \cite{16}, Legget \cite{17}, Douglas \cite{18}  have published 
important  works on this area. 
Nevertheless, the time issue remained still open.

The beautiful researches  of all above and many other authors \cite{19}-\cite{30}  are 
only  a  very  small sample of the world literature on time's nature. 
However, there still exist very seriously resisting problems in particular 
in quantum theory which make this issue central to the atomic, to the 
nuclear and to the elementary particles  the\-oretical physics \cite{31}-
\cite{40}.

\subsection{Annoying  questions}

   Spectacularly successful results have been achieved in these areas of  
physics  during our century, and a high degree of maturity both 
in  experiment and in theory. Nevertheless, the  nature of time was 
still unclear  and very important questions remained open:
\begin{enumerate}

\item   
  Can we understand the wave packet's decay in absence of interactions.
\item 
  Can we understand the reduction of the wave function in the framework 
  of the Schroedinger equation?
\item 
  Can we derive rigorously quantum statistical 
   mechanics (QSM), in particular, the 
  Boltzmann factor  from quantum field theory (QFT)?
\item  
  Can we explain the microscopic and the macroscopic irreversibility 
  of many phenomena starting   from QFT?
\item 
    Why is there a tunnel effect?
\item 
  Why is there an ergodicity?
\item   Why is there a Poincare returning.
\item
   Why quarks are not directly observable? 
\item
  Have the non-locality, related to the Bell theory \cite{41}, and the 
  interpretation of the Aspect et al.  experiment \cite{42}
  to do with the topology of the time?, etc, etc, $\ldots$.  
\end{enumerate}

   After the discovery of Einstein's relativity it became clear by 
the Lorentz transformation that time and space are  interwoven in  
the  Minkowski  space-time, $ M^4$. 
 Another  important  recognition provided by relativity was that each 3-space 
point is  associated with its own time (event), the proper-time. 

Accordingly, one would expect that these facts should, normally, impose 
the replacement in modern physics of the  universal Newtonian 
time by the new Einsteinian time. 
This would  make justice to Dirac's early proposal \cite{9} that 
every particle in the many-particle Schroedinger equation should have 
its own time  variable. Dirac's proposal, being related to the topology 
  of the space-time has not yet found the place in physics which it deserves. 

To shed some light on a few  aspects  of  the  above problems of 
paramount importance for the future physics  is  the hope and the 
purpose of  the present work. 

\subsection{The mighty Newton}

It is important to note that, whichever is the topology adopted 
for the time, a transformation like

\begin{equation}
\{ x' = L_x(x,t), t'=L_t(x,t)\}, \:x\in X\subset \Rset^1
\end{equation}
induces on the Minkowski space, the Cartesian  product space
$M^4 = i\Rset^1  \times \Rset^3$, the topology of that time. 

In the same way the  Lorentz  transformation $ (x,t)\to (x',t')$ 
induces the topology of the Newtonian time, $t$, on each space point, $x'$, in 
the neighbourhood  associated  with  that  time, $t$, and  space point, $x$.  

  Space-time topologies resulting from solutions of the Einstein field 
equations will be mentioned  here only occasionally. However, we cannot 
tacitly bypass the fact that in general relativity the proper-time is a 
function of the Newtonian time. For example, in the Schwartzschild metric
\begin{equation}
ds^2 = c^2 (1-\frac{r_g}{r})dt^2 - r^2 (\sin\theta d\phi + d\theta ^2) -
\frac{dr^2}{1-\frac{r_g}{r}},
\end{equation}
the time variable takes values as $t \in T = \Rset^1$  \cite{44}.   

Also in the general relativity, for example \cite{45}, one reads: ''Any 
monotonic pa\-rameter, increasing from the past to the future 
(i.e. $t\in [-\infty,+\infty]$) 
might be used to measure time on the world line of a material particle''. 
This is  clearly correct for a macroscopic theory. Is it correct for 
the discontinuous quantum phenomena? 

  Nevertheless, this attitude reflects the view of some researchers according 
to which time had nothing to do with the  fundamental interactions and with the 
changes induced by them in the different neighbourhoods of the universe. 

This attitude is rejected in the present work.

\subsection{Quanta and their self-determination - The interaction 
    proper-time neighbourhood (IPN)}

In view of these facts one may speculate, if not reasonably  conjecture, 
that the well-known paradoxes in relativity and quantum theory, as well 
as the possibility for their appearance in physical theories are due to 
the space-time topology imposed by the Newtonian time. 

Despite the obvious necessity to replace the Newtonian  time and its topology 
by IPNs  to be defined precisely below for  each  event, the Newtonian 
time remaines until today generally  dominant  in classical and in 
quantum theory. 

The present paper is dealing with the derivation of some consequences  of  a  
new type of time topology discovered earlier \cite{1} and taken as 
the basis in this work.  
{\it 
The new topology derives from the fundamental observation teaching us 
that no time would be definable, if nothing changed in the universe} \cite{47}.

Since the universe for a non-interacting, structureless particle is the 
particle itself, no time   exists for it. 
Moreover, since the nuclear and sub-nuclear interactions factually are, each 
one, of finite duration, i.e., they are related to finite changes 
of the observables involved in the interaction, it is clear that  IPN 
cannot be identified with the Newtonian time. 
Because the latter is homeomorphic to the whole $\Rset^1$, while IPN $\in T 
\subset \Rset^1$, and $ T_4$ is disconnected. 

It is also important to observe, that  the time for, e.g., a nucleon  
is related to its corresponding interaction, and it does change as long as  
the  interaction  lasts.  Just this time is used in our work, in the 
equations of Schroedinger, of Dirac and of QFT in connection with 
problems of nuclear and sub-nuclear interactions. 
This time can flow within the corresponding IPN as long as the  
interaction is going on in the rest reference 
frame of the interacting particle.

  On the contrary, for an observer the reaction time ($t'$) may, but must not, 
flow further, depending, according to Lorentz transformation above,
on whether  he changes its  position ($x'$) or not with respect to the rest 
frame of the particle.
This stresses the importance of the interaction for the changes in any system.                           

The nucleon reaction time, for example, cannot be  identified  with the 
universal time which consists, according to our chrono-topology of 
the union of the maps of all individual interactions  occuring  in  
the entire observable universe. 

 On the other hand, the free-field quantum equations of physics,
mathematically so instructive, are   relevant physically only as 
approximations to real phenomena. 
Time-dependent quantum equations without interactions do not 
supply us with any information concerning physical changes.

   We must stress, however, that  macroscopic motion, and in  particular  
inertial motions, are correctly expressed either in terms of the Newtonian 
time or in terms of unions of large numbers of {\it IPN}s ($\bigcup_{
\kappa,\lambda\in \Zset^+} \tau_{\kappa\lambda}$), deriving 
from interactions in the observable neighbourhood of the universe.

   It seems that the way to pave for general relativity towards quantization 
is to redefine the space-time  topology  by taking  into account  
the  topology of  the IPNs  and to reformulate the field equations 
in the topology of $t\in{\cal T}_4$. 
In such a case the quantization of the theory can  most easily be 
carried out by means the field action-integral quantization.
Before going into the detailed presentation of our  chrono-topology, we want  
to  review some important time topologies used in the past or proposed 
recently to describe the phenomena  or to explain  the interpretational  
problems  in quantum theory. 

   To make precise the description and to facilitate  the  understanding, it 
is expedient  to  give first some notation and some  definitions  from 
general topology which are required for the presentation  of  the results. 

It should by no means be understood as  a  substitute for a reading of 
a book on general topology  which  is recommended to the more 
interested reader. 

\subsection{Time-space topology in physics}

Let a set $\cal T$, called the space, be given together with a family 
$\{\tau\}$ of subsets $\tau\subseteq {\cal T}$ and  together with the 
empty set $\emptyset$. 
The elements  of  $\cal T$  are called points of the space and the 
elements are called  open sets. 

\begin{defi}
A pair  $( {\cal T},\tau)$  of  $\cal T$  and $\tau$ represents 
a topological space, if  the following conditions are satisfied \cite{46}: 
\begin{itemize}
\item $\emptyset \in \tau $ and $\cal T$ $\in \tau$.
\item If $U_1 \in \tau $ and $U_2 \in \tau $, then $U_1 \cap U_2 \in \tau$.
\item If  ${\cal  A} = \{A_1 ,A_2,\ldots \}$ is a family of elements of $\tau$
 and $I$ is  a subset of the index set $\Zset^+$  such  that $A_i\in \tau,
\forall i\in I,$ then $\cup_{i\in I} A_i\in \tau$.
\end{itemize}
\end{defi}

It is clear that the intersection $\cap_i A_i$ of a finite subset $\{
A_i  i\in I\subseteq  \Zset^+\}$ of open subsets  is open.

\begin{defi}
A space, $\cal T$, is called regular if  and  only  if  for every 
$x\in {\cal T}$  and every neighbourhood  $\cal V$ of $x$ in a  fixed  
subbase  $\cal P$ there exists a neighbourhood $U$ of $x$ such that ${\cal U} 
\subset {\cal  V}$, 
where $\ cal U$ is the closure of $U$. 
\end{defi}

The topological spaces may be ordered in a hierarchy according to the 
restrictions which are imposed on them. These  restrictions are called 
{\it axioms of separation}. 
Here are the axioms of separation concerning the fundamental interactions 
in physics: 

\begin{defi}
\begin{enumerate}
\item
  A topological space, $\cal T$, is called a ${\cal T}_0$-space, 
if for every pair of distinct points $t_1$, $t_2 \in {\cal T}$  there exists 
an open $\tau '$  containing exactly one of these points.
\item 
 A topological space, $\cal T$, is called a ${\cal T}_1$-space, if for every 
pair of distinct points $t_1, t_2 \in {\cal T}$ 
there exists an open  $\tau '\subset {\cal T}$  such that either 
$t_1\in \tau ', t_2 \not\in \tau '$ or $t_1 \not\in \tau ', t_2 \in \tau '$.
\item 
  A topological space, $\cal T$, is called a  ${\cal T}_2$-space, or  a  
Hausdorff space, if for every pair of distinct points $t_1,t_2\in {\cal T}$ 
there  exist open sets $\tau_1, \tau_2\subset{\cal T}$ such that 
$t_1 \in \tau, t_2\in \tau_2$ and $\tau_1 \cap \tau_2 = \emptyset$.
 \item 
  A topological space, $\cal T$, is called a ${\cal T}_3$-space or a 
  regular space, if it is a ${\cal T}_1$-space and for every  $t\in \tau$ 
  and for every closed set ${\cal F} \in {\cal T}$ such that $t\not\in \cal F$ 
  there exist open sets $\tau_1, \tau_2$  such that $t\in\tau_2, {\cal F} \in
  \tau_2$ and $\tau_2 \cap \tau_2 = \emptyset$. 

\item 
 A topological space, $\cal T$, is called a ${\cal T}_4$-space or a 
  normal space, if ${\cal T}$ is a ${\cal T}_1$-space and for every pair 
 of disjoint closed subsets $\tau_1 \subset U, \tau_2 \in V$ 
         and $U\cap V=\emptyset$.
Clearly, a ${\cal T}_4$-space is a ${\cal T}_3$-space so that the hierarchy 
 holds:
$$ {\cal T}_0 \Rightarrow {\cal T}_1 \Rightarrow {\cal T}_2 
\Rightarrow {\cal T}_3  \Rightarrow {\cal T}_4 .$$
\end{enumerate}
\end{defi}

\subsection{Nature's choice }

There are still  the  axioms  of  separation  for  the  spaces 
${\cal T}_{3 \: 1/2}, {\cal T}_5, {\cal T}_6$ 
whose definitions are not given here. 
The  topology of the  time  considered  in this paper  is just that 
of  ${\cal T}_4$ Fig. (\ref{fig_3_1}).  
This is the time topology generated by 
distinct, finite interactions.

This paper is divided in 12 sections. 
In  Sec. \ref{section_2} a few time topologies are presented and discussed 
from the point of view of their relevance to the physics of nuclear 
and sub-nuclear particle systems. 

In Sec. \ref{section_3} some aspects  are briefly presented of the relation of the 
time to the energy changes. 
Although this relation looks rather trivial it is, nevertheless,
the basis for the chrono-topology developed and  whose main definitions are 
given in this section. The chrono-topology allows to solve some of 
the paradoxes of the quantum theory. For example, the puzzle of the 
flowing time. In our chrono-topology an answer to the question is
possible, why time is felt as flowing. 
The feeling of the flowing time is generated by the disconnectedness 
as an application of Zermolo's theorem on well-ordering, 
and by the limited discrimination power of the human neural sensors. 
If time were continuous, then a succession of discrete sensations would not
exist and, consequently, an ordering, generating the feeling of flowing, on the
biological level,would be impossible.

   In Sec. \ref{section_4}  we continue with the further presentation of  
the structure properties of the present time topology. 
We define there the microscopic and the macroscopic system-times.  
They are important in integrations of the S-matrix and of the time 
evolution operator. 

    An extremely surprising fact is that the reality and the additivity 
conditions of some elementary  solutions of  the classical Liouville equation  
imply quantization of the time. Although the conditions of the 
derivation (constant  forces) of these elementary solutions 
are rather special, we cannot overlook that the result is typical.  
This is presented  in Sec. \ref{section_5}.
An interesting {\it novum} is the calculation of Planck's constant, 
$\hbar \approx 1\cdot 10^{-34} Js$ from the expression for the time
quantization.

   It is a clear experimental fact that physical interactions imply finite 
changes in observables. The time as a map of these changes must also be 
created in finite amounts. 
It is surprising that this fact has for long times escaped our attentions. 
If  the observer moves with respect to the interacting elementary particles,
while they interact, the time and the space coordinates appear to him as 
quantized.  Examples are shown in Sec. \ref{section_6}.

   Sec. \ref{section_7} makes exploitation of the chrono-topology. A fundamental
proposition is demonstrated therein.  

 The main results obtained with the help of  this section are as follows:
\begin{enumerate}

\item A time evolution operator is obtained which, in Penrose's 
      terminology, exhibits ${\bf U}+{\bf R}$ properties. 
      It is unitary or non unitary, depending on the kind of quantization of 
      the field action-integral.
\item The Boltzmann factor, $\exp[-E\delta(\tau)/\hbar]$  is derived 
      directly from QFT. This is  tantamount  to  the derivation  of  
      QSM  from  QFT in Minkowski's  space, $\bar{M}^4_1 \subset M^4$.
\item The quantization of the field action-integral  spontaneously  
      renormalizes  the time integration of the interaction Hamiltonian.
\item The renormalization of the action yields the possibility
  (Sec. \ref{section_8}) 
      for a natural explanation of the wave function  reduction  in  
      the framework of the Schroedinger equation. The existence of microscopic 
      and of macroscopic irreversibility in the framework of QFT has 
      been demonstrated.
\item The famous  Schroedinger's cat is, finally, dead. 
      Not because of the poison and the radioactivity, but simply 
      because he always was alive before he died. This is the result 
      of Sec. \ref{section_9}.
\item The wave packet, as anything else, cannot evolve in absence of an 
      interaction (Sec. \ref{section_10}).  It can decay only  inside  the IPN, and 
      this  is of  finite duration.
\end{enumerate}
  As a byproduct, Einstein's spectacular insight and insistence along his 
  life-time,  according  to  which  quantum  theory {\it per se } is  not 
a  statistical theory, follows spontaneously. We find that the  
statistical character (Born's hypothesis  about  the wave function) comes  
about only in the framework of the chrono-topology of the fundamental 
interactions.

    Finally, in Sec. \ref{section_11}  
the results are discussed an the perspectives  
for further developments are sketched.

\setcounter{equation}{0}
\setcounter{table}{0}
\setcounter{figure}{0}
\section{Classical definitions of time and their topological structures}
\label{section_2}

If the entire universe consisted  of  one  single, structurless particle, e.g., 
an electron, then the idea of time would be for a 'foreign' observer
neither  definable  nor  useful.  
Motion would be, on  the basis  of our  familiar  physical 
criteria, unobservable and meaningless. The particle would be describable by 
its intrinsic characteristics, mass, spin, charge, etc., and  no
change whatsoever would be possible. 
In particular no change of the particle energy would 
be possible. 

If  the entire universe consisted  of non-interacting structureless particles,
then  the idea of time would again  be undefinable and  the motion, if any,
would be  unobservable  by  an  observer in  the frame of  reference of any  
particle (due to the absence of quanta created and emitted by means of any 
interactions). 

If the particles do interact, then  messages  between  them 
conveying physical characteristics 
exist, and a new parameter is required for the description of their 
changes. By mapping the changes in particular subsets of $\Rset^1$  we get a particular parameter for 
each IPN, $\tau$. 
However, interaction means change of values of the physical observables
and exchange of parts of them 
between  the interacting particles. Even in the simplest 
form of interaction, in the elastic scattering, a change does occur 
in the linear momentum.
Moreover, transfer of physical characteristics implies in any case energy changes 
inside the universe of the particles. Consequently, it appears that 
associated with 
any energy  change is a time laps. 
This association has not the character of a causal relationship.This becomes clear 
from the fact that, if no  description of the phenomenon is desired, 
then there is no 
need for a time variable to be defined. 
Conversely, it is empirically clear that no time laps is observed,
if no energy change - 
and more generally - no physical change takes place.

\subsection{The Aristotelian time} 

Historically, the first and most extensive (15 pages) 
scientific discussion on the 
nature of time is published by Aristotle in his book $\Phi\upsilon\sigma\iota\kappa\alpha$ \cite{2}.
Aristotle considered the time as a  set  of '$\nu\upsilon\nu$' (now).This set of '$\nu\upsilon\nu$' may 
be defined by anybody, anywhere, anyhow  and  at  any  time  relative to other 
people's  '$\nu\upsilon\nu$' . Consequently, between two  '$\nu\upsilon\nu$' there may exist any number 
of other peoples' $\nu\upsilon\nu$' . The union of  '$\nu\upsilon\nu$' is dense in a subset of $\Rset^1$ . Each 
'$\nu\upsilon\nu$'  as a feeling or as a product of thinking  is an
open interval, $\tau$.

More precisely, let $\Rset^1$ be the time axis and  $\cal O$ the family  of  all sets $\tau\in\Rset^1$  with 
the property that for every $t\in\tau$ there exists  an $\epsilon>0$, 
such that $(t-\epsilon, t+\epsilon)\in\tau$.  
The family, $\cal O$, of  sets has the properties:

\begin{itemize}
\item[O1)] $\emptyset\in\tau$ and $\Rset^1\in{\cal O}$.
\item[O2)]   If $\tau_1\in{\cal O}$ and $\tau_2\in{\cal O}$ then $\tau_1\cap\tau_2\in{\cal O}$.
\item[O3)]   If  $A\in{\cal O}$ then $\cup A\in{\cal O}$.
\end{itemize}

This makes clear that Aristotle conceived time  as  a  set of open
intervals, because in $\tau$, as in $\Rset^1$,
between any two points there exist infinitely many points ( Hausdorff). 
The topology of the Aristotelian time is the natural topology of $\Rset^1$ .

The similarity of this time topology to the topology of the Newtonian  time is 
obvious. It is remarkable that the Aristotelian time does not have the dynamics 
of a flowing, because the '$\nu\upsilon\nu$' is static.

\subsection{ The Newtonian time}

The best known time in physics is the Newtonian universal time. It finds still 
today use generally in science and in particular in relativity and in quantum 
theory. 

The  topological propeties of this time structure may be summarized in that
$t$ is (as in the Aristotelian time topology)  a continuous function
with the topology of 
$\Rset^1, \:t\in [-\infty , +\infty ]$ .  Although the topology of the Newtonian time is that 
of $\Rset^1$, one assignes to it an additional characteristic property. 
It consists in that 
time incorporates the germ of dynamics in the  most general sense: It
is considered in physics as continuously {\it flowing} \cite{48}. 

However, there is no indication experimental  or  theoretical,
that this flowing is 
real like, for example, that of a fluid. On the contrary,
there exist indications that  
the flowing of time is a property subject to the  anthropic principle. The time is, according to 
relativity, not more and not less flowing than space itself. 

The admission to quantum theory of the never demonstrated continuous
flowing of time having the 
properties of the topology of $\Rset^1$, is the source of a number of
paradoxes 
and interpretational problems in quantum physics. 
One most  prominent paradoxe is the decay of the wave packet in time, which in 
the absence of interactions cannot be explained. Another puzzle for
contemporary 
physics is the time reversal invariance of the fundamental equations of physics 
and the simultaneous irreversibility of almost all phenomena in the macrocosmos.

Having in mind our ${\cal T}_4$-topology  of the time for quantum systems it is easy to 
contemplate the way the Newtonian time is generated:
Let us consider two point sets $P_1$  and $P_2$  such that $P_1, P_2\in\Rset$ 
and $P_1\cap P_2 = \emptyset$.
These two sets are for the human senses distinct. Consider next large numbers of 
sets $\{ P_j\mid j\in I'\in\Zset^+\}$, so that for many pairs of sets
we observe $P_j\cap P_{j+1}\not= \emptyset$, but 
there are still some for which  $P_i\cap P_{i+1}= \emptyset$. 
In this case the human senses still
see some gaps  in the union   $\cup_{j\in I'}P_j$.

If we take a still larger number $I$ of sets $\{ P_j\mid j\in I\supset I'\}$
 such that there exists no partition 
$\{ P_i'\mid P_i' \cap  P_j' = \emptyset, \forall i\in I\}$ 
of $\cup_{j\in I}P_j$, then $P_i\cap P_{i+1} \not= \emptyset, \forall i\in I$.

If we now identify the collection of the maps of the changes of all 
physical observables 
with the union $\cup_{j\in I}P_j$, then this union can be densely embedded in 
$\Rset^1$ and the 
continuity of  the Newtonian time emerges.
    
\begin{defi}
\label{def_2_1}
We call Newtonian or universal time, $N_t$, the union
$$
N_t = \bigcup_{j\in I}P_j \;\mbox{with}\; P_i\bigcap P_{i+1}\not=
\emptyset \;\mbox{and}\; P_i\setminus P_{i+1}\not= \emptyset
\forall i\in \Zset^+. 
$$
\end{defi}

{\bf Remark 2.1}

If  $N_t = [-\infty , +\infty]$ then we write $N_t\subseteq \Rset^1$.

Historically, the idea of the continuous time emerged from the
biological structure 
of the human senses and from the empirical laws of mechanics of the
macrocosmos. 
By this it is meant that the observers have the ability to a certain extent 
to distinguish adjacent but distinct changes of physical observables. This ability,
however, has definite limits valid for all human observers and the
subjective continuity 
of the time becomes conscious.

{\bf Remark 2.2}

It is a very typical feature of our chrono-topology the fact 
that time is felt as 
flowing. This is due to the disconnectedness, to the Zermelo theorem on
well-ordering and 
to the discrimination by  the neural sensors. If time were 
continuous, then a succession of discrete sensations would not exist
and, consequently, 
an order generating the feeling of flowing,  on the biological level, 
would be impossible.

\subsection{The Einsteinian time}
 
The Einsteinian time is essentially identical to the Newtonian time. The 
usual view
is that every space-time point (event) is assigned its own proper-time. This is strictly speaking 
not true for the proper-time,

$$
d\tau = \sqrt{c^2dt^2 - dx^2 - dy^2 - dz^2}
$$
                                         
or

$$
\tau - \tau_0 = \sqrt{c^2(t-t_0)^2 - (x-x_0)^2 - (y-y_0)^2 - (z-z_0)^2 },
$$
                  
where $t\in\Rset^1$ is the Newtonian universal time.

\begin{figure}[htbp]
\begin{center}
\leavevmode
\epsfxsize=0.5\textwidth
\epsffile{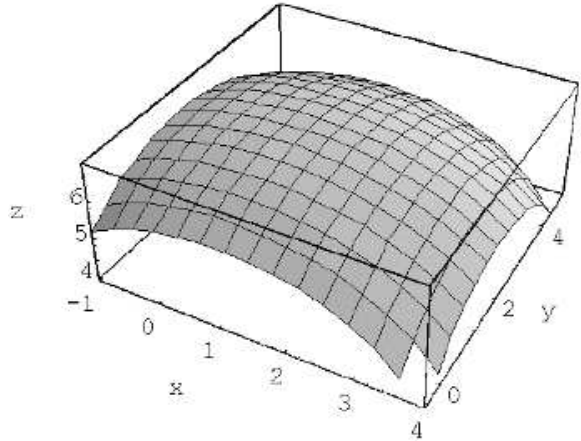}
\end{center}
\caption{A  sector of a  2-dimensional manifold embedded in Minkowski's 
space-time.
All points of the surface have the same proper-time $\tau =4, \tau_0 =
3,  t = 5, t_0 = 1,
x_0/c=1, y_0/c = 2$ and $z_0/c=3$.
Obviously all events on the surface correspond to
the same Newtonian time.}
\label{Fig.2.1}
\end{figure}

\begin{figure}[htbp]
\begin{center}
\leavevmode
\epsfxsize=0.5\textwidth
\epsffile{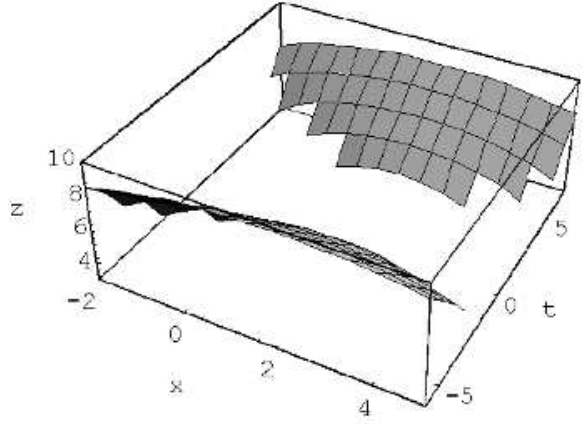}
\end{center}
\caption{Two sectors   of a 2-dimensional  manifold  embedded  in  Minkowski's 
space-time having the same proper-time $\tau = 4$. Here the variables
are $x, z, t\cdot y =2$,
while the constants are  $\tau_0 = 3, t_0 = 1, x_0/c = 2, y_0/c = 1,
z_0/c = 3$.  All
events of the two sectors correspond to the same proper time, $\tau$,  
and $y$ values.}
\label{Fig.2.2}
\end{figure}

It is seen from this very expression for the proper-time that the same proper-time  
$\tau$ corresponds to all space-time points $(t,x,y,z)$  
of the manifold (Figs. \ref{Fig.2.1}, \ref{Fig.2.2}), defined 
by 

$$
\tau = \sqrt{(t-t_0)^2 - ((x-x_0)^2 - (y-y_0)^2 - (z-z_0)^2)/2} = \mbox{const}. 
$$

The source of the Einsteinian time in special relativity is the classical equation 
of uniform motion \cite{44}

\begin{equation}
s = v t \quad\mbox{or}\quad c^2 t^2 - x^2 - y^2 - z^2 = 0, \; (m_0^2 = 0), 
\label{2.1}                  
\end{equation}

where $c$ is the velocity of  light. Einstein in writing down the
 above equation  giving 
the propagation distance, $s$, of a light wave front has nothing stated
about  the magnitude of $s$,
or the length of the time interval, $[0,t]$. 

The minimum operationally measurable time in connection with the  light  wave 
propagation is the time corresponding to the distance of one wave
length, $\lambda$. This time, $\tau$, is 
related to the wave length by $\tau = \lambda / c = 1/f$, where $f$ is the wave frequency. 
The time $\tau$ stands  in a simple relation with the interaction duration
causing the energy 
change, $\Delta E$, and with the emission of a corresponding photon.
More generally, 
the de Broglie wave length of an emitted particle, e.g., of an electron in 
$\beta$-decay  is  
related  to  the energy released during the emission interaction. 

For the $\beta$-particle of rest mass, $m_0$ , which, of course, does
not move on the light cone, there holds

\begin{equation}
c^2t^2 - x^2 - y^2 - z^2 = s'^2  > 0 \;\mbox{or}\; c^2   - v^2 = s
>0, \;s = s'/t.
\label{2.2}
\end{equation}

Multiplying both sides by $c^2\gamma^2 m_0^2$, where $v$ is the
velocity of a quantum, one gets 

\begin{equation}
c^4\gamma^2 m_0^2 = \gamma^2 v^2c^2m_0^2 + c^2\gamma^2s^2m_0^2.
\label{2.3}
\end{equation}

   The term on the lhs  represents the total energy of the particle, 
whilst the first 
term on the rhs is equal to $c^2 p^2$ . 
By introducing the definitions of $\gamma = (1-\beta^2)^{-1/2}, \beta = v/c$ 
and of $s$ the energy-momentum relation emerges,  

\begin{equation}                                      
  E^2  = c^2 p^2 + c^4  m_0^2.                                            
\label{2.4}
\end{equation}

This may be considered as a clear indication that a quntization of the general rela-
tivity  might be obtained by formulating the field equations  inside an IPN,
$\tau$ , \cite{47}. 

Also, it is important to notice thereby that $\tau\in T\subset \Rset^1$. 
In other words the
relativity time, $t$, in Minkowski 
space (as the Cartesian product, $M^4 = i\Rset^1\times  \Rset^3$  
of  the Euclidean 
space, $\Rset^1$ and $i\Rset^3$ ) is homeomorphic 
to the Newtonian time ($ t\in  \Rset^1$ ). 

We shall abandon the view that time, $t$, in relativity during every
fundamental interaction takes 
values from the whole $\Rset^1$. Instead, it will be assumed throughout that 
$t\in\tau$.

\subsection{The Wheeler's foam space-time topology}

   An interesting space-time topolology was proposed by Wheeler \cite{wheeler}
   in the framework of the geometrodynamics, a name coined by
   Einstein.
 Wheeler followed Einstein's vision according to which all quantum phenomena -
waves and particles - should be traced back to geometrical properties
of a superspace-time. This superspace-time is multiply connected and
its topology varies as a consequence of geometrical dynamical quantum
fluctuations.
   The space-time structure envisaged by Wheeler very much resemles
   the topology of our superspace-time, but there are many substantial
   differences. These differences originate from both, the way of
   generating the superspace-time and the constants characterizing its
structure.   

   Since Wheeler does not give a formal characterization of the topology in
   terms of its topological properties, we give 
  in Table \ref{Tabelle2_1} the main properties
   extracted from  \cite{wheeler} in comparison with the corresponding
   properties of our superspace-time.

\begin{table}[htb]
\label{Tabelle2_1}
\caption{Properties of two superspace-times.}
\begin{tabular}{llll}
\hline\\
 n &   Property    &   Wheeler superspace-time  &   Present work \\
\hline\\
 1 &   Origin    &   Quantum     &       Fundamental   \\
   &             &    fluctuations &      interactions\\ \hline \\
 2 &   Symbolic  &       --       &  $({\cal T}^{\kappa 1}_4 \oplus
                                           {\cal T}^{\kappa 2}_4 \oplus
                                  \ldots  {\cal T}^{\kappa \Lambda }_4 ) \times
                                  R^3 $ \\
   & definition  &                 & \\ \hline \\
 3  &  Geometry  &  Neither unique nor &    Every sub-sheet is \\
    &          &     classical. It fluctuates  &   is a sub-manifold of \\
    &          &     everywhere with &   Minkowski's or of \\
    &          &     amplitudes comparable &  Riemann's space-\\
    &          &     to Planck's length between &  time. It is a mul- \\
    &          &   configurations of various & tiply disconnected \\
    &          &    sub-microscopic          &  geometry. \\
    &          &    curvatures and different &  $ \tau_{\kappa\lambda}\times R^3 \subset \bar{M}^4_{\kappa\lambda}$. \\
    &          &     topologies.          &  \\
\hline\\
 4  & Topological  &  The geometries which &   Multiply  disconnected,  \\
    &               &  appear with high pro- & locally Hausdorff. \\
    &               &  babilities are multi-  &  \\
    &               &  ply connected Haus-    &  \\
   &                &  dorff geometries   &  \\
  &                 &  (foam structure).         &  \\
\hline\\
 5 &  Dimensions &  $\quad\quad\infty^{\infty ^3} $ &  $\infty^{3K}\;\forall K \in \Zset^+$\\
\hline\\
 6 &  Time topology &       Newtonian         &  $\quad\quad{\cal T}_4$  \\
\hline\\
\end{tabular}
\end{table}

\subsection{The stochastically branching space-time model} 

   Interesting from the point of view of the present work is the model of the 
stochastically branching space-time discussed by Douglas \cite{18}.  
This model of space-time is inspired by the Many-World  interpretation of quantum 
mechanics \cite{49}. 

The  main features of this time topological space 
$({\cal Y}, t)\subset\Rset^2$  are: 
The space-time is constructed as the  Cartesian product of certain
open or semi-open 
time intervals $W_i (\alpha, \beta ),\; i= 1, 2,\ldots ,5$, where
$(\alpha, \beta )$ are real,and  $\Rset^3$ is the Euclidean 
space.
The sets $W_i$  are defined by:

\begin{eqnarray*}
    W_1(\alpha, \beta ) &=& \{(t,0)\mid \alpha < t < \beta \leq 0\} \\
    W_2(\alpha, \beta ) &=& \{(t,1)\mid   0 \leq\alpha < t < \beta \} \\
    W_3(\alpha, \beta ) &=& \{( ,0)\mid \alpha < t <0\}\cup \{(t,1)\mid
    0\leq t < < \beta \} \\                      
    W_4(\alpha, \beta ) &=& \{(t,-1)\mid  0 \leq \alpha < t < \beta \} \\
    W_5(\alpha, \beta ) &=& \{(t,0)\mid \alpha < t <0 \}\cup
    \{(t,-1)\mid 0\leq t < \beta \}
\end{eqnarray*}

The main properties of this space-time are:
\begin{itemize}
\item[i)]   It is locally Euclidean.
\item[ii)]  It is not Hausdorff.
\item[iii)]  It associates probabilistic properties with the  topology of the space-time.
\item[iv)]  It is based on the Many-World interpretation of quantum mechanics. 
\item[v)]   The time  topology cannot accommodate in general physical 
interactions, because they take place  in the present $t=0$ of  the  
rest frame of reference. 
This point in the model is not uniquely defined.
\end{itemize}

   The above given sets have  topological cuts at  $t=0$.  
more precisely:
 $ W_1(\alpha, \beta )$ represents the lower section of the time (past) 
and it has no largest element.
 $ W_2(\alpha, \beta )$ and  $ W_4(\alpha, \beta )$  represent the upper sections (futures) and they have 
no smallest elements.
$ W_3(\alpha, \beta )$ and $ W_5(\alpha, \beta )$ consist of two lower sections  and two upper sections. 
The lower sections have no largest and the upper sections have smallest elements.

  This time topological space $({\cal Y}, \tau)$ has the structure in the simplest case

\begin{equation}
{\cal Y} = \{(t, 0)\mid t<0\}\cup \{(t, 1)\mid t\geq 0\}\cup 
\{(t, -1)\mid t\geq 0\}\subset  \Rset^2 .
\label{2.6}
\end{equation}                   

\begin{figure}[htbp]
\begin{center}
\leavevmode
\epsfxsize=0.5\textwidth
\epsffile{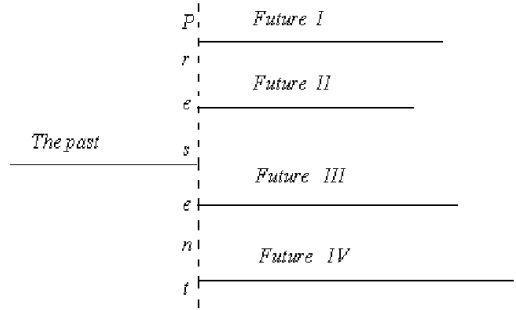}
\end{center}
\caption{The simplest case of Douglas' time-topological space. The
fundamental 
time-dependent differential equations of physics (Schroedinger, Dirac,
quantum field theories) are not unique  at  the 'present, $t=0$. The time 
line 
has a cut at $t=0$, and is for $t = 0 + \epsilon$   multivalued.}
\label{Fig_2_3}
\end{figure}

It is  seen from the above  that the  'present', which is  the only
directly observable 
part of the time consists, of a cut in $\Rset^1$, because the 
past of the model has 
no largest element, and the future  has a  least element $(t=0)$.  
This topological 
structure of the 'present'  makes  ambiguous  the solution of the fundamental 
time-dependent differential equations of physics at $t=0$ 
describing transitions from $t<0$ to $t>0$ (Fig. \ref{Fig_2_3} ).

It should be pointed out that all experimental physical measuring processes 
take place at the 'present' of  any  time topology. Just this point is
not uniquely defined  
in Douglas time topology. On the other hand, of course, measurements in
the past 
time or  in  future states of a given quantum system are, 
in general, impossible. 

The existence of the cut  implies that between any particular open set
$\{(0, t)\mid t\leq -0\}$ and the present, $t=0$, of the open  sets 
$\{(t, n)\mid t\geq 0, n = \pm 1, \pm2, \ldots \}$ 
there must exist  neighbourhoods of time points which do not contain
the 'present'  $t=0$ for particles coming from the past. 

Also, the transition from the 'past' to the 'present', 
at which a future state is 
generated, happens during a time interval of measure equal to zero. This is not 
in agreement with  the experimental  evidence according  to which interactions 
are associated with a finite duration. 

\setcounter{equation}{0}
\setcounter{table}{0}
\setcounter{figure}{0}
\section{Chrono-topology and space-time of the fundamental quantum 
  interactions }
\label{section_3}

We mentioned in various occasions in the foregoing  sections the reasons 
for which the Newtonian universal time must be replaced by the appropriate 
time-space topology. 
We are going to discuss in this section more in depth the time problem and 
its consequences for the quantum processes resulting from the 
$\Rset^1$ topology assumes in the past.

\subsection{Operational meaning of the commutation relations}

   The idea that time is related to  energy changes is not new. 
Already Schroedinger 
and  Pauli considered  the relation  of  the  time with the energy as a direct 
consequence of the commutation relations  
\begin{equation}
     [ x, p  ] = i\hbar I  \delta_{\mu\nu} \: \mu, \nu = 1, 2, 3.                            
\end{equation}

However, the  position  coordinate, $x$, and the conjugate  momentum,
  $ p_x$, are related, despite their independence in the sense of 
the mathematical analysis of the 
phase space mechanics, not just  by  the commutation  relations. 
There  is  still  an  other reciprocal physical relationship:  
The change, $\Delta  x$, of  the  
position variable, $x$, of a particle generates its momentum, $p_x$.
The converse is  also  true. The  change  of  the momentum, $p_x$, 
(or even its mere existence) 
of a particle necessarily implies change, $\Delta x$, of its position, $x$. 
This  mutual 
relationship has not been sufficiently emphasized although its existence 
is quite 
evident.  This relationship  will prove very instructive in the
following considerations about the generation of  time.

   The commutation relation between the energy operator $H$ and the 
time operator $t$ is 
\cite{50}-\cite{51}
\begin{equation}     [t,H] = i\hbar I.                                                    
\end{equation}

The ''generating'' relationship between energy change and time change 
is apparent 
here, as it was in (3.1), for the pair  $(x, p_x)$. 

   In quite a similar way, the change, $\Delta E$, of the energy of a 
particle generates the time 
laps, $\Delta t$, which is appropriate for the description 
of this particular event \cite{1}.  

 A further analogy between (3.1) and (3.2) of great importance for the 
understanding of the nature of the time is the following: 
The result of applying (3.1) 
on a wave function  is  to  describe the creation of a quantum 
pertaining to the particle having the momentum, $p_x$. 

   Similarly, the application of (3.2) on a wave  function creates 
a quantum pertaining to the particle having the  energy $E$. 

   It is appropriate to emphasize that every time neighbourhood pertains to the 
particle in its rest  frame subject to the corresponding interaction. 
It would not 
be  in agreement with relativity, if the same  time  neighbourhood 
would be used 
universally to describe the evolution of other interactions at 
different points of the space. 

\subsection{Change and  time }

By  ''mixing''  the time and the space variables, as it happens in the Lorentz  
transformation,
we do not yet fully eliminate the classical, 
absolute character of  the 
time. Such should  be achieved better by attaching to every one 
act of elementary 
energy changing interaction its own time neighbourhood. It takes values 
exactly as long as the interaction is going on. 

Considering that in a many  particle  system  each  particle's history is  
described by its own set of time neighbourhoods - each one starting and ending 
with the starting and the ending of the corresponding interaction (causing 
associated changes  
in observables of the respective  particle) - it is  not  obvious  
at  first sight, which one of the many ''pieces'' of time 
(which, by the way, clearly may overlap partially or entirely, in the sense 
of the relativistic simultaneity) would  
be  appropriate to descricribe the set of particles as a physical system. 
This difficulty is avoided by introducing  the notion of  the IPN.  In 
conformity with the above ideas we shall prove the following

{\bf Proposition 3.1}

   The changes $(\Delta x',\Delta t')$ of the coordinates $(x',t')$ in 
observer's  moving reference system of an event $(x,t)$ in its rest 
system of reference are linear functions of the 
changes  $(\Delta x, \Delta t)$. 

{\bf Proof}

   Consider the Lorentz transformations:
\begin{equation}
\label{3_3a}
        x' = \gamma (x - vt)                       
\end{equation}
and                                                                                                    
\begin{equation}
\label{3_3b}
       t' = \gamma (t - (\beta /c)x).
\end{equation}
where $\gamma = 1 / \sqrt{(1-\beta^2)}, \beta = v/c.$

   Let  $x = 0 $ in (\ref{3_3a}).  Any change $\Delta t$, of the time $t$,
is a linear function of the change $\Delta x$ of $x$.  
The converse is also true: It follows from (\ref{3_3b}) that the change  
$\Delta t'$, of the time $t'$, 
for $t = 0$ is a linear function of  the change, $\Delta x$, of the 
space variable, $x$, and  vice versa.

Therefore:
\begin{equation}
\label{3.4a}
      \Delta x' = - \gamma  v\Delta t,        
\end{equation}
\begin{equation}
\label{3.4b}
      \Delta t' = - \gamma (\beta   /c)  x    
\end{equation}
and the proof is complete.

{\bf Remark 3.1}

   This obvious and rather trivial result is known to many people since 
almost one century.  However, its special meaning seems to have 
escaped hitherto our attention:  
If we convene to consider the coordinate $x$  as an observable, 
then (\ref{3.4b}) is a 
regular, continuous map of the change of an observable to a linear set, 
the IPN.

\begin{table}[htb]
\caption{Orders of magnitude of the IPNs for QED and QCD following
                 from (\protect\ref{3.4a}), (\protect\ref{3.4b}) 
and the magnitudes of atoms and nuclei.}
\begin{center}
\begin{tabular}{lccc}
\hline 
 Theory & $\beta$ & approx. radius $[m]$ & IPN diameter $\delta (\tau)\: [s]$\\
\hline 
  QED   &    .1   &    $10^{-10}$ & $10^{-19}$ \\    
  QCD   &    .1   &    $10^{-15}$ &  $10^{-24}$ \\ \hline
\end{tabular}
\end{center}
\label{Tabelle_3_1}
\end{table}

   In addition, $\Delta x$  represents in physics the displacement of, e.g., 
a particle. By generalizing this to any observable change one obtains a 
map of the changes onto the time-space. This is a generalization of
{\bf Proposition 3.1}.

\subsection{The construction of the time-space topology}

   The following {\bf Definition 3.1} and {\bf Definition 3.2} are 
considered as the  two axioms of the present new chrono-topology
developed in this work. 
    
 {\bf Axiom I.} 

{\it 
All time definitions, classical or quantal, are based on some process
implementing a change, natural or technical and generates time neighbourhoods.
The generated IPN's are regular, into-maps of just these changes.}

 {\bf Axiom II.} 

{\it 
Every  fundamental interaction is associated  with (different among them, 
but) finite changes of the involved physical observables. The changes of the 
observables  have intrinsic the   random character, as to their embedment
in  the Newtonian time. 
They start at irregular Newtonian times and have, within
limits, stochastically distributed 
durations. They may be thought of as embedded in the Newtonian universal time,
$\Rset^1$, but their union has not the topology $\Rset^1$.   
}
  
{\bf  Axiom III.}

{\it 
   The elements of an empty set, $\emptyset$, of a class of observable sets 
$\{ O_{\lambda}| \lambda \in Z^+\}$ are not 
observable, and their values are identically equal to zero.  
}

\begin{figure}[htbp]
\begin{center}
\leavevmode
\epsfxsize=0.5\textwidth
\epsffile{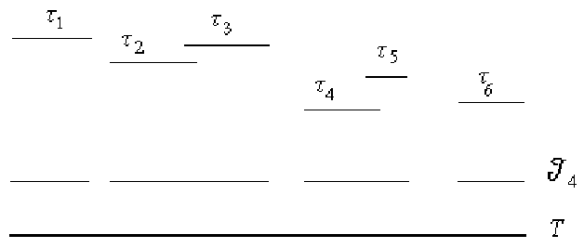}
\end{center}
\caption{ Representation  of  six  IPNs $\{\tau_i | i=1, 2, \ldots 6\}$, 
the union, ${\cal T}_4$, of their projections, and a subset, 
$T\subset \Rset^1$, 
of  the Newtonian  universal  time, $\Rset^1$, in which ${\cal T}_4 $ may 
be considered as  embedded. }
\label{fig_3_1}
\end{figure}

 Here is the principal definition of the interaction proper-time 
neighbourhood, the IPN:

\begin{defi}
\label{Definition_3_1}

       Let $O$ be an observable characterizing  one or both of  a given 
pair of interacting quanta. 
     Let  $\Delta O$  be the corresponding change due to a fundamental
     interaction. 
    We define the IPN (interaction proper-time neighbourhood) 
as the regular and continuous map:
\begin{equation}
\tau_{\lambda} =  IPN  =  f: \Delta O\rightarrow \tau_{\lambda} =
f(\Delta O)\in {\cal T}_4.
\end{equation}                  
    IPN  is  the  time  ''quantum'' of  the  process
    corresponding  to the fundamental interaction under
    consideration, characteristic of  and proper to that 
interaction and only to that.
\end{defi}

\subsection{The many-folded super space-time}

\begin{defi}
\label{Definition_3_2}

\begin{enumerate}
\item   Let $K\times \Lambda_K $  pairs of quanta  interact.
\item   Let $\{T_K| \:\kappa \in [1, K]\equiv I_K \subset \Zset^+\}$ 
         be a family of subsets $T_K \subset \Rset^1$   such that
         $\{ T_K\cap T_{K'} = \emptyset \: | \: \forall (\kappa, \kappa '] \in I_K 
\subset \Zset^+\}.$
\item   Let   $\{ \tau_{\kappa \lambda_{\kappa}} \in T_K, \:\forall
\lambda_{\kappa}  \in [1, \Lambda_K]\equiv I_{\Lambda_K} \}$ 
     be a family of IPNs such that  
$$\{ \tau_{\kappa \lambda_{\kappa}}
\cap \tau_{\kappa \lambda_{\kappa}'} = \emptyset \mbox{ for}\: \lambda_{\kappa}
\neq \lambda_\kappa '\}.$$
\end{enumerate}

    We define:

\begin{enumerate}
\item   The $\Lambda_{\kappa}$-fold, disconnected time-space  by 
\begin{equation}
{\cal T}^{(\Lambda\kappa)}_4 = \tau_{\kappa 1} \oplus
                           \tau_{\kappa 2} \oplus \ldots \oplus
                           \tau_{\kappa \Lambda_\kappa}.
\end{equation}
$\{ \delta(\tau_{\kappa \lambda_{\kappa}})\} $ may be thought as the random
absolute values of vectors orthogonal at every point of Riemann space-like
super-surfaces.

\item   The   $\Lambda_\kappa$-fold, disconnected  super-space-time 
in the sense of  $\lambda_\kappa$-fold Riemann super-space-time  by
\begin{equation}
\bar{M}^4_{\Lambda_\kappa} = i\left( \tau_1 \oplus  \tau_2 \oplus
               \ldots \oplus \tau_{\Lambda_\kappa}\right) \times \Rset^3,
\end{equation}
where $\Rset^3 $ is a 3-dimensional Riemann space.

\rm
\end{enumerate}
\end{defi}

The formulation of  a physical theory in terms of generalized 
random and infinitely 
divisible fields requires  space-time structures of the above form.  

   To make this clear, let us consider  one single IPN, $\tau_1$, 
and the corresponding space-time,  $\bar{M}^4_{\kappa 1}$.  
The lower index signifies that  $\bar{M}^4_{\kappa 1} = i\tau_{\kappa 1 }
\times \Rset^3$, 
and this space-time is {\it simple} in time, i.e., a subset of a Riemann space.
If $\Rset^3$ is flat, then $\bar{M}^4_{\kappa 1}$ becomes a 
subset of the Minkowski space. 

    If  there are two different  IPNs, ${(\tau_{\kappa 1}, \tau_{\kappa 2})}$ 
such that on 
the one hand  $\tau_{\kappa 1}\cap \tau_{\kappa 2} = \emptyset$ and on 
the other hand their projections $\pi_1, \pi_2$  into $T_\kappa$  satisfy
$\pi_1 \subseteq \pi_2$, then the 
corresponding space-time is  $\bar{M}^4_2 = i( \tau_{\kappa 1 } \oplus  
\tau_{\kappa 2}) \times \Rset^3$.   This space-time is {\it two-fold in 
time}.
In case $\Rset^3 = E^3$, the Euklidian $3$-space,
$\bar{M}^4_2 $ is not a subset of Minkowski's space anymore.

 It is said in terms of relativistic simultaneity  
fully or partly simultaneous according to the relations 
$$
(\pi_1\subseteq \pi_2) \wedge (\pi_1\supseteq \pi_2)\quad \mbox{or} \quad 
(\pi_1\subseteq \pi_2) \vee \pi_2 \subseteq \pi_1)
$$   respectively. 

   More generally, if  $\lambda_\kappa$  IPNs  satisfy 
$$\tau_{\kappa \lambda_{\kappa}} \cap \tau_{\kappa \lambda_{\kappa}'} 
= \emptyset, \quad\forall (\lambda_{\kappa}, \lambda_{\kappa}')
\in I_\kappa $$                                                     
and their projections  into  $T_\kappa$

$$
(\pi_{\lambda_{\kappa}} \subseteq \pi_{\lambda_{\kappa}'} ) \wedge 
(\pi_{\lambda_{\kappa}} \supseteq \pi_{\lambda_{\kappa}'})
\quad  \mbox{or} \quad
(\pi_{\lambda_{\kappa}} \subseteq \pi_{\lambda_{\kappa}'} ) 
\vee ( \pi_{\lambda_{\kappa}'}  \subseteq  
\pi_{\lambda_{\kappa}} ), \quad\forall (\lambda_{\kappa}, \lambda_{\kappa}') 
\in I_\kappa,
$$
then the strucure of $\bar{M}^4_{\kappa\lambda_{\kappa}}$ is even higher.

\begin{figure}[htbp]
\begin{center}
\leavevmode
\epsfxsize=0.5\textwidth
\epsffile{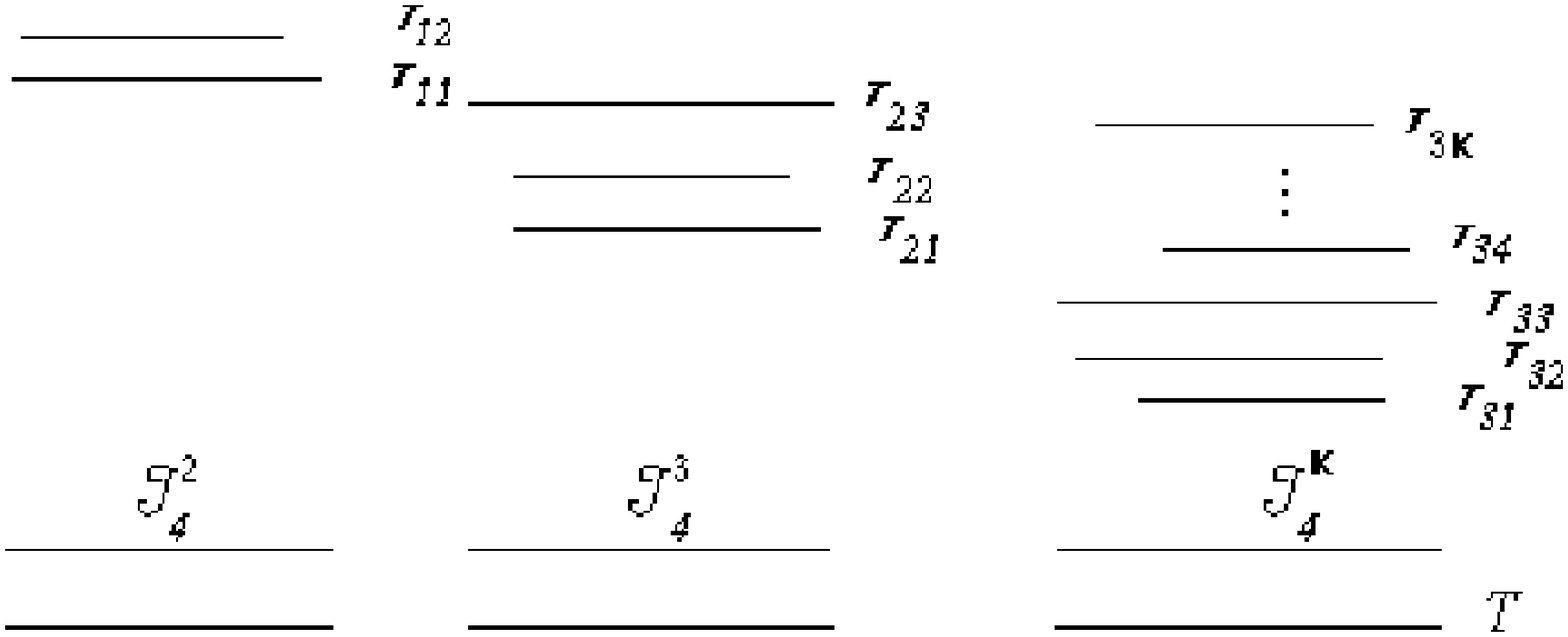}
\end{center}
\caption{Three types of many-folded time topological time-spaces, two-fold,
${\cal T}^2_4$, 
three-fold, ${\cal T}^3_4$  and $\kappa$-fold, ${\cal T}^\kappa_4$. 
These time-spaces give rise to the creation of the space-times,
$\bar{M}^4_{12}, \bar{M}^4_{23}, \bar{M}^4_{3\kappa}$. 
}
\label{fig_3_2}
\end{figure}

In a  $\lambda_\kappa$-fold in time space-time the decomposition 
of  a divisible field $\cal L$ in up to ${\lambda_{\kappa}}$ 
 terms is possible without interfering neither with the 
definition of the function 
notion nor with the conservation laws of physics cases in which 
$f(x)\ne 2 f(x)$.
  An illustration of our time-space  $K = 4, (\Lambda_1= 1,
\Lambda_2 = 2, \Lambda_3 = 2, \Lambda_4 = 1)$, 
is given in Fig \ref{fig_3_1}, 
while the case $K = 3, (\Lambda_1 = 2, \Lambda_2 = 3, \Lambda_3 = \kappa)$
time-space is shown in Fig. \ref{fig_3_2}  .

It is important that  the time in, e.g., the rest frame of a particle  
is related to its 
corresponding interaction.  If to all IPNs were given the properties 
of one single 
IPN, the time-space would lose its randomness. 

We put just this time in the  equations of  Schroedinger, of Dirac and of 
QFT in connection 
 with problems of  nuclear and  sub-nuclear intereactions. 
   The time change within an IPN cannot generate the impression of
   flowing: i) it escapes the discrimination power of the human sensors, and
ii) There is one single IPN and no ordering is feasible.

 On the contrary, for a moving observer the reaction 
time may flow or not flow further  
depending, according to (\ref{3_3b}) above, on whether the particle 
changes either its 
position, $x$, or its time, $t$, or both, or any other of its observables.                        
 Hence, it is clear that the particle  reaction time cannot  be 
identified with the universal time 
which is the union of the maps of all observable changes, occurring  in  the 
entire observable universe. 

{\bf Remark 3.2 }

The factor ${\cal T}_4$  determines the 
structure of the new  space-time $\bar{M}^4_\kappa$. 

    The space-time, $\bar{M}^4_{\kappa\lambda_{\kappa}}$,  
$\lambda_{\kappa}$-fold in time is 
the natural space-time for the application 
of the theory of the generalized and inifinitely divisible fields. 

{\bf Remark 3.3}

   Time-dependent quantum equations not including  interactions  
do not supply us  with any physical information regarding the  
evolution of the particle system. For example, an electron moving 
in vacuum without interaction is described by 
free-field quantum time-dependent equations, but it does not exist, it is
not observable.

  However, the situation is still more complex: 
The kind of time topology Nature 
chooses in every individual case of interacting particle systems, 
depends on the 
number of  the interacting  particle  pairs and on whether  
the interactions are 
partially or totally simultaneous in the sense of relativity.   
One easily  realizes  based 
on our definition of the time that the topological space, ${\cal T}_4$, tends 
to the  space with the 
natural topolgy of $\Rset^1$, if the number of the interacting particles 
becomes very large 
and the intersections of the adjacent IPNs are not empty anymore \cite{47}. 
More precisely:
\begin{equation}
\{ ({\cal T}_4 = \mbox{\it  natural topology of } T_{\Lambda} \rightarrow
\Rset^1 )\wedge( \bar{M}^4_{\Lambda} \rightarrow M^4) \mbox{ for } \Lambda\rightarrow \Zset^+\}
\end{equation}
$\bar{M}^4_{\Lambda} $
 {\it  is the physical space-time created by the  dynamics and yields the scenery for 
the evolution of the dynamical  particle systems.}

$M^4$,  Minkowki's space-time, is a mathematical 
object representing the limit 
of $\bar{M}^4_{\Lambda} $ for an infinity of interacting particles,
such that $\{\tau_{\lambda} | \:\forall \lambda \in \Lambda \rightarrow \Zset^+\} $  
is  a covering basis of $\Rset^1$.

\subsection{Randomness and covariance}

The randomness of the IPNs  in a 
set of  interacting particles is a direct consequence 
of the randomness of the interaction durations.  Also the randomness of the physical 
fields as well as the resolution of some paradoxes in quantum theory can  be 
understood on the basis of the topology  of  ${\cal T }_4 $  
in  the rest  frame of  the     interacting particles.

\begin{enumerate}
\item The topology of the time-space in the reference frame of a moving observer is 
determined through Lorentz transforming  the time-space ${\cal T }_4 $
  to the time of moving 
observers with respect to interacting particles. 

\item  All functions $f(x)$   of the space-time coordinates 
$(x^0, x^1, x^2, x^3) \in \bar{M}^4_\kappa$     
bear the random character of the time topology of ${\cal T }_4 $.
\item  In the topology of space-times resulting from  ${\cal T }_4 $ 
the time and the space do not 
flow. The domain of the coordinates  $ \{ x^\mu | \mu = 0, 1, 2, 3\}$  
is  compact  in  the subsets  of  
the disconnected space-time $\bar{M}^4_\kappa$.  
\end{enumerate}

  The above observations give rise to the question as  
to the covariance of  the 
fundamental equations of  QFT in the super space-time. It is not difficult to see that  
this aspect of the theory does not  suffer any important change.

 Since every  single physical process  takes place  inside its  respective IPN, it  is 
sufficient to verify that the equations of quantum mechanics and of QFT  remain 
covariant  for the Poincare transformation  group within the space-time   for  
$\bar{M}^4_\kappa$ for $\kappa =1$. 
The Noether theorem is valid in every IPN and the proof is   identical to the 
usual proof in the Minkowski space-time and it is omitted. 

\subsection{Chrono-topology and irreversibility considerations}

 The chrono-topology opens  new possibilities 
for the investigation of the $\bf U$ and $\bf R$ 
kinds of time evolution.  We continue here the the examination of these aspects.
\begin{itemize}  
\item[i)]  The fundamental equations of physics - including interactions - 
      as well as the 
        phenomena described by them are time-reversal  invariant on  every single
        IPN, $\tau$.  The conservation laws are valid for $\bf U$ processes. 
All phenomena   are time reversible inside one and the same IPN, 
$\tau$, during $\bf U$ time evolution.
\item[ii)]   But (attention!) the event that the  {\it time-reversed
    interaction action-integral} 
 equals the action-integral of the (factual) reverse 
interaction has a zero probability  measure. 
\end{itemize}

The probability measures for these processes have the following 
properties:

\begin{itemize}
\item[i)]   The measure, $\mu_{Direct}$, 
 for the direct process is associated with a mathematically 
 realizable and physically possible process.
\item[ii)]  The measure, $\mu_{Time-reversed}$, 
 for the time-reversed process is associated with a 
 mathematically realizable process wich is  physically impossible.
\item[iii)]  The measure, $\mu_{Reverse \: interaction}$, 
  for the reverse interaction corresponds to a 
  process both mathematically and physically possible. It is important to 
realize that a time reversed and a factually reverse reaction do not take
place in the same $\tau$.

The combinations of the measures have the properties:  

\end{itemize}

\begin{eqnarray*}
\mu_{Direct}, \mu_{Time-reversed}, \mu_{Reverse \: interaction} & > & 0, \\
 \mbox{\it Probability measure} \{ \mu_{Direct} = \mu_{Reverse \: interaction} \} & = & 0, \\
\mbox{\it Probability measure} \{  \mu_{Direct} = \mu_{Time-reversed} \} & = & 
1. 
\end{eqnarray*}

  These relations can become more clear with the help of three, generally, 
different well-ordered IPNs
IPNs $\{\tau_1\succ \tau_2 \succ\tau_3\}$.
Suppose that the direct and the time reversed reactions take place for
$t \in \tau_2$. Since the factually reverse reaction cannot proceed 
simultaneously  with the direct reaction, it will take place either in $\tau_1
\succ \tau_2$ or in $\tau_3 \prec \tau_2$.
$$     
\{ \mbox {\it TIME-REVERSED action} \quad 
\int dtH(t),  \quad t \in \tau_2 \} \neq $$
$$ \{ \mbox{\it action }\quad  \int dtH(t) \quad 
\mbox{\it of the REVERSE
  interaction } \quad 
(t\in \tau_1 \succ \tau_2)
\vee ( t\in \tau_3 \prec \tau_2)\} . $$   

    The above relation (i.e., ''{\it time-reversed process action''
is different from the ''action of the factually  reverse  process}'') 
holds  true, because the IPNs  
$$ \{ \tau_1, \tau_2, \tau_3\} $$
may be different in two respects:
\begin{enumerate}
\item As sets. 
\item As set diameters, 
$\{ \delta(\tau_{\lambda}), \lambda = 1, 2, 3\}.$  
\end{enumerate}

On the other hand, the 
ranges of any functions  in $ \{\tau_1, \tau_2, \tau_3\}$   are, 
with  high probability, different at least 
for two reasons:
     
\begin{itemize}
\item[i)] $\delta(\tau_{\lambda}), \lambda = 1,2,3$, as numbers:  
 {\it Probability measure}
         $\{\delta ( \tau_i)=\delta (\tau_j), j\neq i\} = 0 $,  and 
\item[ii)] $ \tau_{\lambda}, \lambda =\:1, 2, 3 $  as point sets: 
{\it Probability measure}
            $\left\{  \{\tau_i\} \cap  \{\tau_j\}  = 
\emptyset\right\} = 1, j\neq i.$
\end{itemize}

\subsection{Planck time and chrono-topology}

 Despite the differences between our space-time topology in  conception and in 
construction method and  the space-time foam of S.~Hawking \cite{23}  there is, nevertheless, a 
certain resemblance at the limit $\delta(\tau_{\lambda})\rightarrow  \mbox{\it Planck time }
\:\forall \lambda \in \Zset^+$,  when the interactions become very fast.

If the ''foam'' time intervals had all the Planck time magnitude 
they would loose the random character.

 If  the observers lived in $\tau$, it would be 
impossible to compare $\tau_j$ with 
$\tau_i$  for $i\neq j$, 
because each $\tau$  is its own unit in the rest frame. However, 
such a comparison is for 
the human observers perfectly possible, because our senses are 
exposed to quanta 
coming from many different, but, more or less, overlapping interactions in 
$T\subset \Rset^1$, due to our ability to observe (almost) simultaneously 
more than one physical changes.

    The time-space topology ${\cal T}_4$  introduced above bears {\it 
intrinsically the random 
character of the IPNs}. 
It is this property that imposes randomness to every function 
of the time. An important observation is that the 
randomness can be perceived by the 
observers, because they are living in the background 
of the Newtonian time which 
has the topology of  $\Rset^1 $. 

   Some examples of functions defined in $\tau$  becoming random are:

\begin{itemize}
\item[i)]   The {\it space-time coordinates for the moving observer of a particle system}. 
  The observers are almost in all cases moving with respect to the interacting elementary 
  particles, so that observation is mediated by  Lorenz transformations.

\item[ii)]  All {\it observables expressed as functions of the space-time coordinates} 
  in the  {\it rest frame of reference of the observer}.

\item[iii)] The {\it components of the quantum fields}  which become generalized random 
fields.
\item[iv)]  The {\it Hamiltonian and the Lagrangian densities} become generalized random 
and infinitely divisible fields, thus  admitting the representation 
$$ F(\phi(x), \partial \phi(x))=F(\phi(x_1),\partial \phi(x_1))+ F(\phi(x_2),\partial \phi(x_2))
+\ldots + F(\phi(x_{\lambda}), \partial \phi(x_{\lambda})), $$
$$ \kappa = 2, 3,\ldots \mbox{ for } x\in \bar{M} _\kappa^4.$$
for $\kappa \in \Zset^+ $   and with  {\it probability distributions independent of} $\kappa$.
\item[v)]   The {\it metric tensor} $g_{\mu \nu} $ of the {\it
    space-time} in general relativity.

    The IPNs, as maps of  finite observables' 
changes  through interactions,  they 
are each one compact in $T\subset \Rset^1 $, 
and their  set diameters are empirically  inversely  proportional to 
the strength of the interaction.
\end{itemize}

\setcounter{equation}{0}
\setcounter{table}{0}
\setcounter{figure}{0}
\section{Microscopic and macroscopic system-time}
\label{section_4}

   The considerations of the foregoing sections make clear that any
   time variable defined to describe a microscopic system will be
   conceived as the union of IPNs $\{\tau\}$ generated during the
   numerous interactions of atomic,  nuclear or sub-nuclear constituents
   of the system under observation.
   It becomes,  therefore,  clear that every sytem of particles has its
   own macroscopic time given by the union in question. If two
   different particle system S and S' have equal numbers N=N' of
   identical particles,  interacting via identical forces,  they may or
   may not have identical microscopic IPNs.Because the changes of
   the observables and, consequently, their maps, the interaction
   proper-time neighbourhoods, are random.

If the numbers of the interactions $i, i'\in I\subset \Zset^+$ in the two
systems become very large, then the microscopic system time variables $t$ and $t'$
of the respective systems of particles will be with high probability
the congruent time neighbourhoods $\tau$ and $\tau'$. They take values in
the union of the IPNs $\{ \tau \in \bigcup_{i\in I} \tau_{i} $ and 
$\tau' \in \bigcup_{i'\in I'} \tau'_{i'}$ pertaining to the set $A$ of 
interactions between the  set $B$ of particles in the systems
(${\cal T}_4 $ in \ref{fig_3_1}):

\subsection{The microscopic system-time}
\label{subsec4}

$A=\{\alpha\}$ is the set of the numbers of interactions taking place in one 
single particle system with IPN projections in $\Rset^1$ partially or totally
overlapping. 
$B=\{\beta\}$ is the set of the numbers of interacting particle pairs with
disjunct IPN projections in $\Rset^1$.

\begin{defi}
\label{Definition4_1}
Every closed microscopic system of interacting particles has its own
microscopic system time given by:
 
{\bf Microscopic system time} :=
\begin{equation}
\label{4_1}
\tau \in\bigcup_{\alpha \in A}                               
                               \bigcup_{\beta\in B}\tau^{(\alpha, \beta)}
\end{equation}
                               = {\bf Union of all factual IPNs for a small
                                 number of elements  in the index sets 
                                 $A =\{\alpha\}$ and 
                                 $B=\{\beta\}$.}

\end{defi}

There are possible topological cuts and gaps in the set of 
individual IPNs, if the system consists of a small number of
particles.

\subsection{Macroscopic system-time}

If $\{A\}$ and $\{B\}$ are families of  sets as defined in \ref{subsec4}, then
we give

\begin{defi}
\label{Definition4_2}
  Every closed system of interacting quantum systems of particles has
its own macroscopic system-time given by

{\bf Macroscopic system-time} =
\begin{eqnarray}
\label{4_2}
                             T := \bigcup_{A\in \{A\}}\bigcup_{B\in
                                 \{B\}}\left\{ \bigcup_{\alpha\in
                                 A}\bigcup_{\beta\in B} \tau^{(\alpha,
                                 \beta )} \right\} 
\end{eqnarray}
                                 = Union of all factual IPNs with
                                 large number of elements in the index
                                 sets $\{A\}$ and $\{B\}$.
\end{defi}

{\bf Remark 4.1}

\begin{figure}[htbp]
\begin{center}
\leavevmode
\epsfxsize=0.5\textwidth
\epsffile{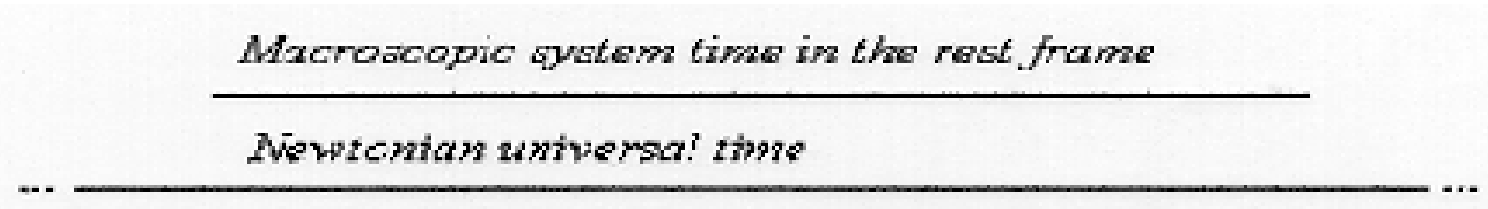}
\end{center}
\label{Fig41} 
\caption{The macroscopic system-time $T\subset \Rset^1$ is a subset dense
   in $\Rset^1$  with a high probability measure.}
\end{figure}

   The facility with which a moving observer describes the macrocosmos
   by means of a continuous time variable is based in its embedment in
   the universe at large. The collection of all observable
   interactions in the universe generates the Newtonian universal time
   despite the countability of the sets {A} and {B}. An important part
   plays here the limited time discrimination power of the observer for
   sucsessive IPNs (\ref{Fig41}).

\subsection{The relationship of our chrono-topology with the Newtonian time}

On the basis of the above defined universal time and of the relativity theory
it is possible to decide about the priority, the simultaneity and the
posteriority of a number of events.
We consider an observer observing simultaneously an interacting
particle together with the interacting particles of all observable (by him)
systems in the universe.

If the number of the interacting particles in other particle systems
perceived simultaneously by the observer, or interacting with the
particle under consideration  and if the corresponding number of 
IPNs is very large,  the previously disconnected
time-space acquires the topology of the universal time, i.e., 
the union becomes homeomorphic to an interval of $T_4 \subseteq \Rset^1$.
Hence, in case '$N$ is very large', $T$ consits entirely of partially overlapping 
IPNs, and a continuous macroscopic time variable emerges,
 $t \in T\subseteq \Rset^1$.
This is identical with a subset of macroscopic or universal 
Newtonian time.

Next, suppose there is a neighbourhood in the universe from which not
all physical events of the universe are observable from observer's
rest frame of reference. The physical changes of observales in
observer's neighbourhood allows to him to define his macroscopic time
by means of (\ref{4_2}).

If the distinct physical observable changes in the neighbourhoods, not
observable from his rest frame, do not proceed in the same way
(different distributions of energies or temperatures) as in his own
neighbourhood, then the time topologies in these different neighbourhoods
of the universe may be quite different and the times
may 'flow' in different ways.

Hence, the macroscopic times defined in two different space-time
neighbourhoods of the universe may, in principle,  differ from
one another. This opens the question about the possibility to define one
single time for the entire universe in cosmology.

It seems to us that one cannot cunstructively define one single macroscopic
time in all neighbourhoods of the universe. If
this is so, the time definition in cosmology might be questionable.

\setcounter{equation}{0}
\setcounter{table}{0}
\setcounter{figure}{0}
\section{Planck's constant from Liouville's time quantization}
\label{section_5}

    Empirically, all fundamental interactions are of finite
    durations. Consequently the interaction proper-time neighbourhoods,
    being defined as regular injective maps of the changes of the observables
    implied by the respective interactions, have in every
    particular case a finite diameter.
   Also, considered from the point of view of the number of the
   exchanged  quanta, the interaction time  is factually  finite.

 From these facts it becomes clear that the time variable
 for atomic and sub-atomic reactions can be only a
 sectionally continuous function of the changes of the
 observables.

\subsection{Liouville reality and form invariance of the distribution function
    with respect to the number of particles} 

It is interesting to show that this intuitive truth can be
demonstrated also
in a formal way in the framework of the classical theory of the
Liouville equation. 
The fact that the Planck constant is determined from the
solution of Liouville's equation in the present section 
it may be considered as an indirect verification of the above ideas.
For this purpose we shall prove first the following 

{\bf Proposition 5.1 }

\begin{enumerate}
\item  Let $PS = P\times Q$ be the phase space.

\item  Let $\{p^{(n)}\in P, q^{(n)}\in Q, n=1, 2, \ldots N\} $ 
be the phase space coordinates of an $N$-particle system 
interacting via given constant forces
$ \{F^{(n)}|n=1, 2, \ldots N\subset \Zset^+\}$.

\item Let, further:

\begin{equation}
\label{5_1}
g(q, p, t) = \sum_{n=1}^N\left[i \lambda \epsilon_n t - 
  \mu_n {\cal F}^{(n)} \cdot q^{(n)} +    \mu_n  \left( 
\frac{p^{(n)}}{\sqrt{m_n}} - \nu_n {\cal F}^{(n)} \right)^2 \right], 
\end{equation}

\begin{equation}
\label{5_1_0_1}
{\cal L} = \partial_t + \sum_{n=1}^N ( \frac{p^{(n)}}{m_n} \cdot
  \nabla^{q^{(n)}} + {\cal F}^{(n)} \cdot \nabla^{p^{(n)}}), 
\end{equation}

\begin{eqnarray}
\label{5_1_0}
p^{(n)}\cdot {\cal F}^{(n)} -  {\cal F}^{(n)}  \cdot p^{(n)} &=& 0, \\
\sum_{n=1}^N \frac{\mu_n \nu_n}{\sqrt{m_n}} (p^{(n)} F^{(n)} 
+ F^{(n)}  p^{(n)}  ) & = & 0,  \\
Im\{ f(g)\}&=& 0, \mbox{ reality},\label{real} \\
f(g_1)\cdot f(g_2) &=& f(g_1+g_2)  \mbox{  additivity}, 
\label{addi}\\
\epsilon_n & = & \frac{1}{i\lambda} \mu_n \nu_n ({\cal F}^{(n)})^2/\sqrt{m_n}. 
\end{eqnarray}

Then,

\begin{enumerate}
\item  $ g(q, p, t)$ and $f(g)$ satisfy

\begin{equation}
{\cal L} g(q, p, t) = 0, 
\end{equation}

\begin{equation}
\label{5_2}
{\cal L}f(g) = 0. 
\end{equation}

\item The time variable $t$ takes values in ${\cal T}_4$ 
which are given for $\lambda = \hbar^{-1}$ by
\begin{equation}
\label{5_3}
\epsilon_n t = 2 \hbar\pi k_n, n\in \Zset^+ \:\mbox{and} \:k_n \in Z, 
n=1, 2, \ldots N.
\end{equation}

\end{enumerate}

\end{enumerate}

{\bf Proof}

Application of ${\cal L}$  on $g(q, p, t)$     gives

\begin{equation}
\sum_{n=1}^N \langle 
i \lambda\epsilon_n - \mu_n  \frac{{\cal F}^{(n)}}{\sqrt{m_n}}
 \cdot p^{(n)}
+ \frac{1}{2} \mu_n\bigg[ \frac{{\cal F}^{(n)}}{\sqrt{m_n}} 
 \cdot(  \frac{p^{(n)}}{\sqrt{m_n}}
 - \nu_n {\cal F}^{(n)})
+ (  \frac{p^{(n)}}{\sqrt{m_n}} - \nu_n {\cal F}^{(n)}) \cdot 
\frac{{\cal F}^{(n)}}{\sqrt{m_n}}\bigg] 
\rangle.
\end{equation}

 From \ref{5_1}(c) and from \ref{5_4} we infer (5.1g). This shows that
\begin{equation}
\label{5_4}
{\cal L} g(q, p, t) =0.
\end{equation}

From \ref{5_1}(d) we conclude that $f(g)$ satisfies \ref{5_2}.

\subsection{The Liouville time quantization}
\label{sub5_1}

Next, we put 
\begin{equation}
\label{5_5}
f(g) = C \exp \{ g(q, p, t) \}, 
\end{equation}

where $C\in\Rset$ and from (\ref{5_1}e, f) it follows that 
        
\begin{equation}
\label{5_6}
       t_n = 2\pi\hbar k_n/\epsilon_n,   \quad 
n\in \Zset^+ \quad \mbox{and}\quad  k_n \in \Zset .
\end{equation}

Since $\lambda$ is an arbitrary constant and it has the physical
dimension of an inverse action, we have put $\lambda = \hbar^{-1}$, and
(\ref{5_3}) is a quantization condition for the time.
   It will shown in \ref{subsection_c}  on the basis of experimental 
   data that equation (\ref{5_3})  can be used to determine the
   value of the constant $\lambda$ of (\ref{5_1}) and to find that it equals the
   Planck constant. This is very exciting, because
   Liouville's equation is a classical equation.

Equation (\ref{5_6}) can be used to obtain the system-time. If we sum both
sides of it over $n$, we  find

\begin{equation}
\label{5_7}
       E_N t_N = 2\pi\hbar K_N, \quad K_N \in \Zset^+,
\end{equation}
where we defined:
\begin{equation}
E_N t=\sum_n^N \epsilon_n t_n / t_N,
\end{equation}
\begin{equation}
t_N = \sum_n^N t_n
\end{equation}
and 
\begin{equation}
K_N \le \sum_{n=1}^N | k_n|.
\end{equation}
We may take the total action in the system either as positive
or as negative (see below (\ref{7_10})), but we must assume that the total
energy, $E_N$, is conserved. Equation (\ref{5_7}) shows that the total energy
being conserved, the time $t_N$ can change in steps according to $K_N$.
This completes the proof of {\bf Proposition 5.1}

{\it \bf Corollary 5.1}

Classical particle systems under constant forces are 
subject to quantized action.

{\bf Corollary 5.2}

Time cannot flow continuously. If $E_N$ is conserved, and $K_N$
 is kept constant, then time is generated as neighbourhoods 
$\{\tau_\lambda\}$ and it does not flow.

{\bf Corollary 5.3}

Time can flow for constant $E_N$, only if various partitions 
$\{ k_n\}$ occur with the same $K_N$ in accordance with (\ref{5_4}).

{\bf Corollary 5.4}

If time changes, it does so in steps, $\Delta t$, (Liouvillian time)
at least as large as 
\begin{equation}
\Delta t\le 2\pi \hbar \Delta K_N / E_N.
\end{equation}

{\bf Corollary 5.5}

Quantum processes are the faster (short $\delta t$) the higher their 
energies $E_N$ are.

{\bf Corollary 5.6}

$g(q, p, t) $ is form invariant with respect to the number, $N$, of particles changes.

\begin{figure}[htbp]
\label{fig_5_1}
\begin{center}
\leavevmode
\epsfxsize=0.5\textwidth
\epsffile{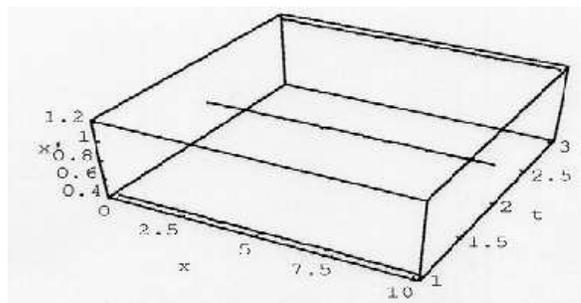}
\end{center}
\caption{$x'$-coordinate $(x'=\gamma(x-vt))$ of a space-time sector with three values
of the Liouville quantized time as seen by a moving observer with
velocity $v = 2.75\times 10^8 m/s$, The three space lines are reminiscent 
of a space having the form of strings. The length of the string  is proportional
to $x$, the rest frame coordinate of the interacting particle.
The width of the string depends on the energy, $\epsilon_n$-fluctuations, if any.}
\label{Fig.5.1}
\end{figure}

\begin{figure}[htbp]
\label{fig_5_2}
\begin{center}
\leavevmode
\epsfxsize=0.5\textwidth
\epsffile{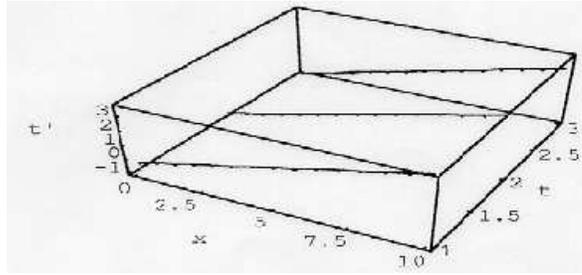}
\end{center}
\caption{The three lines, $t'$, $(t'=\gamma(t - \beta/c\cdot v))$ give the time for a 
moving observer in a space-time sector resulting from a Liouville point time-spectrum. 
The velocity of the observer frame of reference is $ v = 2.75\times 10^8 m/s$.
The three lines remind us of a space having the form of strings or fibers.}
\label{Fig.5.2}
\end{figure}

The initial question 'why time changes by steps' is, thus, answered in the
above classical non-relativistic theory of the Liouville
equation. The reason for this unexpected fact is that this theory
describes observable phenomena only, if two conditions are satisfied:

\begin{itemize}
\item[i)] The energy distribution function of the particle system is
  real
\item[ii)] This distribution function is {\em form-invariant} with respect
  to the number of the described particles.
\end {itemize}
The second condition (\ref{addi}) is equivalent to: The action integral of the 
sum of two
particle systems interactiong by means of constant forces is equal to 
the sum of the action integrals of the separate particle systems, whose
particles interact also by means of constant forces.

These two conditions (\ref{real}, \ref{addi}) 
imposed on the distribution function entail the
quantization of the time which then changes by steps.
These conclusions, if taken at face value, modify our picture of the
time even in a classical theory of atomic systems. 
They indicate that quantization
is a fundamental property of the elements of matter as well as of
radiation (Planck black-body radiation).


\begin{figure}[htbp]
\begin{center}
\leavevmode
\epsfxsize=0.5\textwidth
\epsffile{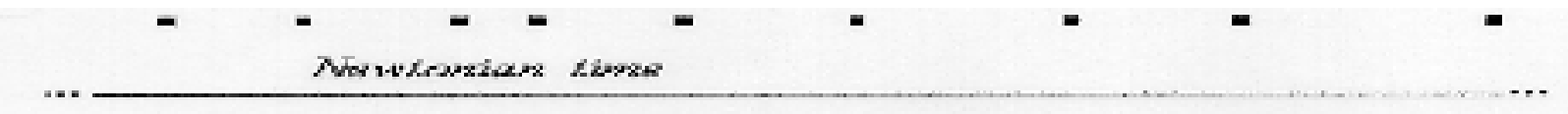}
\end{center}
\label{fig_5_3}
\caption{ The time spectrum resulting from the additivity and from the
reality conditions for the elementary Liouvilian distribution
functions. Constant interaction forces and conserved total energy, $E_N$, are
assumed. If the particle energies, ${\epsilon}_n$ 
fluctuate, then, obviously, the discrete point time-values change to time
neighbourhoods of fluctuating measure.}
\end{figure}

\subsection{Determination of Planck's constant as a verification of
chrono-topology }
\label{subsection_c}

   There are several methods for the determination of the value of
   Planck's constant. It is natural to expect - as it really is - that this
   constant which is
   the ''trade mark'' par excellence of quantum physics, should be the result of
   calculation based typically on some genuine quantum phenomena and on
   data following from them. 

   It is just because of this fact  that we consider it as extemely surprising 
that the
   value of Planck's constant follows from a classical theory. This theory
is  Liouville's equation supplemented with the reality condition and
   the form invariance requirement of the particle distribution function.

We consider this fact not as an accident and we believe that it
entitles us to think about it as an experimental verification of our
chrono-topology. We use for the calculation of the $(\lambda = ) \hbar$-value
expression (\ref{5_6}) 
\begin{equation}
\hbar = \frac{ \epsilon_1 \bar{\tau_1} }{ 2 \pi k_1}
\label{5_18}
\end{equation}
We also consider that the energy $\epsilon_1$ is the thermal (translational)
energy of the gas molecules
\begin{equation}
\epsilon_1 = \frac{3}{2} k_B T,
\label{5_19}
\end{equation}
where $ k_B  = 1.38 \times 10^{-23} J/K$ is the Boltzmann constant. We take
as temperature of the gas $T=300 K$. We have still to find $\bar{\tau_1} $
and to spcify the quantum number $k_n$. We consider first the atomic
hydrogen gas, put $k_n = 1$ and take for the atoms' average interaction range 
$R_{\mbox{Atomic}} \approx 1.45 \times 10^{-10} m$. The average interaction
time, $ \bar{\tau_1} $, is given by
\begin{equation}
\bar{\tau_1}  = \frac{ 2 R_{\mbox{Atom.}}}{\sqrt 2 \epsilon_1 /
M_{\mbox{Atom.}}},
\label{5_20}
\end{equation}
where the denominator is the average thermal velocity of the atoms with the mass
(chemical atomic mass unit $= 1.007593$)
\begin{eqnarray}
M_{\mbox{Atom.}} &=& (1.007593 \times 1.65969 \times 10^{-27} + 9.1083 \times
10^{-31}) kg \nonumber \\
&=& 1.6732 \times  10^{-27} kg.
\label{5_21}
\end{eqnarray}

The above parameters lead for the atomic hydrogen gas to the value

\begin{eqnarray}
\hbar &=& \frac{ \frac{3}{2} k_B 300 \frac{ 3 \times 10^{-10} }{
\sqrt{ 2 \frac{3}{2} k_B 300/(1.6732 \times 10^{-27})}}}{2 \pi k_1}
= 1.0525 \times 10^{-34} Js, 
\label{5_22}\\
\hbar &=& 1.0525 \times 10^{-34} Js \mbox{\bf from atomic hydrogen}
\nonumber
\end{eqnarray}
which is not far from the experimental value.

If we consider molecular hydrogen gas, $M_{\mbox{Molec.}} = 2 M_{\mbox{Atom.}},
R_{\mbox{Molec.}} = 2.05 \times  10^{-10} m, K_2 = 2,$
then we find
\begin{equation}
\hbar 1.0522 \times 10^{-34} Js   \mbox{\bf from diatomic hydrogen.}
\end{equation}

\setcounter{equation}{0}
\setcounter{table}{0}
\setcounter{figure}{0}
\section{Examples of quantum space-time 
{\normalsize $ \bar{M}_1^4 $ }-Impli\-ca\-tions of the chrono-to\-pology}

\label{section_6}

  In the preceding section conditions have been given under  which  time 
cannot change continuously for a system of particles of  given constant
total energy interacting via external forces. This formal proof is 
supporting the experimental evidence that time is a  map of changes  in   
observables involved in the fundamental interactions.
 
\subsection{About the time flow in nature}

   It follows, therefore, 
that at least under the above restrictive conditions time
 is quantized. Using the Lorentz transformation it is easy to show that 
for a moving observer space is also quantized in the form of $\bar{M}_1^4$. 

   To see this let us suppose the time is indeed generally  quantized and 
observe the motion of a particle. As long as a time quantum  is 'flowing' 
we see the particle moving, i.e., changing its position  in space. 
At the end of the time quantum flow, 
and before the beginning of the next time quantum flow,  
no position change  is observable. Because, if motion were observable, it would 
be used to define an other time quantum flow which is contrary to 
the assumption according to which the next time quantum flow 
has  not yet started. 
            
But  if a space position change can be observed only during the time quantum 
flow which can be due only to some fundamental interaction, then, clearly, space
 is observable as quantized. This can be made clear more formally. We can 
calculate exactly  the structure of a subset of  the space, 
$\bar{M}^4_1$, 
provided a set of IPNs is given.  

    We consider the Lorentz transformation

\begin{equation}
\label{6_1_a}
   x' = \gamma (x - vt )  
\end{equation}
\begin{equation}
\label{6_1_b}
  t' = (t - \beta/cx) 
\end{equation}                        
in which $(x, t)$ are particle coordinates in the rest frame of
reference and $(x', t')\in \bar{M}_{1}^{4}$ 
 are the coordinates observed by a moving observer. This transformation 
is necessary in view of the fact that almost all interacting particles 
move with respect to the observer.

\begin{figure}[htbp]
\begin{center}
\leavevmode
\epsfxsize=0.5\textwidth
\epsffile{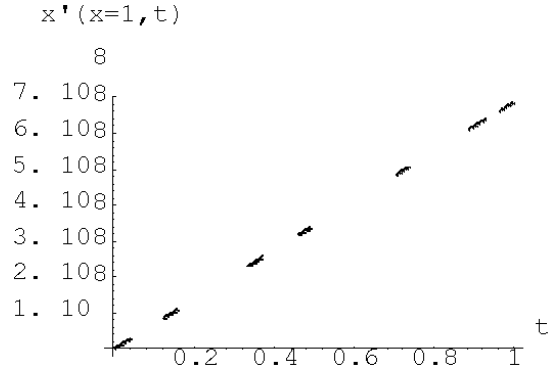}
\end{center}
\caption{ The coordinate $x'$ according to (\protect\ref{6_1_a}a, b) in the   
space-time  topology created by a set of 7  
disconnected IPNs $\{\tau_{\lambda} | \lambda = 1, 2, \ldots 7 \} $  
due to 7  fundamental interaction events in rest at $x=1$. The 
$\{\tau_{\lambda}\}$ -  
and the $x'$-neighbourhoods   obtainable as  orthogonal  projections  on  the  corresponding axes 
are represented in the graph as embedded in the $T$- and in the  $\Rset^1$-axes  for $t$ and $x'$ 
respectively. 
The measures of  and the gaps between the $t$ - and the $x'$-neighbourhoods  are random. 
Seen by a moving observer and generated by Lorentz transformation with  $v/c=0.916667$. 
Arbitrary scale factors are used.}
\label{Fig.6.1}
\end{figure}

An example, of how a  coordinate $x' \in \bar{M}^4_1 $   
appears for fixed $x$ in the rest frame to a moving observer, 
is given in Fig. \ref{Fig.6.1}.
In this example arbitrary scale factors are used. 
 
It is seen, therefore, that the quantization of the time implies also the 
quantization of the space by means of the Lorentz transformation. This,
however, is not the only consequence for the observer.
The discontinuity of the time and the fact that time is embedable
in $R^1$ produces psychologically for the observer an application of
Zermolo's {\sl well-ordering theorem} on the generated time quanta, i.e.,
\begin{equation}
\{ \tau_{\lambda} \succ \tau_{\lambda + 1}, \forall \lambda Z^+\}
\end{equation}

This runs as follows: Every new $ \tau_{\lambda ''} \prec \\tau_{\lambda '}$ 
observed is added (psychologically) to the set $\{  \tau_{\lambda} ,
\lambda < \lambda '\}$, and this creates the impression of a flow.

Such a flow is not physical, does not exist outside the observer's mind, and
time does not flow in full agreement with Relativity.

\subsection{The diameter of the IPNs}

  In actual experiments we have to put the interaction times whose orders  are 
estimated from experience in Table 6.1  independent of Table 3.1 :

\begin{table}[h]
\caption{  Orders of  magnitude of the     
in the fundamental interactions according to evidence}
\begin{center}
\begin{tabular}{ll}
\hline 
Measure of IPN in QED & $10^{-18} s$ to $10^{-14} s$. \\
\hline 
Measure of IPN in QCD & $10^{-44} s$ to $10^{-18} s$. \\
\hline
\end{tabular}
\end{center}
\label{Table_6_1}
\end{table}

\subsection{Interactions and observability}

 Hence, the subset of the Minkowski space 
($x=\mbox{const}$) $\bar{M}^4_1$  for moving observers  is 
quantized with respect to the rest frame of the interacting particles.  In view of the 
above, we prove the following

\pagebreak

{\bf Proposition  6.1}

    An non-interacting particle at x=0 in its rest frame of reference is unobservable by 
all moving observers.

{\bf Proof}

   From (6.1) and from $x=0$, it follows that $x' = -\gamma v t $. On the other
hand, if $C$ is a constant, the map of the empty set
\begin{equation}
\emptyset \rightarrow C \emptyset = \emptyset, \quad\forall C \in Rset^1
\end{equation}
is a regular, continuous map of the empty set onto itself. 
Since $t\in \tau_{\lambda} = \emptyset$ it follows that
$x' \in \emptyset$.
The set of the particle coordinates for the moving observer is empty
and no particle coordinate is observable. Since the constant $C$ may be any
number, we may put $C = \gamma$, and the proof is complete.

{\bf Corollary 6.1}

 No space-time neighbourhood is observable, 
if there are no interacting particles in it. An application of this corollary
is the non-observability of the quarks in the asymptotic freedom state.

\subsection{Sheets of space-time}

   For non-fixed x-coordinates of a particle in a frame of reference, 
with respect to 
which the observer moves, a projection of the space $\bar{M}^4_1$   
appears as two distinct  sets of world sheets  with  widths of the 
orders given in Table \ref{Table_6_1}  and random 
values. The two sets of world sheets lie in two planes forming the angle 
$\theta =\tan^{-1}(1/\gamma )$.

\begin{figure}[htbp]
\begin{center}
\leavevmode
\epsfxsize=0.5\textwidth
\epsffile{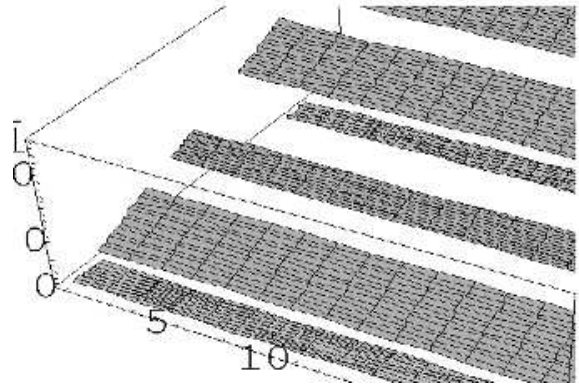}
\end{center}
\caption{ 
 World-sheet  topology of  $\bar{M}^4_1$   created by four fundamental interaction events  in the rest 
frame of the particle system as  observed (after Lorentz
transformation) by a moving  observer.The widths of the sheets equal
to their respective IPNs.The widths of the lower set of sheets equal
to the gaps between successive interactions. The IPNs are randomly embedded in the Newtonian $t$-axis and in the Euclidean $x$- and $x'$-axes. 
In this  graph the ratio of the velocities  is  $v/c=0.996667$ and the scale factors are arbitrary.}
\label{Fig_6_2}
\end{figure}

\begin{figure}[htbp]
\begin{center}
\leavevmode
\epsfxsize=0.5\textwidth
\epsffile{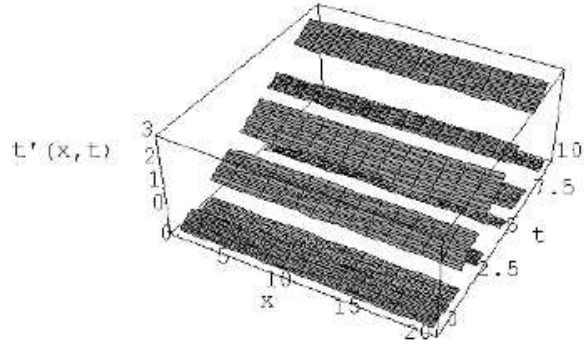}
\end{center}
\label{Fig_6_3}
\caption{ $\bar{M}^4_1$ space-time sheet  
topology due  to five fundamental interaction events in the rest
 frame generated through Lorentz transformation and seen by  a moving observer
(Data as in Fig. \protect\ref{Fig_6_2}). 
It is evident, how strongly the topology
depends on observer's motion on the number of the interactions taking
place in the neighbourhood of the observed system in its 
rest frame of reference.}
\end{figure}

\begin{figure}[htbp]
\begin{center}
\leavevmode
\epsfxsize=0.5\textwidth
\epsffile{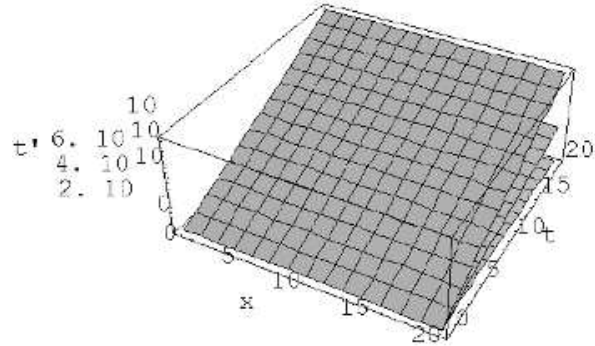}
\end{center}
\label{Fig_6_4}
\caption{Three  world  sheets giving  the time for a moving observer  with  
three different velocity ratios, $\beta=v/c $, with respect to the interacting  pair of 
particles due to one single IPN. Arbitrary $x$ and $t$ scale factors. The $t'$ 
scale factor  follows  from the Lorentz transformation.}
\end{figure}

\begin{figure}[htbp]
\begin{center}
\leavevmode
\epsfxsize=0.5\textwidth
\epsffile{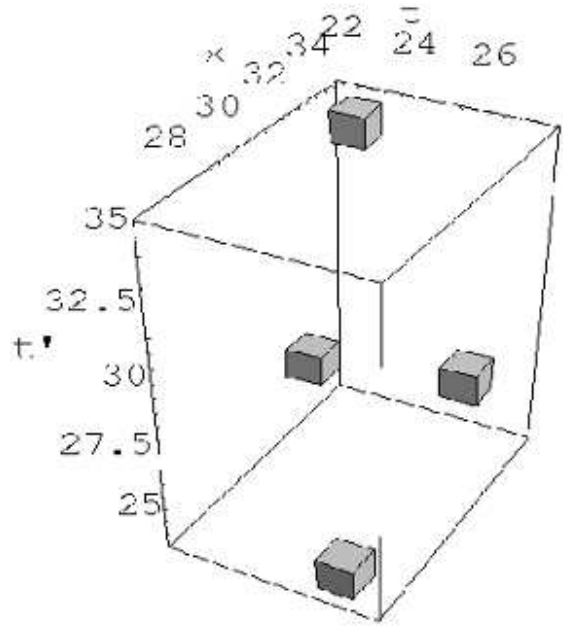}
\end{center}
\label{Fig_6_5}
\caption{Space-time neighbourhoods $\{S_i' \in\bar{M}^4_1| i=1, 2, 3, 4\} $
created by four random IPNs in the rest frame of reference at the  fixed  point $(y,z)$ 
and in the time  neighbourhoods $\{\tau_i|  i=1, 2, 3, 4\}$ 
(arbitrary scale factors).}
\end{figure}

\begin{figure}[htbp]
\begin{center}
\leavevmode
\epsfxsize=0.5\textwidth
\epsffile{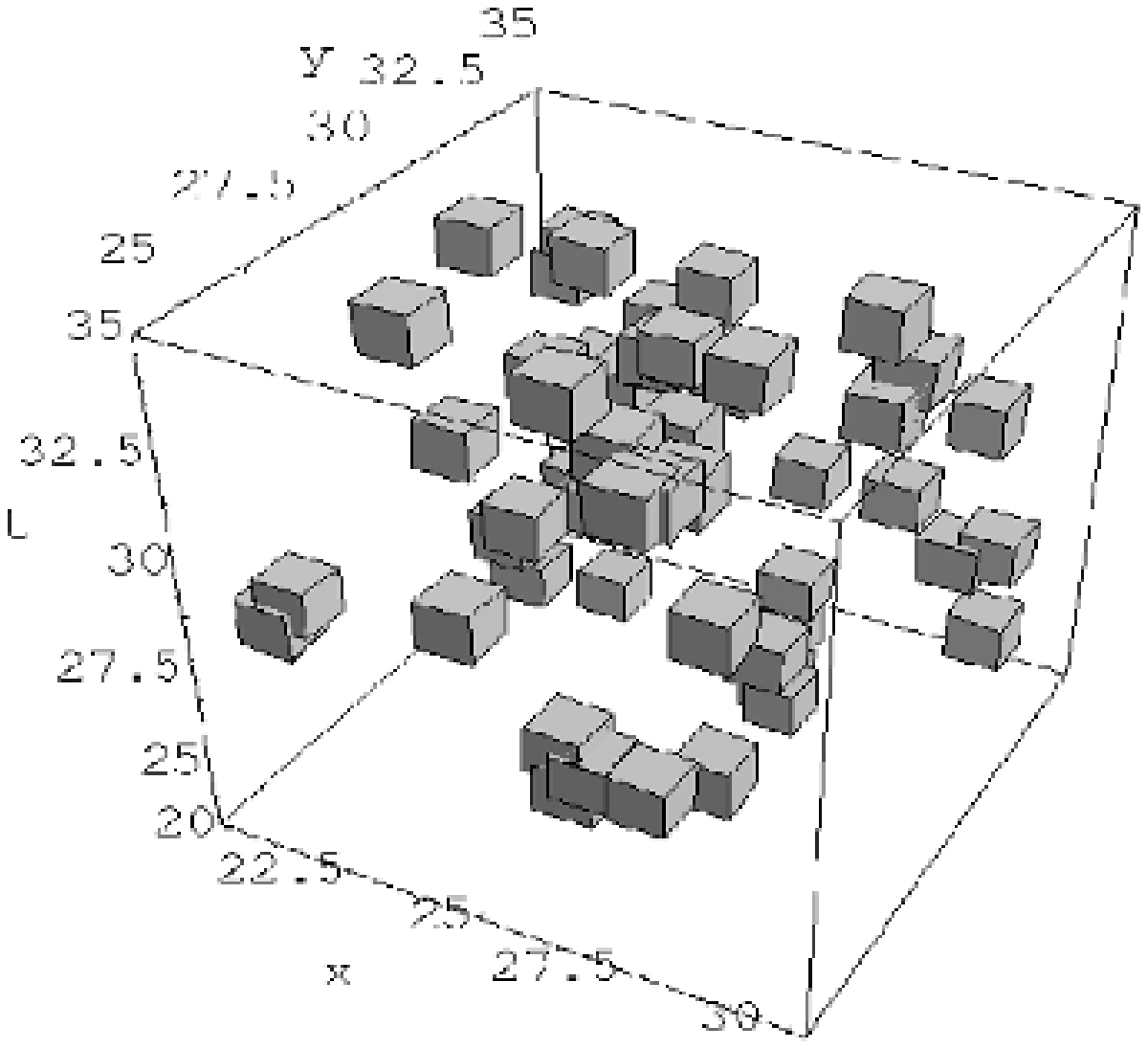}
\end{center}
\label{Fig_6_6}
\caption{  Fifty $ \bar{M}^4_1$  space-time neighbourhoods 
created  by  an  equal  number 
of  fundamental  interactions in the rest frame of  the interacting  particles. 
The continuity is  generated through overlapping of  adjacent neighbourhoods. 
For small numbers of partially simultaneous interactions $ \bar{M}^4_1$  is disconnected.}
\end{figure}

\setcounter{equation}{0}
\setcounter{table}{0}
\setcounter{figure}{0}
\section{Chrono-topology, QFT and irreversibility }

\label{section_7}

  Roger Penrose in his celebrated books ''Emperor's  New  Mind''  and ''Shadows of the Mind'' 
has analyzed virtually all problems of contemporary Physics. 
A particularly important position takes among them the problem 
of the reduction of the wave function. He - as now we may see - very 
clearly states that ''something new must be discovered to solve these 
problems''.
We believe that this is the 
new time topology, the chrono-topology. Let us start it in a slightly 
different way:
There are two simple, since long puzzling, fundamental facts  in physics:  

\subsection{Boltzmann versus Schroedinger}

\begin{itemize}
\item
  The one fact is that Quantum Statistical Mechanics  (QSM)  
is based on essentially, 
the expression ''$ \exp[-E/kT]$ ''  or its various formal expressions.

\item  The second fact is that quantum field theories (QFT)  produced 
hitherto,  instead,   the factor  ''$ \exp[-i Et/\hbar ] $''. 
\end{itemize}

  The deeper reason for the difference of these two kinds of theories 
resides in the spacetime topology \cite{1}. 
QFT sits in Minkowski's topology, $( ict, x, y, z) \in M^4$, 
QSM is based in the Euclidean topology, $(t, x, y, z) \in R^4$ \cite{52}.

Many prominent authors have constructed highly  sophisticated 
theories aiming  at  paving the way for QSM to move from $R^4$ to $M^4$. 
However, a simple problem 
remains still unsolved: {\it The derivation  of  the  statistical, 
or  the Boltzmann factor,
$\exp[-E/kT]$, from QFT in  Minkowski's topology.}

\begin{enumerate}

\item N.~N.~Bogolubov introduces in his books \cite{54} the factor
in question in a directly {\it ad hoc} way.

\item On the other hand, J.~Glimm  and A.~Jaffe  \cite{53}  obtain the 
factor $\exp[-E/kT]  $
from  the Gibbs ensemble, i.e., from an extraneous theory.  

\item Similarly, Itzykson and Zuber \cite{52} consider the expression  
$\exp[-\beta(H - \mu N)]  $ in
their book on QFT as   {\it a priori } given, where $\beta = 1/kT$   and  $\mu$ 
is the chemical potential.
\end{enumerate}

\subsection{Time and temperature - The Wick rotation problem}

 Other authors apply the Wick  rotation, $ t\rightarrow t' = -it$, 
on the evolution operator,
with a view of coming closer to the desired exponent, $-E/kT$. 
This means, of course, among other things, that:

\begin{itemize}
\item[(i)] Time becomes complex, i.e., foreign to the physical reality.

\item[(ii)] The imaginary part of the universal Newtonian time, is 
put $Im \{t\} = \beta=1/kT$.
It can hardly be physically related with a temperature 
(which is not as universal as 
the Newtonian time is) and with the Boltzmann constant of gases  ( In fact, 
the temperature is not related to $(Im \{t\})^{-1}$, but to 
$\langle \delta(\tau)\rangle^{-1} $, 
where $\langle \delta(\tau)\rangle $   is the average interaction time in the 
system of the particles under consideration, \cite{57}).  And,  
\item[(iii)]   the Wick rotation, 
$\{ t\rightarrow t' = -it | \sqrt{1-(v/c)^2} \rightarrow
\sqrt{1+(v/c)^2}\} $, implies, if $c$ is a 
reference-frame-independent universal constant, 
detrimental consequences  for special relativity, because it makes $\gamma$
smaller than $1$. 

For general relativity a complex time variable leads 
to metrics not deriving from gravitational fields \cite{44}.

Despite this fact, Hawking bases his discussion of the black holes 
on the transformation $t \rightarrow i\tau $ \cite{63}.
If the analytic continuation were valid in this particular case from the
point of view of physics, then the Minkowski space,  
and more generally the hyperbolicity would not be necessary for relativity. 
Relativity, however, cannot be formulated in spaces with posotive 
definite metrics.

\end{itemize}

This fact rules out the Wick rotation as a 
means for deriving the statistical factor $\exp[-E/kT]$  in QFT.

   There are also many other, more elaborated approaches \cite{55},  
to the problem of bringing together QFT and QSM in the same space-time.

    In order to give our main  application of the 
chrono-topology  we need  some definitions.
   The chrono - topology (helps  to eliminate  some of the paradoxes and) 
is based on a very simple and obvious  observation that a time  not associated 
with physical changes is neither needed nor definable. 
It  leads  to the question:   How does change the Hamiltonian for 
$t \in \tau$?

\subsection{Time and observability}

   It is interesting to note an unexpected 
consequence of the chrono-topology:  It 
concerns the variation of the Hamiltonian and the Lagrangian densities for 
$t \in \tau$. 
To clarify the situation we give 
 
\begin{defi}
\label{Definition_7_1}
   An  IPN, $\tau_\lambda$,   is an injective  map of the $\lambda$-th 
change, $\Delta O^j_\lambda$, of the $j$-th  observable, $O^j$,   in $\Rset^1$
caused by a fundamental interaction:
\begin{equation}
\label{7_1}
\Delta O^j_\lambda \rightarrow f( \Delta O^j_\lambda ) =\tau_\lambda
\in {\cal T}_\Lambda \subset \Rset^1. 
\end{equation}
\end{defi}

{\bf Remark 7.1}

    The index ''j'' in   $\Delta O^j_\lambda $ has been omitted in 
$ {\cal T}_\Lambda  = \bigcup_{\lambda=1}^\Lambda \tau_\lambda$,  
because the knowlegde 
is  not generally desired, 
from which  interaction comes a particular observable 
change. 

   However, it is very important to observe that 
{\it \bf a  physical change  is not  observable
inside an IPN}, $\tau_\lambda$. Because otherwise, a $\tau'_\lambda \subset 
\tau_\lambda$     would  be definable, contrary to 
the assumption that no $\tau_\lambda$ is divisible.
This is necessary, since there is no experimental evidence for the 
possibility to stop a started fundamental interaction.   

Consequently, 
every change, $\Delta O_\lambda $, of a physical observable 
covers the entire IPN, $\tau_\lambda$, which is the map 
of $\Delta O_\lambda $   and it can be observed at the end of its 
creation.  This state of affairs gives seemingly rise to a 
contradiction:  

\begin{itemize}
\item[i)]  For $t\in\tau_{\lambda}  H(t) , L(t)= \mbox{Constant.} $
\item[ii)] For $t\not\in \tau_{\lambda}  H(t) ,L(t)=  \mbox{Zero.}$
\end{itemize}

The  way out  from  this dilemma is that $ \tau_\lambda$  is  open and  
$\bar{\tau_\lambda}$ is its  closure. Then $H(t)$
and  $L(t)$  change for $t\in \bar{\tau_\lambda} \setminus Int \tau_\lambda$.  

{\bf Remark 7.2}
\begin{equation}
\label{7_2}
\mbox{\it  Measure of} \quad [\bar{\tau_\lambda}\setminus Int \:\tau ]=0.           
\end{equation} 

This result is of impor\-tance to the integra\-tion of time-or\-dered 
pro\-ducts  of  non-com\-muting opera\-tors in QFT. Because of (\ref{7_1}) 
the operators are time-independent in the interval of integration, \cite{57}.

{\bf Remark 7.3}

{\it  {\bf Remark 7.1}  and  {\bf Proposition 6.1}} offer a 
possibility to understand, that quarks are 
not directly observable due to their asymptotic freedom.  
Particles, in general, are observable only during or after interaction.
They are not observable, if they do not interact.

\subsection{Randomness in QFT}

\begin{defi}
\label{def_7_2}
    A field ${\cal L}={\cal L}( \phi(x, t), \partial \phi(x, t)) \in \Rset^1 $
  is called a generalized random field, if  for  
   ${\cal L} < \xi \in \Rset^1$   a probability $P(\xi)$ is given 
such that the conditions are fulfilled:
\begin{itemize}
\item[1.]       $P(\xi_1 )  = P(\xi_2 )$,   if   $\xi_1 =\xi_2$,
\item[2.]      $\lim_{\xi\to -\infty} P(\xi) =0 $  and  
             $\lim_{\xi\to \infty} P(\xi) =1 $,                                 
 \label{def_7_2.2}
\item[3.]    $\lim_{\xi\to a-0} P(\xi) = P(a).$
\end{itemize}       .
\end{defi}

{\bf Remark  7.4}

The limits $(-\infty, \infty)$ in Definition \ref{def_7_2}
 above  must in our case be replaced by some finite 
numbers $(a, b)$, because the field, ${\cal L}$, does not become infinite.

\begin{defi}
\label{def_7_3}
    A generalized random field, $\cal L$, is called infinitely divisible,
if for every $\Lambda \in \Zset^+ $the decomposition is possible \cite{56}
\begin{equation}
{\cal L} = {\cal L}_1 + {\cal L}_2 + \ldots + {\cal L}_{\Lambda \kappa}, \quad
\forall \kappa \in \Zset^+  \mbox{and} x \in \bar{M}^4_{\Lambda_\kappa}
\end{equation}
in which the   $\{{\cal L}_{\lambda}( \phi(x,t), \partial \phi(x,t))\} $ 
are mutually independent, 
have identical probability distributions, $\{P(\xi_\lambda )\}$ 
and are different  from  zero only  in  their  corresponding 
IPNs  $\tau_\lambda \in {\cal T}_4$.
\end{defi}

{\bf Remark 7.5 }

According to \cite{57}  the decomposition of the field Lagrangian density 
into an 
arbitrary  number of  identical terms with identical probability measures at 
any point of the space-time is mathematically perfect in the framework of the
theory of the generalized and infinitly divisible fields. 

   However, from the physical point of view such a decomposition would violate 
all conservation laws in the Minkowski or in the Euclidean space-time. 
That impossibility  disappears in the framework of our many-folded 
space-time  $\bar{M}^4_\kappa$ for 
every $\kappa \in \Zset^+$, since in every IPN the conservation laws 
hold separately.

\begin{eqnarray}
\label{7_4_a}
\mbox{1) }{\cal L}  =& 
{\cal L}_1+{\cal L}_2 \quad\quad\quad\quad &\mbox{in}\: \bar{M}_{\kappa 2}^4 \\
\label{7_4_b}
\mbox{2) }{\cal L}  =& 
{\cal L}_1+{\cal L}_2 +{\cal L}_3\quad\quad & \mbox{in}\: \bar{M}_{\kappa 3}^4 \\
\label{7_4_c}
\mbox{3) }{\cal L}  =& \:\:{\cal L}_1+{\cal L}_2 + \ldots +
{\cal L}_{\Lambda_{\kappa }}
\: &\mbox{in}\: \bar{M}_{\kappa\Lambda_{\kappa}}^4. \\
& \ldots &\nonumber\\
\end{eqnarray}

{\bf Remark  7.6}

    The range of $\{{\cal L }_{\lambda}\}$  
is determined by the domains both of 
$x$ and $t$. The domain of $x$ does not depend only on  the problem 
at hand but also on 
the  velocity of the motion 
of the observer with respect to the rest frame of the 
interacting particle,  since the space topology  depends  on the topology of 
the time. The domain of $t$ depends on the nature of the interaction. 

   Hence, in the defining condition (\ref{def_7_2}) above of the 
generalized random field 
the limit  value $\bar{\xi}$   of  $\xi_\lambda$, for which the conditions
\begin{itemize}
\item[1)]  
\begin{equation}
 {\cal L}_\lambda (\phi(x, t), \partial \phi (x, t)] > \xi_\lambda
\quad \mbox{for}\quad   t\in \tau_\lambda 
\end{equation}

and 
\item[2)] $$\lim_{\xi\to \bar{\xi}} P(\xi)=1 $$
\end{itemize}                                       
are fulfilled, is not infinite.

The chrono-topology  induced by the injective maps in the Definition 3.1 of 
time consists of IPNs that are structured as follows:

\begin{itemize}
 
\item[ (i)]  For systems with few IPNs  is  time given by the union 
\begin{equation}
\label{7_5}
 {\cal T}_{Disconnected} = \bigcup_\lambda \tau_\lambda, 
\quad\lambda \in \Lambda.
\end{equation}
 $\Lambda$ is not very large and  are disconnectedly and 
randomly embedded in the Newtonian time, $\Rset^1 $. 
All observers live in $T\subset \Rset^1$, but not the atomic, the nuclear 
and the sub-nuclear particles; they ''live'' in ${\cal T}_{Disconnected}$ 
  as defined in their rest frame of reference. 

   In order to observe them, a Lorentz transformation  is required.
The time-space for  atomic, nuclear and sub-nuclear particles  
is the chrono-topological space satisfying the 
separation axioms of ${\cal T}_4$ described in the introduction.

\item[(ii)] For systems with  $\Lambda $ very large the system-time 
topology  may change radically in the rest frame of reference 
of the interacting particles. 

\end{itemize}

   The union    
\begin{equation}
\label{7_6}
{\cal T}_{\Lambda} = \bigcup_\lambda \tau_\lambda, 
\quad\lambda \in \Lambda
\end{equation}
may become  equal to the sum of some disconnected spaces 
${\cal T}_{Disconnected}$ and 
of some partitions  $\{{\cal P}_{\Lambda_\kappa}\in \Rset^1 \}$ 
dense  in disconnected subsets, $T$, of $\Rset^1$. If  the cardinality 
of $\Lambda$ approaches $\aleph _0$, then ${\cal T}_\Lambda$  may  
with high probability, but not certainly, be densely  embedded in 
${\cal P}_\Lambda \in \Rset^1$, and 

\begin{center}
Cardinality $({\cal T}_\Lambda )\to {\bf c}$.
\end{center}

\subsection{ A fundamental quantum proposition}

   After  the above clarifications  we shall prove the following 
{\bf Proposition 7.1}

\begin{itemize}
\item[1.]  Let ${\cal T}_\kappa ={\cal T}_{Disconnected}$  
 be a set of  IPNs due to the  interaction  Hamiltonian density 
${\cal H}(\phi(x, t), \partial \phi(x, t)) \neq 0$ 
on the elements of a partition ${\cal P}_\kappa
\subset {\cal T}_\kappa$ of ${\cal T}_\kappa$,   and
\begin{equation}
\label{7_7}
{\cal H}(\phi(x, t), \partial \phi (x, t)) \equiv 0 \quad\mbox{for}\quad 
t\neq {\cal P}_\kappa
\end{equation}

\item[2.]  Let the Lagrangian  have the form
\begin{equation}
\label{7_8}
{\cal L}(\phi(x, t), \partial \phi  (x, t)) = \pi (x, t) \partial_0 \phi(x, t) 
-{\cal H}(\phi(x, t), \partial \phi(x, t)) 
\end{equation}

for $\tau_\lambda \in {\cal P}_\kappa \subset {\cal T}_\kappa $ 
and
${\cal L}(\phi(x, t), \partial \phi(x, t)) \equiv 0$
for $ t\not\in {\cal P}_\kappa$
with 
$$\pi (x, t) = \frac{\partial}{\partial \partial \phi(x, t)} {\cal L}(
\phi(x, t), \partial_0 (x, t)).$$

\item[3.]  Let $\partial_0 \phi(x, t) dt = [ \phi(x, t+d\tau_\lambda) - \phi(x,
t)] := d\phi(x, s)$   be the path variation  for  
some   $s\in {\cal P}_\kappa$  and  every $x\in\bar{M}_\kappa^4$.

\item[4.]  Let  ${\cal L}(\phi(x, t), \partial \phi(x, t))$  be a generalized,
random and infinitely divisible field,
\begin{equation}
\label{7_9}
 {\cal L}= {\cal L}_1 + {\cal L}_2 + \ldots +{\cal L}_\kappa .
\end{equation}

\item[5.] Let the  field action-integral, $A_\Lambda (S_\kappa)$, 
be quantized  \cite{47} by
\begin{equation}
\label{7_10}
A_\Lambda (S_\kappa) = \int_{S_\kappa}
{\cal L}(\phi(x, t), \partial \phi(x, t)) d^4 x = \pm \hbar \Lambda (j, \sigma),
\end{equation}
where                 
\begin{equation}
\Lambda(n, \sigma) = \pi \left\{ \begin{array}{c}
                        n + 1/2, \sigma =1 \\
                         n, \sigma =2 
                            \end{array} \right\} .
\end{equation}
with $n=0, 1, 2, \ldots $        and   $S_\kappa \subset \bar{M}_\kappa^4$.     

\end{itemize} 
Then the time evolution operator of the system  is 
\begin{eqnarray}
\label{7_12}
{\cal U}({\cal T}_{\kappa})&=&\exp  \langle  [(i\hbar)^{-1}\int _{S_\kappa}
d^4x 
{\cal  H}(\phi(x, t), \partial \phi(x, t)) \pm i\Lambda (j, \sigma)] \\
& & \quad \times [ cos[\Lambda(j,\sigma)] \mp i sin[ \Lambda(j, \sigma)]]\rangle,
\nonumber
\end{eqnarray}
and brakes down into two parts:
\begin{itemize}
\item[(i)]  non  measure preserving (nmp),
\begin{equation}
\label{7_13_a}
{\cal U}_{nmp}({\cal T}_{\kappa}) = \exp [ \frac{(-1)^n}{\hbar} \int H(s) ds +
\Lambda(n, 1)],
\end{equation}
or                                                                              \item[(ii)] unitary (u).
\begin{equation}
\label{7_13_b}
{\cal U}_u ({\cal T}_{\kappa}) = \exp [ +(i\hbar)^{-1} \int H(s) ds \pm 
i\Lambda(n, 2)], \quad s\in {\cal P}_\kappa 
\end{equation}

\end{itemize}

{\bf 
Remark 7.7}

   The proof of the Fundamental Proposition 7.1 will be 
given on the basis of the 
solution of the equation governing our time evolution of the state vector 
$\Psi $ \cite{57}  in 
the time space ${\cal T}_\kappa $, a disconnected subset of $\Rset^1$.

\begin{equation}
\label{7_14}
i\hbar \frac{\partial \Psi}{ \partial t} = H(t)\Psi(t), \quad
t\in {\cal T}_\kappa.
\end{equation}

 The difference between the time evolution equation (\ref{7_14})
in chrono-topology   
and in the conventional  \cite{52} evolution in time proceeding 
in the topology of the 
Newtonian time, $\Rset^1$, is that our topology implies (in Penrose's notation) the process ${\bf U} + {\bf R} $.
The Newtonian time topology cannot physically accommodate infinitely divisible fields,
because it induces the violation of the conservation laws
as a consequence of the infinite divisibility of the Lagrangian and the 
Hamiltonian densities of the field.
Hence, the Newtonian time topology leads only to $\bf U$ evolution  according to                                                                                
\begin{equation}
\label{7_15}
i\hbar \frac{\partial \Psi}{ \partial t} = H(t)\Psi(t), t\in \Rset^1.
\end{equation}

{\bf Proof  of the fundamental proposition 7.1}

  We start the proof by giving the solution of (\ref{7_14})
\begin{eqnarray}
{\cal U}({\cal T}_{\kappa}) &=& \{\exp[ -i\hbar^{-1} \int {\cal H}
(\phi(x'), \partial \phi(x')) {dx'}^4  ]\}, \quad x'\in{\bar{M}^4}_{\kappa}
 \\
  &=& \{\exp[ -i\hbar^{-1} \int H(s) ds ]\}, \quad s\in {\cal P}_{\kappa}
\end{eqnarray}
   We:
\begin{itemize} 
\item[1)]  Combine the above integral with premise 2.
\item[2)] Write the exponential of the sum of the terms as a 
product of  exponentials of the 
    terms of the sum and 
\item[3)] Develop the factor $\exp[ -i\hbar^{-1} \int dx^4 \partial_0 
\phi(x) \pi (x)] $   in a power series.
\end{itemize}

\begin{eqnarray}
\label{7_18}
{\cal U}({\cal T}_{\kappa}) &=& \left\{ 
1 + \sum_{\Lambda_{\kappa} =1}^\infty \frac{ (i\hbar)^{-\Lambda_{\kappa}}}{
\Lambda_{\kappa} !}\prod_{\lambda_{\kappa} =1}^\Lambda
\int_{\bar{M}^4_1} {\cal L}(\phi(x'_{\lambda_{\kappa} }),
\partial\phi(x'_{\lambda_{\kappa} })) dx'_{\lambda_{\kappa} } \right\}\\
& &\times
\exp[(-i\hbar)^{-1}\int_{x'\in \bar{M}^4_1}
{\cal L}(\phi(x'), \partial\phi(x')) d^4x'] \nonumber
\end{eqnarray}

   Next, we use premise 3.  in the form  $\partial_0 \phi(x') d^4x' =
d\phi(q, t')d^3q$, insert it  into (\ref{7_18})  
together  with premise 4.  and put $\Lambda_{\kappa} = n$  in the $n$-th 
term of the partition ${\cal P}_{\Lambda_{\kappa} = n} \forall
n\in \Zset^+$. 

The  $\lambda_{\kappa}$-th integration in (\ref{7_18}) is carried out on the 
$\lambda_{\kappa}$-th sector of the 
$\lambda_{\kappa}$-folded-in-time of the space-time super surface. 
The result is:

\begin{eqnarray}
\label{7_19}
{\cal U}({\cal T}_{\Lambda_\kappa}) &=& \left\{
1 + \sum_{\lambda_{\kappa} =1}^\infty \frac{ (i\hbar)^{-\lambda_{\kappa}}}{
\lambda_{\kappa} !}
\prod_{\eta = 1}^{\kappa }
\langle \int_{Q^3} d^3q'
\int_{\phi(\underline{q}, 0)}^{\phi(\overline{q} \delta(\tau_{\eta}))}
d\phi(q', s'_\eta) \pi(q', s'_\eta)\right. \\
& & \left. \times \exp[(-i\hbar)^{-1} \int_{\bar{M}^4_\kappa} {\cal L}
(\phi(x'_{\lambda_{\kappa} }), \partial \phi(x'_{\lambda_{\kappa} })) d^4 
x'_{\lambda_{\kappa} }]\rangle \right\}, \quad s'_\eta\in\tau_\eta. \nonumber
\end{eqnarray}

According to the (\ref{def_7_3})  of the infinitely 
divisible fields $\{{\cal L}_{\lambda_{\kappa}} \}$ have all  
$\lambda_{\kappa}$-independent  probability distributions.  
Using this  property  and omitting, 
therefore, the index $\kappa$, in $\lambda_{\kappa}$ under  the product  sign  
in  (\ref{7_19}) we can sum-up the series. 

   Using again the relation 
\begin{equation}
\pi(x, t)\partial_0 \phi(x, t) = {\cal L}(\phi(x, t), \partial  \phi (x, t)) + 
{\cal H}(\phi(x, t), \partial  \phi (x, t))
\end{equation}

the result after sumation is 

\begin{eqnarray}
{\cal U}({\cal T}_{\Lambda_{\kappa} }) &=&
\exp [ (i\hbar)^{-1} \int_{Q^3} d^3q' 
d^3q \int_{\phi(\underline{q}, 0)}^{(\overline{q}, \delta(\tau))}
d\phi(q' , s'_{\eta}) \pi
(q', s'_\eta ) ] \nonumber  \\
& & \times
\exp[(-i\hbar)^{-1}  \int_{\bar{M}^4_\kappa} {\cal L} (\phi(x''), \partial 
\phi (x'')) d^4x'' )] \\
&=& \exp\langle (i\hbar)^{-1}  \int_{\bar{M}^4_\kappa}
d^4x'' [ {\cal L} (\phi(x, t), \partial \phi (x, t)) +
{\cal H}(\phi(x, t), \partial \phi  (x, t))]\nonumber \\
& & \times
\exp[(-i\hbar)^{-1} \int_{\bar{M}^4_\kappa} {\cal L} (\phi(x'), \partial\phi
(x')) d^4x' \rangle \\
&=& \exp\langle (i\hbar)^{-1}  \int_{\bar{M}^4_\kappa}[
{\cal L} (\phi(x, t), \partial \phi(x, t)) + {\cal H}(\phi(x), \partial
\phi(x)) d^4x] \nonumber \\
& &\times
\{ \cos [ \hbar^{-1} \int_{\bar{M}^4_\kappa}
{\cal L}  (\phi(x'), \partial \phi
(x')) d^4x' ] \nonumber \\
& & - i \sin [\hbar^{-1} \int_{\bar{M}^4_{\kappa}}{\cal L}  (\phi(x'), \partial
 \phi (x')) d^4x' )]\} \rangle 
\end{eqnarray}

Or, after separation of the real from the imaginary part in the exponent,
we get 
the fundamental formula for the time evolution operator  which is a realization 
of both quantum dynamical processes ${\bf U}$ and ${\bf R}$.
\begin{eqnarray}
\label{7_20}
{\cal U}({\cal T}_{\Lambda_{\kappa})  } &=&
\exp\langle\{ (i\hbar)^{-1}  \int_{\bar{M}^4_\kappa}  d^4x
[ {\cal L}  (\phi(x, t), \partial \phi  (x, t)) 
+
{\cal H}(\phi(x, t), \partial \phi  (x, t))]\nonumber \\
 & &\times \cos [ \hbar^{-1}  \int_{\bar{M}^4_\kappa}
 {\cal L}  (\phi(x'), \partial \phi (x')) d^4x'] \nonumber \\
&+& 
\{(-\hbar)^{-1} \int_{\bar{M}^4_\kappa} d^4x
[ {\cal L}  (\phi(x, t), \partial  \phi (x, t)) +
{\cal H}(\phi(x, t), \partial  \phi (x, t))]\nonumber \\
& &\times
\{ \mp \sin [ \hbar^{-1} \int_{\bar{M}^4_\kappa}
{\cal L}  (\phi(x'), \partial \phi (x')) d^4x']\} \} \rangle .  
\end{eqnarray}

\subsection{Unitary ${\bf U}$ and Reduction ${\bf R}$  operators}

Next, we apply the quantization condition (\ref{7_10}) 
on the action integral in the
''cos''  and ''sin'' expressions of (\ref{7_20}). 
The result is
\begin{eqnarray}
\label{7_21}
{\cal U}({\cal T}_{\Lambda_{\kappa}} ) &=&
\exp\langle (i\hbar)^{-1}  
\int_{\bar{M}^4_\kappa} d^4x
[{\cal H}(\phi(x, t), \partial \phi  (x, t))\pm i\Lambda(j, \sigma)] \nonumber \\
& & \times \{\cos [\Lambda(j, \sigma)] \mp i\sin[ \Lambda(j, \sigma)]\} \rangle .
\end{eqnarray}

   Remembering (\ref{7_10}) we see that  (\ref{7_21}) can be 
written as the product of two exponential factors \cite{14}, \cite{63}: 
\begin{itemize}
\item[a)]   The  ${\bf U}$  factor  represents the unitary 
part of the evolution, $ {\cal U}_u( {\cal T}_{{\kappa}} ) $

\item[b)]   The  ${\bf R}$  factor  representing the incoherent evolution,
$ {\cal U}_{nmp}( {\cal T}_{{\kappa}} ) $
\end{itemize}

This completes  the proof of the  Fundamental Proposition 7.1.

{\bf Remark 7.8}

   This is the result contemplated by Prigogine and 
Penrose in their 
publications. The most interesting feature of  (\ref{7_21}) 
is that it exhibits simultaneously ${\bf U}  + {\bf R} $ properties,
just like those postulated  by  Penrose and  required for 
the implementation of the wave function reduction.

{\bf Remark 7.9}  

A new feature of the evolution operator (\ref{7_21}) is the spontaneous 
$\Lambda(n, \sigma) $ renormalization of the action 
integral in the exponent by
means of the term $\Lambda(n,\sigma)$, see also (\ref{7_23}) and (\ref{7_24}) 
below.

\subsection{Random fields and functional integral approach}

{\bf Remark 7.10}

    Each integral in the series (\ref{7_19}) is finite, because $\phi$ is a 
solution of the Euler-Lagrange differential  equation satisfying the 
appropriate boundary conditions. The series 
converges, because  each term is the  
power of a  definite integral.  The proof  follows  by majorization.

{\bf Remark  7.11}

The continuous  parameter $s_{k}'$ characterizes the  integration  path 
$\d\phi(q, s) $    in 
which are the space $[\phi(t), \phi(t+\tau)] $
for the initial and the final values of $\phi(t)$  for $s_{k}' = t $, and 
for $ s_{k}' = \delta(\tau)$   respectively, 
while $\phi$    coincides at  these values  of $s$  
with the lower and the upper limits of the  $\phi$-path integration.

{\bf Corollary 7.1 }

    The contribution ${\cal U}_{\lambda_{\kappa}}( {\cal T}_{\kappa}) $, 
to the evolution operator of the  path integral in  
 the expansion (\ref{7_19}) vanishes in the limit $\lambda_{\kappa}\to \infty$.

\hspace{1cm}

{\bf  Proof }

     The $\lambda_{\kappa} $-th term in equation (\ref{7_19})  is

\begin{eqnarray}
\label{7_22}
{\cal U}_{{\kappa}}( {\cal T}_{\kappa})  &=&
\frac{(i\hbar)^{-\lambda_{\kappa}}}{ \lambda_{\kappa} !}
\prod_{\eta=1}^{\lambda_{\kappa}}
\int_{Q^3}
d^3q \int_{\phi(\underline{q}, 0)}^{\phi(\overline{q}, \delta(\tau_{\eta}))}
d\phi( q, s_\eta) \pi(q, s_\eta) \nonumber \\
    &\times &
\exp [(-i\hbar)^{-1} 
\int_{\bar{M}^4_{\kappa}}
{\cal L} (\phi(\phi(x'), \partial \phi(x'))d^4x' ]
\end{eqnarray}

The integral over $d\phi$  sums values along all paths and 
the integral over $dq^3$   sums the function $\phi(q,s)$-values 
along every selected individual paths between the two 
fixed limit function values 
$\phi(\underline{q}, 0)$ and $\phi(\overline{q}, \delta(\tau))$ .

   The measures in the integrals of the  product are 
well-defined  and they exist on the compact supports:
\begin{itemize}
\item[i)] $\left[ \phi(0), \phi(\delta(\tau_{\lambda_{\kappa}})) \right]$, and
\item[ii)]   $Q^3 \subset R^3$.
\end{itemize}
    Since the  factors $\{ f_\eta \}$ of the product 
$$
\prod_{\eta=1}^{\lambda_{\kappa}}
\int_{Q^3}
d^3q \int_{\phi(\underline{q}, 0)}^{\phi(\overline{q}, \delta(\tau_{\eta}))}
d\phi( q, s_\eta) \pi(q, s_\eta)
$$
are independent of $\eta$, we have $\prod_{\eta=1}^{\lambda_{\kappa}} f^\eta
= f^{\lambda_{\kappa}}$, and since the factor $(\lambda_{\kappa} !)^{-1}$
decreases faster than the power, $f^{\lambda_{\kappa}}$, for 
$\lambda_{\kappa} \to \infty $, 
there holds true: $\lim_{\lambda_{\kappa} \to \infty}  
f^{\lambda_{\kappa}} / \lambda_{\kappa} !  = 0$,
and the functional integral does not contribute to the evolution operator.
This completes the proof of {\bf Corollary 7.1}.

\hspace{1cm}

{\bf Remark 7.12}

     The ''phase'' factor in (\ref{7_22}) is identical to the 
one in the Feynman path integral.   
Differences appear in the functions to be integrated over $d^3q$  and 
over $d\phi(q,s)$. 
The following correspondances with the Feynman path integral  
are intringuing:

{\bf Remark 7.13}

In the limit, $\lim_{\lambda_{\kappa} \to \infty} $, the integrals become functional 
integrals with the measures 
\begin{eqnarray}
  Dq    &\Rightarrow & \prod_{\lambda_{\kappa} =1 }^{\infty} 
dq_{\lambda_{\kappa}}, \quad
 q_{\lambda_{\kappa}} \in Q^3 \subset R^3 \\
\mbox{and} & & \nonumber \\
 Dp    &\Rightarrow & \prod_{\eta}^{\infty} \pi(\phi(q, s_{\eta}), 
\partial \phi(q_{\eta}, s)) d\phi(q, s_{\eta}) 
\end{eqnarray}

for $\phi(q, s_{\eta}), \pi(\phi(q, s), s_{\eta}), 
\partial \phi (q, s_{\eta}) \in L^2$.

\subsection{Functional approach and uncertainty principle}

    In any way the contributions of these  integrals, 
similar to the Feynman path integrals, are zero because of the 
normalization factor, $\lambda_{\kappa} !$   
in the limit  $\lambda_{\kappa} \to \infty $ 
({\bf Corollary 7.1}).

   The functional integral features in (\ref{7_18}) appear 
similar or in a  slightly  more  
general form, and the theory yields in case $\lambda_{\kappa} = n, 
\forall n \in \Zset^+$   the exponential form 
of the Hamiltonian well-known from QFT. The similarities and the differences 
between the present theory and the Feynman path integral are shown in 
\ref{Table_7_1}.

\begin{table}
\label{Table_7_1}
\caption{ Comparison of the Feynman path integral properties with those 
of   the present theory and some statistical and quantum properties.}
\begin{tabular}{lccc}
\hline \\
 & Feynman && Present work \\
\hline \\
Spatial measure: & $Dq = \prod_{k=1}^{\infty} dq_k $ & $\Leftrightarrow $ &  
               $ \prod_{k=1}^{\infty} d^3q_k $\\
Functional measure: & $Dp = \prod_{k=1}^{\infty} dp_k $ & $ \Leftrightarrow $ &
$ \prod_{k=1}^{\infty} \pi(\phi(q_k, s), \partial \phi (q_k, s))d\phi(q_k,s)$ \\

Normalization:    & ? & $\Leftrightarrow $  & $  1/k !$ \\
Uncertainty Principle:&   no & &                 yes \\
Gibbs ensemble:   &       no & &                 yes \\
\hline
\end{tabular}
\end{table}

 In the case $j=(2n+1)/2$ in (\ref{7_21}) the exponential becomes  real 
and has a form reducible to the form known from QSM.

   We,  thus,  have two evolution operators:  One preserving the norm and one 
changing it.

    The canonical momentum, $\pi (q,s)$, enters as a weight factor - 
not as differential under integration. This makes the integration measures 
compatible with Heisenberg's Uncertainty Principle in case of 
quantization of the field action.

\subsection{Statistical ensembles and temperature in QFT}

   The measure preserving evolution is implemented through  
\begin{equation}
\label{7_23}
{\cal U}_u (\delta(\tau), 0) = \exp[ (i\hbar)^{-1} 
\int_{\bar{M}^4_\kappa} d^4x {\cal H}(\phi(x), \partial \phi (x)) \mp i\pi n ], 
n=0, 1, 2, \ldots .
\end{equation}

If the evolution does not preserve the norm of the state vector,
then it is given by 
\begin{eqnarray}
\label{7_24}
{\cal U}_{nmp}(\delta(\tau), 0) & = & \exp[\langle \frac{(-1)^n}{\hbar} 
\int_{S_\lambda} {\cal H}(\phi(x), \partial \phi (x) ) \mp \pi(n+1/2)\rangle],
 \\
& & n = 0, 1, 2, 3, \ldots  .
\end{eqnarray}

    If the state vector, $\Psi $, is expanded  in a series of 
eigenstates of  the Hamiltonian, 
and ${\cal U}_u (\delta(\tau), 0)$  or ${\cal U}_{nmp}(\delta(\tau), 0)$  acts  on $\Psi $, 
then (\ref{7_23}) and (\ref{7_24}) become respectively:
\begin{equation}
\label{7_25}
\exp[ -i \sum_{\lambda =1}^\Lambda E_\lambda \cdot 
\delta(\tau_\lambda)/\hbar \mp i\pi n ],
\end{equation}
and

\begin{equation}
\label{7_26}
\exp[ (-1)^n \sum_{\lambda =1}^\Lambda  E_\lambda \cdot 
\delta(\tau_\lambda)/\hbar \mp \pi (n + 1/2)]. 
\end{equation}

{\bf Corollary 7.2}

   The temperature of a system of particles interacting via a fundamental interaction 
is inversely proportional to the average diameter of the interaction 
proper-time neighbourhoods 
$\langle \delta(\tau)\rangle_{\kappa} $  defined  by 

\begin{equation}
\langle \delta(\tau)\rangle_{\kappa}  = 
\Lambda^{-1} _\kappa \sum_{\lambda =1}^\Lambda
\delta(\tau_{\lambda_{\kappa}})
\end{equation}

\hspace{3cm}

{\bf  Proof}

   We divide and  multiply  the sum in the exponent of                               
$$ \exp[  (-1)^n 
\sum_{\lambda_{\kappa} =1}^{\Lambda_{\kappa}} E_{\lambda} \cdot 
\delta(\tau_{\lambda_{\kappa}}) /\hbar +\pi (n+1/2)] $$
                                          
by $ \sum_{\lambda_\kappa =1}^{\Lambda_\kappa} \delta(\tau_{\lambda_{\kappa}})$   
and we  write for the time averaged energy per particle the expression

$$ \langle \delta(\tau_{\lambda_{\kappa}}) 
\rangle_{\kappa} \cdot \langle E \rangle_{\kappa} \cdot \Lambda_{\kappa} = 
\left( \frac{\sum_{\lambda_{\kappa} =1}^{\Lambda_{\kappa}}  
E_{\lambda_{\kappa}} 
\cdot \delta(\tau_{\lambda_{\kappa}})}{  
 \sum_{\lambda_{\kappa} =1}^{\Lambda_{\kappa}}
\delta(\tau_{\lambda_{\kappa}}) /\Lambda_{\kappa} }\right) \cdot 
 \sum_{\lambda_{\kappa} =1}^{\Lambda_{\kappa}}
 \delta(\tau_{\lambda_{\kappa}}) /\Lambda_{\kappa} .
$$

     The factors (\ref{7_25}, \ref{7_26}) entering the state vector after  the 
action of  $ {\cal U}_{nmp} (\delta (\tau), 0) $  are essentially  
the  Boltzmann  statistical  factors \cite{47}, if the system 
temperature is defined by

\begin{equation}
T_\kappa  = \frac{\hbar }{\langle \delta (\tau) \rangle_\kappa  k_B}, 
\nonumber
\end{equation}

where   $ k_B$  is the Boltzmann constant.

    The average frequency  of  the collisions, $f_{coll}$, is given by 

\begin{equation}
 f_{coll} = \langle \delta (\tau ) \rangle_{\kappa}^{-1} 
\nonumber
\end{equation}

and it allows to give a definition of the temperature 
in the framework of quantum field theory.

   This completes the proof of  {\bf Corollary 7.2}.

{\bf  Remark 7.13}

   The main features of the present temperature definition are:  
\begin{itemize}
\item[ i)] Relativity is  respected by avoiding the Wick  rotation.  
\item[ii)] The canonical  ensemble  follows from the QFT. 
\item[iii)] The temperature is related to the collision frequency \cite{57} 
in the framework of QFT.
\end{itemize}

{\bf Remark 7.14}

  It is remarkable that the averaging of the energy is by the structure of the theory
over the time these energies are possessed by the particles,
but not over the number of the particles. 
This is quite natural, because, if an energy value is possessed during 
zero time by a particle, it contributes zero to the 
average energy of the system.

\subsection{Einstein -- Bohr. Both were right}

Some authors believe that Bohr was right and Einstein wrong or vice versa
( e.g. \cite{12}, \cite{63} ) in 
their disput about the statistical character of quantum mechanics. 

It has been proved in the framework of chrono-topology, that both were right.
The reason for this fact was, that Einstein's statement, according to which
God does not
play dice,  regarded the quantum equations of motion (Schroedinger, Dirac,
etc.) {\it per se }, i.e., inside a single IPN, and in this case
God does not play dice inside $\tau$.  Indeed, the quantum equations are per
construction not statistical. However, the description of the quantum phenomena cannot be done fot $t\in \tau$. The ewason is that observation is conventionally
done in ${\cal T}_{\kappa}$. But for $t\in {\cal T}_{\kappa}$ the 
interpretation of the solutions of the quantum equations becomes necessarily
statistical \cite{47}.

Bohr's statement, on the other hand, regarded 
quantum physics results as a whole, because quantum 
physic's arena is not $\tau$ itself. The physical ''playground'' is rather
${\cal T}_{\kappa} = \bigcup \tau_{\lambda}$, and the space-time, 
$\bar{M}^4_{\kappa\lambda_kappa}$, resulting from it according to Einstein's
relativity.

Moreover since $\{\delta(\tau_{\lambda})\}$ are within limits random numbers,
an averaging  process takes place to give any measured value. 
This is the way in which  the statistical
character of quantum mechanics emerges. So much, as far
as time is concerned.
Regarding the space coordinates, a similar explanation is true:
Since in every interaction the impact parameter is a
random length, a random component is introduced in each space coordinate.
This, again, makes necessary an averaging process on the deterministic
solution of the wave equation.

It becomes, thus, clear \cite{47} that both Bohr and Einstein
were right in their respective statements. Their difference was in their
premises, which can be discerned only in the framework of chrono-topology.

\setcounter{equation}{0}
\setcounter{table}{0}
\setcounter{figure}{0}
\section{Solving the measument problem - The reduction of the state vector}
\label{section_8}
      
In what consists the reduction of the wave function?  According to Einstein 
'Gott wuerfelt nicht'. We, of course, are allowed to 'wuerfeln' and we shall do 
it for a while. 

However, before doing that we shall pay tribute of due  honor  to all  great 
scientists, who came infinitesimally close, in their publications,
to our solution 
of the problem of the present section. They  made  observations  and  remarks extremely approximating 
the solution we provide in this work and they showed to us the 
way for obtaining the solutions of several puzzles in quantum theory.    Of all 
those scientists we shall make particular reference to two whose contributions 
seems of the greatest importance from our point of view: 

   The great forerunners of this work are Ilya Prigogine  and Roger
   Penrose.

\subsection{The irreversibility question}

Prigogine made the following remarkably lucid statements in his by now
famous book ``From being to becoming'' \cite{10} :
{\em 'It is difficult to believe that the observed irreversible processes such as viscosity,
decay  of unstable  particles and  so forth, are simply illusions caused by lack of 
knowledge or by incomplete observation ... .  Therefore,
irreversibility must have 
some basic connection  with the dynamical nature of the system'.}

    Sofar  we fully agree with Prigogine's  philosophy. The
    continuation of his syllogism goes as follows: {\em '... because for
    simple types of 
    dynamical systems the predictions of classical and quantum
    mechanics  
    have  been remarkably well verified'.}
 
  Despite the remarkable verification of quantum mechanics our remarks
  are two:
\begin{itemize}
\item[i)]  The time structure is ill understood in general
  dynamics.The reason is that 
always the interaction time, $\tau_{\lambda}$, has been identified in 
quantum dynamics with the Newtonian 
time having the topology of $\Rset^1$ . 
   We abandon the topology of $\Rset^1$ and  introduce instead the notion of
   chrono-topology 
consisting of unions of IPNs. 
\item[ii)]  The predictions are in fact remarkably well verified, 
except for the description 
of the measurement process and the irreversibility.
\end{itemize}
 
   Continuing, Prigogine concludes-and this was of great help to us:
{\em ' The problems of unity of science and of time are so intimately connected that we 
cannot treat the one without the other'}.
   
\subsection{The reduction question}

 Let us now see Penrose's most clear guidance  in the matter of our problem 
\cite{14}:  {\em 'The quantum measurement problem is to understand, 
how the procedure ${\bf R}$ 
can arise-or effectively arise - as a property of a large-scale  
behavior in  ${\bf U}$-evolving 
quantum systems. The problem is  not solved   
merely by indicating a possible way 
in which an ${\bf R}$-like behavior might conceivably be accommodated. One must have 
a theory providing some understanding of the circumstances under 
which (the illusion ?) ${\bf R}$ comes about'}.

This is exactely the way we followed in constructing our theory which 
simultaneously describes the ${\bf U}$-and the 
${\bf R}$-processes in quantum field theory with exactly 
the same accuracy. Here is, however, an additional aspect: 
In our approach ${\bf R}$ comes about  not only for large-scale systems, but
also for 
single quantum particles, thus 
enabling us to solve the Schroedinger cat's puzzle too.  

   The first and basic idea came to us from \cite{1} and a first derivation has been given 
in \cite{47,57}. In Sec. \ref{section_7} of the present paper  
the full derivation is given.
    Penrose continues:   {\em 'It  appears  that  people  often think of the  precision  of 
quantum theory  as lying in its dynamical equations, namely ${\bf U}$. But ${\bf R}$ itself is also 
very precise in its prediction of probabilities, and unless it can be understood, how it 
comes about, one does not have a satisfactory theory'}.

   In the second statement by Penrose it seems to us that the freedom is contained 
that ${\bf R}$  be or not be a consequence of the same dynamics.    
We show that our theory gives both ${\bf U}$ 
and ${\bf R}$ with exactly the same precision and 
we demonstrate that ${\bf U}$ and ${\bf R}$ come about by means of  quantizing the field action-integral.

\subsection{Experience and expectation}
  
 After the above due clarifications we are ready to 'play  dice' 
and answer the questions entailed by the problem posed at 
the beginning of this section:
In studying the game, we  may  do a small calculation and figure out
what we have 
to expect after throwing a dice.

   The  probability is $1/6$ for getting any number from $1$ to $6$:

 Calculated final state of the dice $=$
\begin{equation}
     F  =  1/6 \times (\mbox{get 1})+1/6 \times
                           (\mbox{get 2})+ \cdots + 1/6 \times
                  (\mbox{get 6}).  
\label{8_1}
\end{equation}

This is, of course, the result of a {\em calculation} of what  is
foreseen. 
There is no relationship whatsoever - causal or acausal - with the future 
decision for doing or  not doing the experiment. It is an {\em empirical} 
statistical 
fact independent of  whether we play  
or not play dice in future. The result of the {\em calculation} (\ref{8_1})
will not 
change after throwing the dice.
The equation remains unaffected, if the dice shows, e.g., $5$ or anything else.
   After having played  dice we know the fact, e.g., and we may
   represent it by,

\begin{equation}  
\mbox{Exp(erimental)Res(ults)} = F_{\mbox{exp.}}
                              =  1 \times \: \mbox{(got 5)},\:  \mbox{and} \: 0 
                              \times \: \mbox{(got all others)},                            
\label{8.2}
\end{equation}

but our equation (\ref{8_1}) remains, of course, unaltered. 
'Calculation' and 'fact' are related only in our brains.
  $F$ in (\ref{8_1} is  a theory-devised construct for  predictions 
  based on empirical data, representing the possibilities for many 
 different (in this case $6$) outcomes of dicing.
   Equation (\ref{8.2})  is of a different character. 
It is constructed to represent a posteriori {\em one single fact}:   
{\em The outcome of one single experiment} 
and there can be no question about any reduction.

   Next, we may make more perfect our theory of playing dice and we
   construct an operator,
${\cal D}$, describing our dice playing. 
We want it to describe the dice-throwing. 
  This will be done by appllying  ${\cal D}$ on $F$. 
The result of this application will, if our theory is a good one, be
equal to $F_{exp}$ . 
It will induce the reduction on the paper, not in Nature.

   If  $F$  and ${\cal D}$ represent exactly the system and our action on it respectively, then 

\begin{equation}
{\cal D} F = 1\times  \mbox{(got 5)}, \quad \mbox{and}\quad  
0\times \mbox{( got all
  others)}\:  = F_{exp}, 
\label{8.3}
\end{equation}         

  ${\cal D}$ describes exactly our way of taking and throwing (the dynamics) the dice in 
the particular experiment above.  It  has  nothing to do with a statistical theory 
(Einstein).

   Let us see  a little more precisely what this means {\em in the particular experiment}: 
It means:
\begin{itemize}
\item[i)] A definite mo\-tion of the hand of the parti\-cular 
ex\-peri\-menta\-list,
implemented through a 
definite preparation and function of his hand-muscle system.  
\item[ii)] A def\-inite motion of his arm,
implemented through a definite preparation and function 
of the arm-muscle system.  
\item[iii)] A definite  elec\-trical  conduc\-tance or polari\-zation  and  
func\-tion of the neu\-ral synapses 
system etc. leading from the brain to the fingers of his hand.
\item[iv)] A certain  preparation and function of his brain,
conscious to a certain degree of the 
programme to be carried out. This degree of consciousness may differ
from one experimentalist  
to an other,
and to an  experimentalist in different experiments.  
\item[v)] A certain  interaction between his 'will' and his brain in order that the latter prepares 
itself and acts.
\end{itemize}

These five steps of preparation  are subject to large  uncertainties,
both macroscopic 
and quantum mechanical.The magnitudes of the uncertainties increase
with increasing  
index  in the above enumeration scheme from i) to v).

 Moreover, 
what is virtually fully undefined is the description in physical terms  of 
the interaction between the 'will' and the brain.
   Hence, the construction of the operator  ${\cal D}$ 
for experiments of the  above  type is 
not an easy task for today's Science and Technology. The difficulty is localized in the lack  of 
knowledge in  the quantum description of the individual human functions. 
   However,
in most physics experiments  participation of human bodyÆs functions at 
the realization of experiment's crucial parts is to a well-defined degree excluded. 
   Also, the human brain is involved only in the preparation of the experiment, 
in the analysis 
and in the interpretation of the {\em ExpRes}.  
   Hence, the construction of the operator  ${\cal D}$ 
in quantum mechanical experiments is 
in  general easier and feasible.

    Similar is the situation in quantum theory. 
Long experience and deep insight  by Euler and Lagrange and by many others
have shown two series of facts:
\begin{itemize} 
\item[i)]  If we construct a certain function, $\cal L$, 
appropriate to the problem at hand and we 
apply a variational principle, 
we derive an equation (Schroedinger) containing some operators 
$\{{\cal D}\}$ which corresponds to our problem.
\item[ii)] The actions of $\{{\cal D}\}$ on a certain function $F=f(\Psi)$
($\Psi$ 
is a wave function) describe 
satisfactorily the {\em ExpRes}, and the construction of our function, 
$\cal L$, is correct.
\end{itemize}

   Hence, if our theory is correct, then we must have:

$$
    {\cal D} f(\Psi) = ExpRes.
$$

 Some authors believe that the construction of  
$\cal D$ is impossible in the framework of 
the theory of Schroedinger's equation in such a way that the above equation is not  
true in the sense of (8.3) and R must come from extraneous agents.  
    We shall  try to examine the actual situation  in  the  framework  of our  present 
chrono-topology. We shall try first to clarify the situation through  the following definitions.

\subsection{Nature is not divisible in classical and quantal}

\begin{defi}

Every experiment in systems ranging from atomic to sub-nuclear is divided into 
two parts:
\begin{itemize}
\item[i)] The  experiment proper which involves one fundamental
  physical {\bf interaction},                                                            
relies on the laws of quantum physics and characterizes 
${\cal D}$ (${\cal D}$-process).
\item[ii)] The process of making a quantum interaction  {\bf visible} 
may  rely either on quantum laws or on laws of classical 
physics or on  both and is not characteristic of $\cal D$ (non-$\cal D$-process).
\item[iii)]  There are many ways, $X\in\{W_i, P_i\mid i = 1, 2, \ldots\}$, 
for implementing an {\em ExpRes} appropriate either to wave 
properties, $W_i$, or to particle properties, $P_i$.
but not simultaneously to both .
\item[iv)]  Part  ii) can be implemented in any one of the possible
  ways, $X\in\{W_i, P_i\mid i = 1, 2, \ldots\}$,
      and, hence, $X$ is  not an uniquely  chacteristic  part of the experiment proper.
\item[v)]  The elements of the set 
$\{ExpRes(X)\}, \forall X\in \{W_i,  P_i\mid i = 1, 2, \ldots\}$, are
equivalent:
\begin{equation}
 ExpRes(X) \Leftrightarrow  ExpRes(Y), \forall (X, Y)\in \{W_i, P_i\mid i = 1,
2, \ldots\}  .           
\label{84}
\end{equation}
\end{itemize}
\label{def81}
\end{defi}

{\bf Remark 8.1} 

According to Definition \ref{def81} an experiment in quantum 
physics consists of a fundamental interaction between two given
quantum entities, on  
the one  hand  a structured or an  
elementary particle, and on the  other  hand,  a  measuring
apparatus, whose
{\bf specifically active  part} may be another structured, 
elementary particle or field. 

{\bf Remark 8.2}

   The process of making the {\em ExpRes}  macroscopically visible
   is a separate step, exterior to the quantum measurement.
   
{\bf Remark 8.3}

     The view that in every quantum physics experiment we have the
     {\bf interaction of a quantum system} with a {\bf classical  apparatus}
     (black box approach) does not correspond 
to  the facts 
 according to the present work. Because the method used for the indication 
of the result of a fundamental interaction is not essential to the interaction 
itself. As a 
rule, the {\em ExpRes} is obtained by means of:  photomultipliers,
scintillators, Wilson chambers, Geiger-Mueller detectors, recoil
detectors, spark detectors and 
other well-known elementary particle detectors.  

The  way  to  magnify  a quantum interaction 
does  not play  an essential  part in the interpretation  {\bf per se}
and to the 
construction of  the operator $\cal D$, as (\ref{84}) makes clear. 

   Having the above clarifications in mind we can see that in
   constructing 
   our operator, $\cal D$, 
implementing the measuring process in a quantum experiment, we do not need 
any input extraneous to the interacting quantum system. One thing, which,
however,
is not extraneous to the quantum interacting system,
is the preparation of the experiment. 
 We must, further, specify, what we understand under 'preparation of the experiment'. 

\begin{defi}

   The preparation of a quantum experiment consists of two processes:

\begin{itemize}
\item[i)] The preparation of the states of the elementary or 
  structured  
particle systems determined to interact with one another  and
subsequently to interact with the active  part of the measuring apparatus.
\item[ii)] Preparation of the active  part of the measuring apparatus and of its 
     state to measure either a  particle property, $P_i$ ,or a  wave
     property, $W_i$.
\end{itemize}
\label{def82}
\end{defi}

\begin{defi}

\begin{itemize}
\item[i)]   A quantum measurement is the experimental determination of one or more quantum 
transitions in the prepared system. The transition may consist  in the change(s) of some observable(s)during 
a fundamental interaction in the prepared quantum system and the active part proper of 
the measuring apparatus.

\item[ii)]  The preparation of  an experiment influences the system to be measured in such away
that it increases the probabilities for one or a few of the possible outcomes, 
constituting the {\em ExpRes} to be determined relative to all other possible outcomes.           
\end{itemize}
\label{def83}
\end{defi}

{\bf Remark 8.4} 

 Accordingly, we understand that the critical part of a quantum
 experiment is an 
interaction between two particles, or between a particle and a field, or  between  two fields 
causing  the  evolution of the system whose an observable is  to  be  determined within 
its corresponding IPN either in the interacting system rest frame of reference or in  
observer's system of reference .

\subsection{Schroedinger's equation produces ${\bf R}$}

{\bf Proposition 8.1}
{\it 

The reduction of the state vector describing  a quantum  
measurement is  effected by the evolution operator ${\cal D}(\tau)$ with the
interaction Hamiltonian, $H(t)$,  
appropriate to the preparation of the  experiment for $t\in \tau$. 
 ${\cal D}(\delta(\tau))$  reduces  the  probability  
amplitudes $\{ C_n(0)\}$ of all components of the  initial state vector 

$$                 
\Psi(x) = \sum_n C_n(0) u_n(x)
$$

representing the system under measurement, except the ones 
$\{C_{\alpha}(0)\mid \alpha = 1, 2, \ldots , K < \infty\}$
corresponding to the observables $\{O_{\alpha}\mid \alpha = 1, 2, \ldots , K < \infty\}$ 
to be obtained in the {\em ExpRes}.
\label{prop81}
}

{\bf Proof}

   The operator ${\cal D}(\delta(\tau))$  
can be taken equal either to evolution operator
   ${\cal U}_{nmp}(\delta(\tau))$  or to 
${\cal U}_{u}(\delta(\tau))$   depending on the case. 
    In the present case we identify ${\cal D}(\delta(\tau)) 
= {\cal U}_{nmp}(\delta(\tau))$. 

\begin{eqnarray}                                       
{\cal U}_{nmp}(\delta(\tau))   &=&\exp \langle (i\hbar)^{-1}
\int_{M_{\kappa}^4} d^4x {\cal H}(\varphi(x, t), \partial\varphi(x, t)\pm
i \Lambda(j, \sigma)) \nonumber \\
& & \times  
(\cos (\Lambda(j, \sigma)) \pm i \sin (\Lambda(j, \sigma)))\rangle )
\label{e85}
\end{eqnarray}

before quantization.   

   The expression for the preparation of the experiment preparation 
comes about 
through  the selection of the appropriate quantum numbers in 
$\Lambda(j(n), \sigma)$ following 
the quantization of the field action-integral (\ref{e85}).
The kind of quantization to be applied becomes clear from 
the expectation to have 
a non-measure-preserving evolution or a unitary evolution,  i.e.,
we expect to measure substantial changes in the relative probability measures of the components characterizing the system before and after the measurement with respect to the remaining components.

Carrying out the multiplication of the quantities in the brakets 
of  the  exponent in (\ref{e85})  we see that the above requirement is
fulfilled 
according to (\ref{7_10}), if one puts

\begin{equation}
\Lambda(n, \sigma) = \pi (n + 1/2), \sigma = 1.
\label{e86}
\end{equation}

From  (\ref{e86}) it follows after application of (\ref{e85}) on the state vector that 

\begin{eqnarray*}
 & &  {\cal U}_{nmp}(\delta(\tau))\Psi(x)\nonumber \\ 
 &=& \exp \langle \left[\frac{(-1)^n}{\hbar}
\int_{M_{\lambda}^4} d^4x {\cal H}(\varphi(x, t), \partial\varphi(x, t) \mp 
\Lambda(j(n), \sigma))\right]\rangle\Psi(x)\\
 &=& \sum_{n=1}^{\infty}   
\exp \langle \left[\frac{(-1)^n}{\hbar}
\int_{M_{\lambda}^4} d^4x {\cal H}(\varphi(x, t), \partial\varphi(x, t)) \mp 
\Lambda(j(n),  \sigma)\right]\rangle C_n(0)u_n(x)\\
 &=& \sum_{n=1}^m   
\exp\langle  \left[(-\hbar)^{-1}
\int_{M_{\lambda}^4} d^4x {\cal H}(\varphi(x, t), \partial\varphi(x, t)) - 
(j(n) + 1/2)\right] \rangle C_n(0)u_n(x)\\
 &+& \sum_{\alpha=1+m}^{m+K}   
\exp \langle \left[ (+\hbar)^{-1}
\int_{M_{\lambda}^4} d^4x {\cal H}(\varphi(x, t), \partial\varphi(x, t)) + 
(\alpha(n) + 1/2)\right] \rangle C_{\alpha}(0)u_{\alpha}(x)\\
&+& \sum_{n=1+m+K}^{\infty}   
\exp \langle  \left[(-\hbar)^{-1}
\int_{M_{\lambda}^4} d^4x {\cal H}(\varphi(x, t), \partial\varphi(x, t)) - 
(j(n) + 1/2)\right]\rangle  C_n(0)u_n(x).
\end{eqnarray*}

   By appropriately choosing the respective action-integral
values, i.e.,  $\{ j(n)\}$ in each class 
of states in the sum, we obtain that only the intermediate sum survives. 
The corresponding 
exponents are the sums of two positive numbers. 
The first and the third sums above become as small as we please by taking the 
differnces in the respective exponents sufficiently small in comparison with the
smallest term in the sum, as implies the preparation of the experiment 
$\alpha \in [ 1+m, n+K]$.

\begin{eqnarray}
{\cal U}_{nmp}(\delta(\tau))\Psi(x) &\approx &
\sum_{\alpha=1+m}^{K}   
exp \left[ (+\hbar)^{-1}
\int_{M_{\lambda}^4} d^4x {\cal H}(\varphi(x, t), \partial\varphi(x, t) + 
(\alpha(n) + 1/2))\right] \nonumber \\
& & \quad\times C_{\alpha}(0)u_{\alpha}(x)
\label{e87}
\end{eqnarray}

     If the set of the orthonormal functions $\{ u_n(x)\}$ are eigefunctions of the energy 
operator, then (\ref{e87}) can be simplified in the form                 
 
\begin{eqnarray*}
{\cal U}_{nmp}(\delta(\tau))\Psi(x)
 &=& \sum_{\alpha = 1+m}^K \exp\left( \hbar^{-1} E_{\alpha} \delta(\tau)
   + (\alpha(n) + 1/2))  \right) C_{\alpha}(0)u_{\alpha}(x)\\
 &=& \sum_{\alpha = 1+m}^K C_{\alpha}(\delta(\tau)) u_{\alpha}(x),  
\end{eqnarray*}

where 

$$
C_{\alpha}(\delta(\tau)) = 
\exp\left( \hbar^{-1} E_{\alpha} \delta(\tau) + (\alpha(n) + 1/2))  \right) C_{\alpha}(0).
$$
   Obviously,  the probability coefficients for the surviving states
   are much larger than the 
rest of them 

\begin{equation}
\mid C_{\alpha}(\delta(\tau))\mid^2 \gg \mid C_{n}(\delta(\tau))\mid^2
\forall \alpha\in [1, K],  \forall n\in\Zset^+ \setminus [1, K].
\label{88}
\end{equation}

and the proof Proposition 8.1 is complete.

{\bf Remark 8.5} 

This is the expected result corresponding to the
 preparation of the experiment and implying the 
reduction, ${\bf R}$, of the wave function after the experiment. It is seen that ${\bf R}$ is an integral
 part of quantum dynamics and it does not need the presence of any extraneous agents.
The numbers $\alpha(n), j(n)$ and $K$ depend  on the preparation  and  the kind of interaction 
in   the  experiment. $\{\alpha(n)$ may be  large, $\{j(n)\}$ are correspondingly of the orders of 
$\{E_n \}$. $K$ is in most experiments equal to  $1$.

{\bf Remark 8.6}  

The novum in the above proof is:
\begin{itemize} 
\item[i)] It is seen that ${\bf R}$ does not imply necessarily 
reduction to one single sate, but to any finite number,
$K$, of final states.

\item[ii)] The reduced states are not fully extinguished! They simply
  become of very small relative probability.

\item[iii)] The result ii) stresses the {\em statistical appearence}
  of quantum theory which is traced back to the {\em chrono-topology}. 

\end{itemize}

\setcounter{equation}{0}
\setcounter{table}{0}
\setcounter{figure}{0}
\section{Schroedinger's cat is only alive before he dies
and only dead after he lived}
\label{section_9}
\subsection{Cats in physics}

    There is a considerable literature about cats in physics. 
To us are known at least 
two famous cats: The first was Lewis Carroll's cat in the Wonderland
of Alice's Adventures.
The other one, equally important for quantum theory, is Schroedinger's cat.

   Are they really so important for  physics?    
We do not know for sure. In any way,
they have been both important for some people for some time. 

 The importance of  Alice's cat consisted in that he was able 
to leave his charmed smile in the air as his signature - 
alike the quarks in the QCD of the  jets.

    Schroedinger's cat  did not smile.  
Some important people wanted  him for some 
decades dead and alive at the same time. We will examine here, 
if his sentence was
just, and, if not, we shall try to revise it.

    Let us see what this means for quantum theory and let us  
briefly recall the story:
There is a hermitically closed room and a glass bottle of an extremely strong poison in it. 
A hammer 
hangs above the bottle and can be activated for its purpose by means
of the decay of  one nucleus from a quantity of a radioactive 
isotope put in the mechanism \cite{49}. 

  The life of the cat has exactly the same chances with the life of
  the (first) 
nucleus to decay, and the state vector of the nucleus consists 
of two components.  The one component stands  for an unstable nucleus 
(before decay) and corresponds to alive cat. 

The other component of the state vector represents a  stable nucleus 
(after decay) and is a proof for the dead cat. 

   The linear superposition of the components is meant to 
represent a natural state of affairs as far as 
the nucleus is concerned. 
But it is considered as a paradoxical business  in cat's case. 

Nevertheless, it is considered as thoroughly plausible that, 
in principle, the same state vector 
represents both the state of the nucleus and cat's  state.  
It is thereby clear that the nucleus 
may consist of  a few (for the Tritium 4)  particles, whilest  the cat
is made of  some $10^{25}$ particles. 

Besides, it has not been possible to demonstrate until today,
that the state vector of    about $10^{25}$ 
particles does not acquire additional (collective) unknown 
properties beyond those of the state 
vector of the Tritium. 
These properties might profoudly change the character of the wave 
function by shifting the values of its importance parameters.

\subsection{The action quantization saves the cat}

Questions like the above will be left aside and we shall apply 
the theory  developed in the 
previous sections to demostrate that  the  cat  can  
be only alive before he dies, because of  the 
poison in the bottle and he can be only dead after he definitely has finished his life
for any reason whatsoever. 

   We do that by means a two-component state vector.  
The component  number   $n=1$ 
represents the unstable nucleus before radioactive decay.
The component  number $n=2$ 
represents the nucleus after the emission of a quantum.

   Further assumptions, e.g., 
about the wave function of the cat will not be done. Also no 
assumptions are necessary about the wave functions of  the hammer, 
of  the bottle, of the 
poison and the mechanical parts of a device interacting with our
quantum system, because they are not a part of the quantum system
whose solution is sought. 

   The quantum mechanically crucial phase of the experiment 
beyond the radioactive decay of 
one of the unstable nuclei is not the engineering mechanism,
but rather the interaction of the radiation quantum with one or 
more nuclei of the detector which will produce the electric pulse,
required for the macroscopic activation of the mechanism.

   We trust that, if one nucleus  decays,
the hammer will move, and the rest will be in 
accordance with the engineering design. 
Our decision about the ''sentence'' of the cat will 
result from the proof of the following

{\bf Proposition 9.1 }

  Let 

\begin{equation}
\label{9_1}
\Psi (x,0) = \sum_{n=1}^2 C_n(0) u_n(x,0)
\end{equation}

{\it 
be the two-state vector at time $t=0$ of a nucleus with two 
possible bound states of  its  
Hamiltonian, $H(t)$, with  binding energies $E_1, E_2 <0$ :
\begin{itemize}
\item[a)] State $n=1$ represents the initial unstable nucleus 
before the radioactive decay ($t=0$)
and is associated with  a very large probability with respect to the state 
$n=2$.
\item[b)] State $n=2$ represents the stable nucleus after the 
radioactive decay ($t=\delta(\tau) $) and is 
associated with a very large  probability with respect to the state $n=1$. 
\end{itemize}

\begin{equation}
\label{9_2}
\left| C_1(t=0) \right| >> \left| C_2(t=0)\right|.
\end{equation}

Then, the transition from state $n=1$  to state $n=2$ 
is effected by means of the  non-measure 
preserving time evolution operator with '' $\pm $''
 in the front of the integral in the exponent:
\begin{equation}
\label{9_3}
{\cal U}_{nmp} (\delta(\tau))  = \exp \langle [\pm \hbar^{-1} 
\int_{\bar{M}^4_\lambda}
d^4x {\cal H}(\phi(x,t), \partial \phi(x,t)) +\Lambda(j,\sigma) ]\rangle, 
\quad \sigma =1
\end{equation}

with the interaction Hamiltonian
\begin{equation}
\label{9_4}
H = \int_{\bar{M}^4_\lambda} d^4x {\cal H}(\phi(x,t), \partial \phi(x,t))
\end{equation}

within the IPN, $\tau_{\lambda}$. The probability for the state $n=1$ at 
$t = \delta(\tau)$, i.e., at the end of the 
nuclear emission process is very small,
while that for the state $n=2$ is very large after 
the transition:
\begin{equation}
\label{9_5}
\left| C_2(t=\delta(\tau)) \right| >> \left| C_1(t=\delta(\tau))\right|.
\end{equation}

} 

{\bf Remark 9.1}

Relations (\ref{9_2}) express the fact that the cat is alive with very high,
virtually $1$, probability and dead with very small, 
virtually zero, probability.
Relations (\ref{9_5}) express the converse case,
namely that the cat is dead with very high, virtually 
$1$, probability and alive with very small, virtually zero, probability.
The seeming small 
uncertainty about ''alive'' and about ''dead'' does not need to confuse us,
because this is the 
natural state of affaires in quantum mechanics.

   For example, according  to  statistical mechanics, and, also, 
to the Poincare recurrence 
theorem, there exists a vanishingly  small probability that the set of atoms,
of which consisted 
Einstein short before his death, 
recombine to give the living Einstein. This probability is 
virtually zero.

{\bf Proof}

    The interaction (\ref{9_4}) (fixed by its structure)  
inducing the transition of the unstable nucleus from the state characterized 
by (\ref{9_2}) to the state characterized by (\ref{9_5}) constitutes the 
''preparation'' of the experiment according to Definitions \ref{def81} and  
\ref{def83}  and gives the appropriate 
value to the field action integral, $+\Lambda(j,1)$.  The sign ''$+$''
with large $\Lambda(j,1) $
corresponds to the ''present'' state of the system,
because it gives to the action integral the maximum value and 
maximum probability to the state. The sign ''$-$'' 
determines the state which is  the ''absent'' state,
because its action, $- \Lambda(j,1)$, gives a very small, virtually zero,
probability.

If we act with the operator \ref{9_3} on the state vector \ref{9_1} we get
\begin{eqnarray}
\sum_{n=1}^{2} &   \exp \langle [\frac{(-1)^n}{\hbar}
\int_{\bar{M}^4_\lambda}
d^4x {\cal H}(\phi(x,t), \partial \phi(x,t)) \mp\Lambda(j(n),\sigma) ]\rangle
C_n(0)u_n(x,0) & \nonumber \\
 =& 
\exp \langle \frac{(-1)^n}{\hbar} \int_{\bar{M}^4_\lambda}
{\cal H}(\phi(x,t), \partial \phi(x,t)) -\Lambda(j(1),\sigma) ]\rangle
C_1(0)u_1(x,0)  &\nonumber \\
&\quad +  \exp \langle +\hbar^{-1} \int_{\bar{M}^4_\lambda}
{\cal H}(\phi(x,t), \partial \phi(x,t)) +\Lambda(j(2),\sigma) ]\rangle
C_2(0)u_2(x,0)  &\\ 
\label{9_6}
  =& 
\exp \rangle  [ -\hbar^{-1}  |E_1|\delta(\tau) -\Lambda(j(1),\sigma)]
\rangle C_1(0) u_1(x,\delta(\tau)) \quad\quad\quad &\mbox{(a)} \nonumber\\
&\quad + \exp \rangle  [ +\hbar^{-1}  |E_2|\delta(\tau) +\Lambda(j(2),\sigma)]
\rangle C_2(0) u_2(x,\delta(\tau))  \quad\quad\quad &\mbox{(b)} \nonumber\\
 & & \nonumber \\
=& C_1 \delta(\tau) u_1(x,\delta(\tau)) + C_2 \delta(\tau) u_2(x,\delta(\tau))
\; \quad\quad\quad \quad\quad &\mbox{(c)} \nonumber\\
 \approx &  C_2 \delta(\tau) u_2(x,\delta(\tau))\;  \quad\quad\quad \quad\quad
& \mbox{(d)}
\label{9_7}
\end{eqnarray}

From \ref{9_7} we conclude by taking  appropriate values 
for $j(1)$ and $j(2)$ that 
\begin{eqnarray}  
| C_2 \delta(\tau) u_2(x,\delta(\tau)) |^2 &=&| C_2 (0) u_2(x,0))|^2
\exp[ +2/\hbar  |E_2|\delta(\tau)+ 2j(2) +1] \nonumber\\
 &\gg &   
| C_1 (0) u_1(x,0))|^2  \exp[ -2/\hbar |E_1| \delta(\tau) -2j(1) +1] 
\label{9_8}
\end{eqnarray}
   From \ref{9_8}  it follows that 
\begin{equation}
| C_2 (  \delta(\tau)) u_2(x, \delta(\tau)) |^2 >>  
| C_1 ( \delta(\tau)) u_1(x, \delta(\tau)) |^2
\label{9_9}
\end{equation}
and this completes the proof of {\bf Proposition 9.1}.

{\bf Remark 9.2}

     The values of $\pm \Lambda(j(n))$  which determine the probabilities of the corresponding 
components of the state vector are determined by the interaction \ref{9_4} effecting the transition 
of the nucleus.  In case it is very large,
it describes a ''clear cut''  state of the system.

{\bf Remark 9.3}

    Equations \ref{9_7}  prove that that the state vector does not evolve any more after 
the transition, because $\delta(\tau), E_1, E_2 $  are  constants.

{\bf Remark 9.4}

    If the probability, $p_{2l} $, of the state for the living 
Schroedinger cat 
after the decay of the 
unstable nucleus  is vanishingly small in comparison with the probability, $p_{2d} \gg  p_{2l}$, 
 for the dead cat, it is experimentally impossible to observe the alive cat after he died with 
very high probability. 

   Hence, there is no Schroedinger's cat paradox in the 
framework of the theory of the Schroedinger equation. 

{\bf Remark 9.5}

The solution given in the present section opens  new aspects for all problems concerning quantum 
systems with two or few levels possible. This is a  novel aspect of {\bf R} and it is 
in full agreement with the statistical nature of the wave function.

\setcounter{equation}{0}
\setcounter{table}{0}
\setcounter{figure}{0}
\section{The non-decay of the wave packet }

\label{section_10}
     Our basic principle in chrono-topology is that interactions are the 
causes of all changes in 
the universe. 
No change whatsoever is possible without an interaction. Accordingly,
fluctuations also should be due to some sort of interaction. 
The fluctuations in the density of a gas are due to particular 
'constellations' of a number of interacting pairs of atoms with
particular values of the scattering angles.  
We propose to consider the wave packet decay  from this point of view.

\subsection{Attempts at avoiding decay}

There exist rather ambitious  mathematical theories making various attempts to
reconcile the  non-decay of the particle with the decay  of the wave
packet representing it. The problem consists in that the wave packet
{\it does} decay in the absence of interactions in  the Newtonian 
time topology, while the particle {\it does not} decay.  

   One of the theories endeavoring to eliminate the decay of the wave
   packet in the
   Newtonian time topology proposes some gravitational agents \cite{58}. 
   These agents 
   result to the restauration of the decaying wave packet implied by
   quantum theory. However, there is no indication about a relation of
   the agents with any physical action directing them exclusively
   towards decaying wave packets and preventing them of acting on
   non-decaying waves too. 
 Another theory postulates for the same purpose the spontaneous
 multiplication of the wave packet (function) by a strongly picked
 Gaussian function in definite time intervals to cancel the calculated
 decay of the wave packet \cite{59}.

These theories refrain from giving in any way neither the locations of
the large number of the required computers to implement the
multiplication of all travelling wave packets in the
universe, corresponding to the particles of the cosmic radiation and to
the beam particles in
the accelarator laboratories or in the nuclear reactors. 
Also they give no indication, if, for example, instead of a mathematical
multiplication of the wave packet, there occurs a physical amplification, e.g., like the
amplification of an electrical pulse in an electronic device. These
are questions which in our view should also be answered, if these
theories claim the qualification of completeness. 

   It is, finally, interesting to note that E.~P.~Wigner proceeded to
   accepting the existence of an influence 
on the unconscious matter  by any living matter in such a way as to change
the {\bf U} action of the time evolution operator to an {\bf R } action \cite{60}.

 All three  explanations of the factual non-decay of the 
particle corresponding to a decaying 
wave packet, make appeal  to effects which physically  are difficult to explain
in the framework of the valid laws. This entitles us to look  for other, 
simpler explanations.

\subsection{The flowing Newtonian time at the root of the confusion}

   The solution of the above problem in the framework of 
our chrono-topology appears in comparison with the above mentioned
very sophisticated theories as a rather trivial excercise. In this
context it would appear surprising, if the wave packet did really decay
in the absence of any interactions. 

    As in the problem of the measurement in quantum mechanics, 
it will be, here too, assumed 
that the decay of the wave packet, whenever it occurs, is described exclusively 
by the evolution operator. To see this we consider an orthonormal and complete set 
of eigenfunctions $\{ u_n\left| n=1, 2, \ldots \right. \} $ 
of the Hamiltonian $H(t)$ and we write for the state 
vector the expression  for $t=0$
\begin{equation}
\Psi(x,0) = \sum_{n=1}^\infty  C_n u_n(x),
\label{10_1} 
\end{equation}
where $x \in \bar{M}^4_1$ is the particle coordinate in its 
rest frame of reference, $S$.

\subsection{Let us see, what happens} 

    The details of the wave packet construction  are described in
    books on quantum mechanics and they are not repeated here. We
    shall, instead, prove only the following

{\bf Proposition 10.1 }
{\it 

\begin{itemize}  
\item[1.]  Let $\Psi(x,0) $  be the wave packet representing a particle 
in the point  $x$  at the time $t=0 $ in its rest-frame of reference.

\item[2.] Let $\{ u_n(x), E_n, \left|  n=1, 2, \ldots \right. \} $     
represent an orthonormal, complete system of eigenfunctions of the Hamiltonian  
${\cal H}(\phi(x), \partial \phi(x)) $ and the corresponding energy eigenvalues.

\item[3.]  Let ${\cal U}_u ({\cal T}_{\Lambda} )$ be  the unitary time 
evolution operator implementing the changes  on 
$\Psi(x,0)$   implied by ${\cal H}(\phi(x), \partial \phi(x)) $.

\item[4.]  The rest-frame of reference of the particle is a macroscopic
material body with which a moving                         
observer's detector may interact via any quanta exchange
and in this way determine the position of the particle without in 
any way disturbing it or its wave packet.
\end{itemize}
}

   Then, 
\begin{itemize}
\item[i)] The  wave packet remains unchanged in its rest-frame of reference without interactions.
\item[ii)] The  wave packet does not decay for an observer moving  with respect to the rest-frame 
of reference of the particle without interactions.
\item[iii)] The wave packet changes  form under an {\bf U} and
- a fortiori - under an ${\bf R}$ interaction.
\end{itemize}

{\bf Proof}
\begin{itemize}
\item[a)] We act on $\Psi(x,0)$  with our unitary time evolution operator, 
${\cal U}_u ({\cal T}_{\Lambda} )$:
\begin{eqnarray}
\label{10_2}
& & {\cal U}_u ({\cal T}_{\Lambda} ) \Psi(x,0)  \nonumber \\
&=& \sum_{n=1}^\infty
\exp \langle  [ (i\hbar)^{-1} 
\int_{\bar{M}^4_{\lambda}}
d^4x {\cal H}(\phi(x,t), \partial \phi(x,t)) 
\pm \Lambda(n, \sigma) ]\rangle
C_n u_n(x) \\
&=& \sum_{n=1}^\infty
\exp \langle  [ (i\hbar)^{-1} 
E_n \delta(\tau) \pm i\Lambda (n, \sigma)] \rangle C_n u_n(x) \nonumber
\end{eqnarray}
where the time integration is taken inside a single IPN, $t\in \tau $. 
We take for convenience the unitary time evolution operator, 
because of the periodic bounrary conditions imposed on the 
orthonormal set $\{ u_n\left| n \in \Zset^+ \right. \}$   used in (\ref{10_2}). 

  Considering   $\{ u_n\left| n \in \Zset^+ \right. \}$ 
 and replacing the sum $\sum_n$  by the integral $\int dn$  we find,  for   
$$ \Lambda(n,\sigma), \sigma =2, $$

and   
$$
E_n = \frac{\hbar^2 n^2}{2m}, n\in \Zset^+, $$
by the usual procedure \cite{61}  the wave packet form 
\begin{equation}
\Psi_{\pm} (x, \delta(\tau)) = (2\pi)_x^{-1}
(\Delta x + \frac{ i\hbar \delta(\tau)}{2m\Delta x} )^{-1/2} \exp [
-\frac{ (x\pm 1)^2}{(\Delta x)^2 + 2i\hbar \delta(\tau) /m}]. 
\label{10_3}
\end{equation}  

Expression (\ref{10_3}) proves that the wave packet cannot 
decay in time in its rest frame of 
reference, since, according to {\it Axiom III}  the IPN, $\tau$, 
without interaction represents simply the epmpty  set, and $\delta(\tau) =0 $.

This proves assertion a).

\begin{figure}[htbp]
\begin{center}
\leavevmode
\epsfxsize=0.5\textwidth
\epsffile{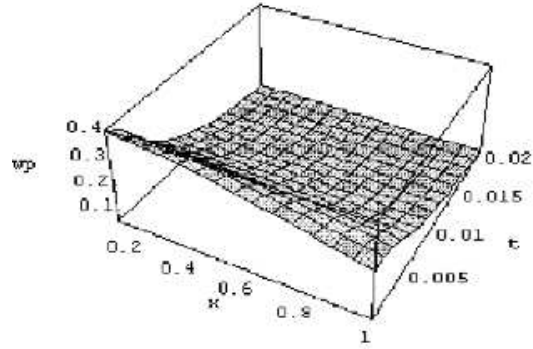}
\end{center}
\caption{A  minimal wave packet decaying in Minkowski's $S$
space-time (lines in the plane  t,wp). In  chrono-topology there 
is  no time-flow  in absence  of  interactions, and, consequently,
the wave packet conserves  its initial form (any constant curve 
in the plane x,wp). Arbitrary scale factors.}
\label{fig_10_1}
\end{figure}

\item[ b)]  

We consider the Lorentz transformation 
with the velocity, $v$, 
of the observer along the 
positive $x$-direction. 
Then, we have the expressions: 
\begin{eqnarray}
\label{10_4}
        x &=& \gamma (x' - vt),  t = \gamma (t' - (\beta /c)x') ,\mbox{(a)}
\nonumber \\
        x'& =& \gamma (x + vt),  t' = \gamma (t + (\beta /c)x) ,\mbox{(b)}
\end{eqnarray}

and we get the transformed wave packet
\begin{eqnarray}
\Psi_{\pm} (x, \delta(\tau))&\rightarrow& \Psi'_{\pm} (x', \delta(\tau '))  
 \nonumber \\
  &=& (2\pi)_x^{-1/2}
(\Delta x + \frac{ i\hbar \gamma(t' +(\beta/c)x') }{2m\Delta x} )^{-1/2} 
 \nonumber \\
& & \quad\times \exp [ -\frac{ \gamma (x' + vt') \pm 1)^2}{ (\Delta x)^2 + 2i\hbar \gamma (t' + (\beta /c)x')/m}]. 
\label{10_5}
\end{eqnarray}  

   Let us now see what are the domains of variation of $x'$ and $t'$: 
In absence of an interaction in S the particle conserves: 
$ x=\rho =$ {\it Constant},  and   $t = 0$. 
For an observer in $S'$ there follows from (\ref{10_4}):
$$  x' = \gamma\rho  = \mbox{ {\it  Constant}, and}\quad 
t' = \beta \cdot\rho\; c = \mbox{ {\it Constant}.}$$

   Hence, the wave packet appears to the moving observer as having a different form, but it does not 
change, it does not decay, if not subject to an interaction. 

This proves b).

\item[c)]  Let $\rho$ be the constant position of the particle in $S$.  
Since the particle is subject to an 
interaction its proper time $t$ takes values in $\tau, t\in \tau$. 
\end{itemize}
We know, in addition, that the diameter of the IPN, $\tau$, is 
finite, $\delta(\tau) = \delta < \infty$.
   
From (\ref{10_4}) we find 
$$ x=\rho, \mbox{and}\; \tau \equiv (0, \delta). $$

Consequently, for constant  $B_i, i= 1, 2, 3, 4$
\begin{eqnarray}
& & -\infty < B_1 = \gamma \rho \le x' \le \gamma (\rho +(\beta/c)\delta ) = B_2 < +\infty\; \quad \mbox{(a)} 
\nonumber \\
&\mbox{and} & \nonumber \\
& & -\infty < B_3 = \gamma \rho \le t' \le \gamma (\delta +(\beta/c)\rho ) = B_4 < +\infty\; \quad \mbox{(b)} 
\label{10_6}
\end{eqnarray}

Since the independent variables $x'$ and $t'$ in (\ref{10_5}) change by finite amounts, 
the wave packet 
changes by a finite amount its form, but it does not decay any more in $S'$, {\it given
that its interaction is of finite duration.}

This proves assertion c) , and the proof  of  {\bf Proposition 10.1}  is complete.

\begin{figure}[htbp]
\begin{center}
\leavevmode
\epsfxsize=0.5\textwidth
\epsffile{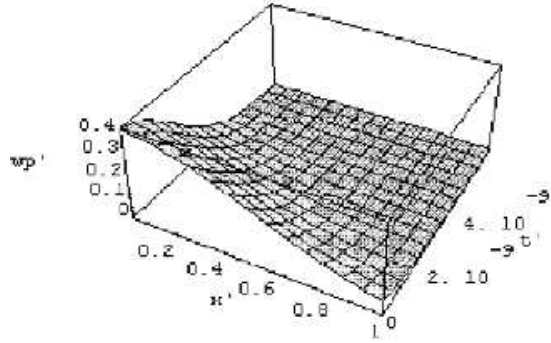}
\end{center}
\caption{The minimal wave packet of Figure \protect\ref{fig_10_1} seen from a
moving  frame of  reference $S'$  in Minkowki's  space-time.
If an interaction is present with, say $\delta(\tau) = 10^{-15} $  seconds,
one can see the wave packet deformation in a  plane ( $t',wp'$).
To appreciate the amount  of  the  wave  packet  change in 
two parallel planes ($x',wp'$) compare $\delta(\tau) $ with  the length, $ 6\cdot 10^{-9}$
seconds, of the $t'$- axis  segment.}
\label{fig_10_2}
\end{figure}

{\bf Remark 10.1}

    The double sign  in  (\ref{10_3}) represents a degree of  freedom  and  it stems from the double 
sign in the quantization of the action integral, $\pm \Lambda $. The shift of the wave 
packet center comes from the non-vanishing action-integral.

{\bf Remark 10.2}

Since the proof does not depend on the form of the wave packet (\ref{10_1}), it follows that
all wave packets do not decay, if not subject to an interaction. 
This is the situation in the case of soliton waves. 

{\bf Remark 10.3}

If the interaction Hamiltonian is of such a structure, that it changes the nature of the particle 
described by the wave packet under discussion, this will follow from
(\ref{10_2}) under the action of  
non-measure preserving evolution operator , ${\cal U}_{nmp} ({\cal T}_{\lambda} )$.
In this case also we shall have one or 
more wave packets behaving according to {\bf Proposition 10.1}.

\setcounter{equation}{0}
\setcounter{table}{0}
\setcounter{figure}{0}
\section{Conclusions and discussion}
\label{section_11}
   
There have been numerous investigations into the nature and the 
structure of time in the 
history of physics  and philosophy.  Already in ancient times, Plato,
Aristotle, St.~Augustine 
have written essays trying to understand, what is that, we call the time. 

   The references given are only a tiny sample of the world literature 
on time.  There have been reported very important and less important 
ideas  on the time issue, to all of whom we cannot do due justice 
with an appropriate reference. 

   Until recent times of our ending millenium, however,
the discussion about the nature of time 
was mainly of a philosophical character and, hence, it could not have
any important repercussions on the 
advances and on the results of physics and technology.   

   With the advent of relativity as well as of  quantum 
theory a vast literature on the time has been, and is still today being,
produced in the framework of theoretical and  of mathematical 
physics.

   In the cases, of these two disciplines of physics, however,
the ideas about the time are 
not so harmless as they were in the case of philosophy: As a
consequence, a considerable 
number of paradoxes and awkward situations arised due,
on the one hand, to the fundamental character of these important disciplines,
and, on the other hand, due to the assumed structure 
of time - as we now believe to be able to see. 
   To be specific, some of the most impressive paradoxes in physics are:

\begin{itemize}
\item[1.]  The measurement problem in quantum physics.
\item[2.]  The long-standing impasse 'Time reversal invariance of
  QFT-Irreversibility  in  macroscopic Nature'.
\item[3.]  The decay of the wave packet in absence of interactions.
\item[4.]  The catastrophies in QFT.
\item[5.]  The Schroedinger cat.
\item[6.]  The twin paradox in relativity.
\item[7.]  The time machines, etc. 
\end{itemize}

    After an examination of the fundamental concepts of physics, we
    believe that all these 
and many other paradoxes are due to the character and the topological structure of the time 
assumed in physics. The topological properties of the time, 
tacitly introduced both in relativity 
and in quantum theory,
 are those of the Newtonian universal time.  One should remember here  
the requirement of relativity according to which every space point has its own time. 
Also, Dirac's  proposal that each particle should have its own 
time variable has not helped to 
abondon the {\em natural topology of the line} as the topology of the
{\em time in atomic and sub-atomic physics}.   

   The trouble is not limited on the topology of time. 
Beyond that, the human physiology and in particular  
the neural structure of the human body 
and of the brain contributed considerably to the confusion of the time mystery.It lead to the 
assignment  to the time of a phantastic property:  {\em The flowing}.

    Let us be a little more precise and explain, 
how the feeling of the time flowing  comes about 
in the framework of our chrono-topology:
   According to our theory of the time structure (Sec. \ref{section_3}) 
is directly related to the 
changes occuring in nature and recorded in any  way by  the nervous  system of the observer. 
Every elementary change generates a time neighbourhood
$\tau_{\lambda}$. The next interaction 
proper-time neighbourhood, $\tau_{\lambda + 1}$, 
necessarily starts  before or after the ending of  the preceding one 
and ends later, if it is to be distinguished from the first.
The successively observed non-overlapping changes create the sense of 
{\bf ordering}. Zermelo's well-ordering theorem finds here 
a practical application in the production of the feeling of the
flowing time.

   The excitations by the successive changes of, 
both the external and the internal world, act in 
succession on the neural system  of the observer,
and give to him the impression of something
running, the apparent time  property of {\bf flowing}.  
   
   On the other hand, the feeling of the time {\bf continuity} 
is created by the finite ability 
of the observer to discriminate closely adjacent 
$\{ \tau_{\lambda} \mid \forall \lambda\in I\subset \Zset^ +\}$ 
of a large number
(some Avogadro-number-interaction-proper-time) of 
successive neighbourhoods, created in observer's brain by his observable
environment (external and internal), when the gaps between
$\tau_{\lambda}$ and $\tau_{\lambda + 1}$  forall $\lambda\in\Zset^+$ 
are large enough.

   Since time, alike the space, does not flow, 
the view that a wave-packet should decay in 
absence of interactions, appears in our chrono-topology as a paradox, 
seen that time changes only as long as the interaction is going on.

   As far as phenomena are concerned, whose durations are larger 
than the interaction proper
time neighbourhood diameter of the intervening fundamental interaction,
the physical  results 
are perfectly described in physics, and experiment  and theory are in 
very  good agreement.

   However, as soon as the distances between successive single 
elementary phenomena approach the limit
of the IPNs perception proper-time, then troubles set in.

   Many, extremely subtle affaires concern the use of relativity:  
Whenever one writes 
an equation containing a time-dependent fundamental interaction,
 $H(t)$, and one 
solves the equation, one does not identify himself with the 
interacting particles and one should 
always discuss the results in terms of observer's coordinates $(x',t')$.

    As we have seen in Sec. \ref{section_6}, 
there are considerable differences between  the topologies 
of the spaces in the rest frame of reference of the interacting particle system, $S$, 
and the space of the observer's moving system of reference, $S'$.  
For example, the time variable, $t$, for a particle in $S$ is  
$t\in\tau\in T\in\Rset$, 
while for the observer  in $S'$ the  time 
variable, $t' = \gamma(t-b/cx)\in T\times\Rset\subset \Rset^ 2$.
The topology of the time determines to an important extent the topology  of  the space-time. 
We find that the space-time topology of the quantum world by no means is identical to the 
topology of the relativity space-time, i.e., the topology  of  the
Minkowski space-time, $M^ 4$. 
The reasons are quite obvious from the point of view of our chrono-topology: 

\begin{itemize}
\item[1.] The Minkowski space-time, $M^ 4 = i\Rset\times\Rset^ 3$, is 
a four-dimensional continuum, including infinity.
It has been conceived as the host space of physical bodies wich may last  
    and travel {\em ad infinitum}, ignoring fundamental interactions 
and their discrete time structure. 
\item[2.] The quantum phenomena, on the contrary, cannot be continuous
  in time {\em per definition}. Fundamental quantum interactions, being
  mediated by quanta 
exchange between the interacting particles, cannot run continuously in
time generated by themselves. 
The quantum world space-time, 
${\bar{M}}_1^4 = i{\cal T}_4 \times \Rset^3\subset M^4$,
is a disconnected space-time with randomly variable 
    diameters, $\{ \delta(\tau_{\lambda}\}$, 
of the interaction proper-time neighbourhods and degree of 
    disconnectedness.
\item[3.] With the understanding that our chrono-topology is the
  natural topology for 
quantum  physics, we conclude that general relativity is a
non-quantizable theory,
because it is a theory based on the Newtonian universal time topology, $N_t$.
\end{itemize}

The characteristic property of the random and infinitely divisible fields is 
related to the decomposition

\begin{equation}
{\cal L} = {\cal L}_1 + {\cal L}_2 + \ldots  
+ {\cal L}_n, \forall n\in\Zset^+ .                              
\label{113}
\end{equation}
    This equation,
being a definition of the infinitely divisible  fields in the framework of the 
generalized random field theory, is mathematically fully clear.  But  from  the  physical point 
of view it is at least difficult to understand. It is physically  unacceptable  that 
a scalar field at a point of the Euclidean or of the Minkowski space-time be equal to the 
sum of any number of identical fields of the same strength and
probability distribution. 
On the contrary, this equation becomes  self-evident  in the framework 
of our chrono-topology of 
the many-folded time-space, in every part of which the conservation 
laws of physics are valid.

   To make this clear let us consider  one single IPN,  $\tau_{\lambda}$,
and the corresponding space-time, 
${\bar{M}}_1^ 4$.  
The lower index signifies that ${\bar{M}}_1^ 4 = \tau_1\times\Rset^ 3$, 
and this space-time {\em is simple in time}.
   If  there are two different  IPNs, such that, on the one hand,  
$\tau_1\cap \tau_2 = \emptyset$ and, on the other 
hand, their projections $\pi_1, \pi_2$ into ${\cal T}_{\kappa}$ 
satisfy $\pi_1\subset\pi_2$, then the corresponding space-time is 
${\bar{M}}^ 4 = (\tau_1 \oplus \tau_2)\times\Rset^3$ .   
This space-time is {\em two-fold in time}. It is said, in terms of the 
relativistic simultaneity, fully or partly simultaneous according to the 
relations satisfied by 
the projections $\{\pi_{\lambda}\}$ of the IPNs $\{\tau_{\lambda}\}$
into ${\cal T}_{\kappa}$

$$
(\pi_1\subseteq \pi_2) \wedge (\pi_1\supseteq \pi_2)\;\mbox{or}\; 
(\pi_1\subseteq \pi_2) \vee (\pi_1\supseteq \pi_2)  
$$ 

respectively.                              
   More generally, if $\lambda_{\kappa}$   IPNs  satisfy 

$$
\tau_{\lambda}\cap \tau_{\lambda'} = \emptyset\; \forall (\lambda_{\kappa},
\lambda'_{\kappa})\in I_{\kappa}, 
$$
                                                     
and their projections  into $T_{\kappa}$ 

$$
(\pi_{\lambda_{\kappa}}\subseteq \pi_{\lambda'_{\kappa}}) \wedge
(\pi_{\lambda_{\kappa}}
\supseteq \pi_{\lambda'_{\kappa}})\;\mbox{or}\; 
(\pi_{\lambda_{\kappa}}\subseteq \pi_{\lambda'_{\kappa}}) \vee (\pi_{\lambda_{\kappa}}\supseteq
\pi_{\lambda'_{\kappa}})\;\forall (\lambda_{\kappa},
\lambda_{\kappa}')\in I_{\kappa}  
$$ 

then the space-time,

\begin{equation}
\overline{M_{\lambda_{\kappa}}^ 4} = i(\tau_1\oplus\cdots \oplus
\tau_{\lambda_{\kappa}}) \times \Rset^ 3,
\label{114}
\end{equation}
                                                                                           
is $\lambda_{\kappa}$-fold in time.

   In view if the above definition the physical meaning of (\ref{113}) becomes perfectly clear:
In a $\lambda_{\kappa}$-fold in time space-time the decomposition of
an infinitely 
divisible field $\cal L$ in up to $\lambda_{\kappa}$ 
 terms is possible without interfering neither with the definition of the notion function nor 
with the conservation laws of physics which remain valid in each one
IPN.

  The chrono-topology allowed to us to demonstrate the existence of a time evolution 
operator which is appropriate for implementing the long time sought reduction of the wave
packet as well as to help explain other paradoxes of quantum theory presented inSec. \ref{section_9} and \ref{section_10}.   

    However, the most important result for us is the insight that Eistein's 
    life-long conviction, according to which the structure of
    quantum theory  
{\em per se} does not justify the 
characterization as a statistical theory. 
In fact, a proof of the statistical character of 
the wave function inside the chrono-topology was posible \cite{57}. In the topology of the 
Newtonian universal time such a proof is impossible.

   The full meaning of this fact becomes clear, only if it is born  
in mind  that  the fundamental 
interactions act  in every case during a finite time duration.  
On  the  contrary,
if the durations of the fundamental interactions were infinite,
no proof of Born's hypothesis 
about the wave function would exist. Because in that case the group represented
by the evolution operators, both ${\cal U}_u$   and ${\cal U}_{nmp}$ ,
would be continuous both in $\tau_{\lambda}$ and in $\Rset$.
Of significant importance we consider the fact that Planck's constant
has been calculated in the framework of chrono-topology. 

The chrono-topology is appropriate - as we believe - to accommodate the solution of 
still more other problems in quantum theory.   As  the most  prominent  problem  we consider 
the elimination of the divergent integrals in the perturbation theory of QFT.In this event
renormalization of the field theories would become 
unnecesary. This view is corroborated by the spontaneous
$\Lambda$-renormalization appearing in our theory. 

    In view of the topological structure of the space-time
    ${\bar{M}}_{\lambda_{\kappa}}^ 4$ 
 in comparison with the
space-times encountered in general relativity based on the Newtonian universal time,
it seems to us very likely that general relativity as it stands is conceived as a non-quantizable theory.

   One way to quantize general relativity is, possibly, to consider it 
{\em ab initio} in the chrono-topology.  
In this   case the field becomes a generalized and infinitely divisible field,
and the quantity to 
quantize will be the the field action-integral  of gravity. This would
lead to an extension ( not  
revision)  of the validity domain of general relativity.
 Since the time integral of the interaction Hamiltonian is spontaneously $\Lambda$-renormalized 
through subtraction of $\hbar\Lambda(n, \sigma)$ from the action integral,
it is expected that
the divergent  integrals 
of the covariant perturbation expansion in QFT might be eliminated by
taking them for $t\in T_{\kappa}$, as it must in our 
chrono-topology instead of taking $t\in \Rset^1$. 
   It is obvious that all theories' action-integrals are quantizable 
and $\Lambda$-renormalizable.
 
Finally, an interesting research topic is the following: Since the
time axis in each $\tau_{\lambda}$ must be orthogonal to $\Rset^3$ it
follows that there are $3^{\infty}$ orientation possibilities of the
light cone of each $\tau_{\lambda}$ at every point $x\in\Rset^3$. This
opens the question as to the extent in which we are allowed to
consider the future cone in analysing the quantum behavior of the
black holes. This question does not arise only in the Minkowski space
but also in Riemann spaces of general relativity. It seems, therefore,
that Penrose's reasoning \cite{63} is the correct one also from the
point of view of our chrono-topology. 

Another important problem is Euclidization of Minkowski's
spacetime. In view of the radical change of the space-time topology,
clearly ${\bar{M}}^4_{\lambda_{\kappa}}$ is not euclidizable. Also it
is very exciting to see whether chronotopology will allow the
existence of singularities in view of the apparent fact that there is
neither one beginning nor one end of the time but instead countable
sets of them both. The essential difference is that in the
conventional understanding of the time $t\in\Rset^1$ while in the
understanding of chrono-topology $t\in\tau_{\lambda}$ with 
$\delta(\tau_{\lambda}) < \infty$.


\newpage
\begin{thebibliography}{99}
\bibitem{1}C. Syros
             Lettere al Nuovo Cim. Vol. 10, N. 16 (1974) p. 718-723.

\bibitem{2}Aristotle $\Phi\upsilon\sigma\iota\kappa\acute{\eta}\varsigma\: 
A\kappa\rho o\acute{\alpha}\sigma\epsilon\omega\varsigma\:
\Delta\:\mbox{ in }\:\Phi\upsilon\sigma\iota\kappa\acute{\alpha}$, 
$\: E\pi i\sigma\tau\eta\mu o\nu\iota\kappa\acute{\eta}
\: E\tau\alpha\iota\rho\epsilon\acute{\iota}\alpha \: \tau\omega\nu\ $
\linebreak
$  E\lambda\lambda\eta\nu\iota\kappa\acute{\omega}\nu \: \Gamma\rho
  \alpha\mu\mu\acute{\alpha}\tau\omega\nu.
  \: \Pi\acute{\alpha}\pi\upsilon\rho o\varsigma$, A.E. (1975)
p.184-214.             
         

\bibitem{3}Isaac Newton
             {\em Principia mathematica philosophiae naturalis. }
             1686.

\bibitem{4}Immanuel Kant
             {\em Kritik der reinen Vernunft. }
             1787.

\bibitem{5}Henri Bergson
             {\em L'evolution creatrice. }
             PUF 1970.

\bibitem{6}A.Eddington
             {\em The nature of the physical world. }
             Cambridge University Press 1928.

\bibitem{7}A. Whitehead
             {\em Process and reality: An Essay in Cosmology. }
             D.R. Griffin and D.W. Sherburn (eds.), Free Press N.Y. 1979.

\bibitem{8}A.Einstein
             {\em The meaning of relativity. }
             1955.

\bibitem{9}P.A.M.Dirac
             Proc.Roy.Soc. London 136 (1932) p. 453.

\bibitem{10}I. Prigogine
             {\em From being to becoming. }
             W.H.Freeman, San Francisco 1980.

\bibitem{12}J.S.Bell
             {\em Speakable and unspeakable in quantum mechanics. }
             Cambridge University Press 1987.

\bibitem{13}S.W.Hawking
             {\em A brief history of time. }
             N.Y., Bantam 1988.

\bibitem{11}S.Weinberg
             {\em The first three minutes. }
             Basic Books N.Y. 1977.

\bibitem{wheeler}J.A. Wheeler
             {\em Einstein's Vision. }
             Springer, 1968, p. 51.

\bibitem{14}R.Penrose
             {\em Shadows of the mind. }
             Oxford University Press, 1994.

\bibitem{15}W.Unruh
             {\em Time gravity and quantum mechanics. }
             In: Time's arrows today, S.F.Savitt (ed), Cambridge University Press 1995.

\bibitem{16}Ph. Stamp
             {\em Time decoherence and reversible masurements. }
             In: Time's arrows today, S.F.Savitt (ed), Cambridge University 
             Press 1995.

\bibitem{17}A. Legget
             {\em Times arrow and quantum measurement problem. }
             In: Time's arrows today, S.F.Savitt (ed), 
             Cambridge University Press 1995.

\bibitem{18}R. Douglas
             {\em Stochastically brancing space-time topology. }
             In: Time's arrows today, S. F. Savitt (ed),  
             Cambridge University Press 1995.

\bibitem{19}A. P. Balachadran and L. Chandar
             {\em Discrete Time from Quantum Physics. }
             SU-4240-579, May 1994.

\bibitem{20}A. Anderson
             {\em The Global Problem of Time. }
             McGill 92-15, hep-th/9205112, 29 May 1992.

\bibitem{21}R.Gambini and P.Mra
             {\em Intrinsic Time and Evolving Hilbert Spaces in Relational Dynamical Systems and Quantum Gravity. }
            hep-th/ 9404169 11 May 1994 .

\bibitem{22}B.Baumgartner
             {\em Postulates for Time-Evolution in Quantum Mechanics. }
             UWThPh- 1992-56.

\bibitem{23}S.W.Hawking, D.Page and C.Pope
             Nucl.Phys. B 170 (1980) p. 283.

\bibitem{24}A.Aspect and R.Grangier
             {\em Experiments on EPR-type correlations with pairs  of visible photons. }
             In: Quantum Concepts in Space and Time,  R.Penrose and C.J.Isham (eds), Oxford University Press (1986).

\bibitem{25}P.Landsberg
             {\em ime in statistical physics and special relativity.}
             Studium Generale, 23 (1970) p. 1108-58.
             {\em The Enigma of Time.}
             Bristol, Adam Hilger Ltd. 1982.

\bibitem{26}Y. Neeman
             {\em CP and CPT violations, entropy and the expanding universe. }
             International Journal of theoretical Physics 3 (1970) p. 1-5.

\bibitem{27}J.Lebowitz and E.Montroll (eds)
             {\em Non-Equilibrium Phenomena I: The Boltzmann Equation. }
             Amsterdam, North-Holland (1983).

\bibitem{28}F.Pohl
             {\em The World at the End of the Time. }
             N.Y., Ballentine 1990.

\bibitem{29}H.Price
             {\em  A point on the arrow of time. }
             Nature 340 (1989) p. 181-2.

\bibitem{30}P.Yourgrau
             {\em The disapearence of Time. }
             Cambridge University Press 1991.

\bibitem{31}I. Prigogine and C. George
             {\em New quantization rules for dissipative systems. }
             Int.J. Quant. Chem. 12 (Suppl. 1) (1977) p. 177-184.

\bibitem{32}H.Araki
             {\em Von Neumann algebras of local observables for free scalar field. }
             J.Math.Phys. 5 (1964) 1-13.

\bibitem{33}G.Baker
             {\em Self-interacting Boson quantum field theory and the thermodynamic limit in d dimensions. }
             J.Math. Phys. 16 (1975) 1324-1346.

\bibitem{34}N.N.Bogoliubov and O.S.Parasiuk
             {\em Ueber die Multiplikation der Kausalfunktionen in der Quantentheorie der Felder. }
             Acta Math. 97 (1957) 227- 266.

\bibitem{35}J.Feldman
             {\em A relativistic Feynman-Kac formula. }
             Nucl.Physics B 52 (1973) 608-614.

\bibitem{36}J.Froelich
             {\em Classical and quantum statistical mechanics in one and two dimensions: Two component Yukawa and Coulomb systems. }
             Comm. Math.Phys. 47 (1976) 233-268.

\bibitem{37}J.Froelich and K.Osterwalder
             {\em Is there a Eucledian field theory for femions?}
             Helv.Phys.Acta 47 (1974) 781-705.

\bibitem{38}G.Gallavotti and A. Martin-Loef
             Comm.Math.Phys. 25 (1972) 87-126.

\bibitem{39}R.Haag and B.Schroer
             {\em Postulates of quantum field theory. }
             J.Math.Phys. 3 (1962) 248-256.

\bibitem{40}J.Lebowitz and O. Penrose
             {\em  Analytic and clustering properties of
                        thermodynamic functions and distribution functions 
                        for classical lattice and continuoum systems. }
             Comm.Math. Phys. 11 (1968) 99-124.

\bibitem{41}J.S.Bell
             {\em On the Einstein Podolski Rosen paradox. }
             Physics, Vol. I, No 3 (1964) 195-200.

\bibitem{42}C.Aspect, Ph. Grangier and G. Roger
             {\em Experimental realization of the 
                         Einstein-Podolski-Rosen-Bohm Gedankenexperiment : A new
                         violation of Bell's inequalities. }
             Phys. Rev. Lett. Vol. 49, No. 2 (1982) p. 91-94.

\bibitem{43}P.Pearle
             {\em Combining stochastic dynamical state-vector with spontaneous localization. }
             Phys.Rev. A39 (1989)857-868.

\bibitem{44}L.Landau and E.Lifchitz
             {\em Theorie des champs. }
             Editions MIR, Moscou, 1970.

\bibitem{45}J.L.Synge
             {\em Relativity-The general Theory. }
             North-Holland Publ. Co. Amsterdam, (1966) p. 105.

\bibitem{46}R.Engelking
             {\em General Topology. }
             Sigma Series in Pure Mathematics, 
              Haldermann Verlag Berlin (1989) p. 36.

\bibitem{47}C. Syros
             {\em The time concept in atomic and sub-atomic 
              systems-Reconciliation of the time-reversal-invariance 
              and the macroscopic arrow of 
                         time. }
             In: Advances in Nuclear Physics, C.Syros and C.Ronchi                                     (eds),  European Commission, Luxemburg, 
                (1995) p. 242-287.

\bibitem{48}F.S.Savitt
             {\em Times Arrows Today. }
             Cambridge University Press 1995.

\bibitem{49}B.S.de Witt and N.Graham (eds)
             {\em The many Worlds Interpretation of Quantum Mechanics. }
             Princeton University Press 1973.

\bibitem{50}E.Schroedinger
             Berliner Berichte, (1931) p 322.

\bibitem{51}W.Pauli
             {\em Handbuch der Physik. }
             Vol.5, Part I, Springer-Verlag,  Berlin 1958.

\bibitem{52}C.Itzykson and J.-B. Zuber
             {\em Quantum Field Theory. }
             McGraw-Hill International Editions, Physics Series,  1988.

\bibitem{53}J.Glimm and A., Jaffe
             {\em Quantum Physics. }
             Springer-Verlag, N.Y. 1981.
             G.Parisi
             {\em Statistical Field Theory}
             Addison-Wesley Publ.Co. Redwood Calif. 1988.
             L.P.Kadanoff, and G.Baym
             {\em Quantum Statistical Mechanics}
             Addison-Wesley Publ. Co.Inc., The Advanced Book Programm, 
             Redwood City, Calif. 1988. 

\bibitem{54}N.N.Bogoliubov
             {\em Lectures in Quantum Statistics. }
             Vol. I, McDonald Technical and Scientific, London (1967).

\bibitem{55}R.Haag, N.M.Hugenholtz, and M.Winnink
             Commun.Math.Phys. 5 (1967) 215.
             D.Buchholz, H.Eyvind, and H.Wichmann
             Comm.Math.Phys. 106 (1986) 321.

\bibitem{56}I.M.Gelfand and N.Ya. Vilenkin
             {\em Generalized Functions, Vol.4. }
             Academic,  N.Y. (1964) p. 238.

\bibitem{57} C.Syros
             Int. J. Mod. Phys. B Vol. 5 (1991) p. 2909-2934.
             C.Syros
             {\em Die Zeitstruktur von sub-atomaren Teilchen und der
             Zusammenhang mit der makroskopischen Zeit.} 
             Physik der Hadronen und Kerne,  K\"oln, 13-17 Maerz
             1995, p.591.
             C.Syros
             {\em The Space-time Topology of Nuclear and Sub-nuclear Reactions.}
             7-th Symposium of the Hellenic Nuclear Physics Society, 
               Athens, 24-25 June, 1996, in press.

\bibitem{58}F.Karolyhazy
             {\em Gravitation and quantum mechanics macroscopic bodies. }
             Nuovo Cimento, A 42 (1966) 390-402.

\bibitem{59}G.C.Ghirardi, A.Rimini, and T.Weber
             {\em Unified Dynamics for microscopic and macroscopic systems. }
             Phys.Rev D34 (1986), p.470.

\bibitem{60}E.P.Wigner
             In: Quantum theory and measurement, (ed. J.A.Wheeler, and  
              W.H. Zurek ), Princeton University Press, (1983).

\bibitem{61}L.I.Schiff
             {\em Quantum Mechanics. }
             McGraw-Hill Publ.Co., Ltd.London 2nd edition (1955), p.55.

\bibitem{63} S. Hawking ,  and R. Penrose, {\em The nature of spacetime}.
             Princeton University Press, Princeton New Jersey,  (1996).
\end{thebibliography}
\end{document}